\documentclass{llncs}
\usepackage[latin1]{inputenc}
\usepackage[hidelinks]{hyperref}
\newcommand{\unioN}{\cup}
\newcommand{\union}[2]{#1 \unioN #2}

\newcommand{\bi}{\begin{itemize}}
\newcommand{\ei}{\end{itemize}}
\newcommand{\be}{\begin{enumerate}}
\newcommand{\ee}{\end{enumerate}}
\newcommand{\bd}{\begin{description}}
\newcommand{\ed}{\end{description}}
\newcommand{\tab}{\hspace*{1cm}}

\newcommand{\mca}[2]{\multicolumn{#1}{@{}l@{}}{#2}}
\newcommand{\ty}[1]{\raisebox{2mm}{\scriptsize:#1}}
\newcommand{\vs}[2]{{^{#1} \! #2}}
\renewcommand{\.}[1]{\!#1\!}
\renewcommand{\*}{\!\cdot\!}

\newcommand{\subterm}{\mathrel{\sphericalangle}}
\newcommand{\psubterm}{\kackrel{\sphericalangle}{\neq}}

\newcommand{\sortdef}{\doteq}
\newcommand{\corresponds}{\mbox{\^=}}

\newcommand{\by}[1]{\!/\!_{#1}}
\newcommand{\rs}[1]{\hspace{-0.3em}\mid_{\hspace{-0.1em}\raisebox{
	-0.1em}{$\scriptscriptstyle #1$}}}

\newcommand{\ovl}[1]{\overline{#1}}

\newcommand{\3}{\ss}

\newcommand{\display}[1]{
	\begin{center} #1 \end{center}
	}

\newcommand{\ds}{\displaystyle}
\newcommand{\tpl}[1]{\langle #1 \rangle}
\newcommand{\COMMENT}[1]{}

\newcommand{\ra}{\rightarrow}
\newcommand{\la}{\leftarrow}
\newcommand{\lra}{\leftrightarrow}
\newcommand{\raa}{\longrightarrow}
\newcommand{\laa}{\longleftarrow}

\newcommand{\Ra}{\Rightarrow}
\newcommand{\La}{\Leftarrow}
\newcommand{\Lra}{\Leftrightarrow}
\newcommand{\Raa}{\Longrightarrow}

\newcommand{\abl}[2]{
	\begin{picture}(0.37,0.0)
	\put(0,0.05){\makebox(0,0)[l]{$#2$}}
	\put(0.20,0.10){\makebox(0,0)[b]{$\scriptscriptstyle #1$}}
	\end{picture}
	}

\newcommand{\abld}[1]{\mathrel{\abl{}{\ra}^{#1}}}
\newcommand{\abldi}{\abld{}}
\newcommand{\ablds}{\abld*}

\newcommand{\Abld}[1]{\mathrel{\abl{}{\lra}^{#1}}}

\newcommand{\Ablds}{\Abld*}

\newcommand{\ablc}[1]{\mathrel{\abl{c}{\ra}^{#1}}}
\newcommand{\ablci}{\ablc{}}

\newcommand{\Ablc}[1]{\mathrel{\abl{c}{\lra}^{#1}}}

\newcommand{\Ablcs}{\Ablc*}

\newcommand{\ablcd}[1]{\mathrel{\abl{c\!d}{\ra}^{#1}}}
\newcommand{\ablcdi}{\ablcd{}}

\newcommand{\Ablcd}[1]{\mathrel{\abl{c\!d}{\lra}^{#1}}}

\newcommand{\Ablcds}{\Ablcd*}

\newcommand{\pecomp}
	{\mathbin{\mathchoice{\diamond\mkern-7mu\cdot\mkern2mu}
			     {\diamond\mkern-7mu\cdot\mkern2mu}
			     {\diamond\mkern-7mu\cdot\mkern2mu}
			     {\diamond\mkern-7mu\cdot\mkern2mu}}}
\newcommand{\pscomp}
	{\mathbin{\mathchoice{\circ\mkern-7mu\cdot\mkern2mu}
			     {\circ\mkern-7mu\cdot\mkern2mu}
			     {\circ\mkern-7mu\cdot\mkern2mu}
			     {\circ\mkern-7mu\cdot\mkern2mu}}}

\newcommand{\bigpecomp}
     {\Diamond\hspace{-0.52em}\raisebox{0.11em}{$\cdot$}\hspace{0.25em}}

\newcommand{\bigpscomp}
     {\bigcirc\hspace{-0.52em}\raisebox{0.11em}{$\cdot$}\hspace{0.25em}}

\newcommand{\N}{I\!\!N}
\renewcommand{\phi}{\varphi}
\newcommand{\bigmid}{\rule[-0.07cm]{0.04cm}{0.4cm}\hspace*{0.1cm}}

\newcommand{\psubset}{\subsetneqq}

\renewcommand{\leq}[0]{\leqslant}
\renewcommand{\geq}[0]{\geqslant}

\newsavebox{\greybox}
\newlength{\greygrain}
\newcommand{\grey}[2]{
\savebox{\greybox}{\multiput(0,0)(0.1,0){#1}
					{\rule{\greygrain}{\greygrain}}}
\multiput(0,0)(0,0.1){#2}{\usebox{\greybox}}
\multiput(0.05,0.05)(0,0.1){#2}{\usebox{\greybox}}
}

\newlength{\outerparindent}
\newcommand{\cit}[2]{\cite{#2}}

\newcommand{\func}[2]{\mathchoice{(#1\.\ra#2)}
				{(#1\.\ra#2)}
				{(#1\ra#2)}
				{(#1\ra#2)}}
\newcommand{\pfunc}[2]{\mathchoice{(#1\.\hookrightarrow#2)}
				{(#1\.\hookrightarrow#2)}
				{(#1\hookrightarrow#2)}
				{(#1\hookrightarrow#2)}}
\newcommand{\TT}{{\cal T}^*}
\spnewtheorem{algorithm}{Algorithm}[theorem]{\bfseries}{\itshape}

\usepackage{amssymb}
\usepackage{jb}

\newcommand{\m}{{\scriptstyle +}}
\newcommand{\ptm}{{\scriptstyle \oplus}}
\newcommand{\eqr}[1]{\ref{#1}}
\newcommand{\eqR}[1]{\ref{#1}}
\newcommand{\eqd}[1]{\label{#1}}
\newcommand{\eqda}[1]{{#1}}
\newcommand{\eqra}[1]{{#1}}
\newcommand{\eqRa}[1]{{#1}}

\newcommand{\eqi}[1]{#1}

\newcommand{\EQi}[1]{#1}

\renewcommand{\lg}{\stackrel{\mbox{\Large\bf .}}{<}}

\renewcommand{\arraystretch}{1.1}

\newlength{\thmspace}
\newcounter{thm}

\newcommand{\THM}[2]{%
	\par%
	\vspace{\thmspace}%
	\refstepcounter{thm}%
	\noindent%
	{#1~\thethm. }%
	#2%
	\par%
	\vspace{\thmspace}%
	}

\newcommand{\PRF}[2]{%
	\par%
	{#1}%
	#2%
	\par%
	\vspace{\thmspace}%
	}

\newcommand{\EXAMPLE}[1]{\THM{\bf Example}{#1}}
\newcommand{\ALGORITHM}[1]{\THM{\bf Algorithm}{#1}}
\newcommand{\COROLLARY}[1]{\THM{\bf Corollary}{#1}}
\newcommand{\LEMMA}[1]{\THM{\bf Lemma}{#1}}
\newcommand{\THEOREM}[1]{\THM{\bf Theorem}{#1}}
\newcommand{\DEFINITION}[1]{\THM{\bf Definition}{#1}}
\newcommand{\PROOF}[1]{\PRF{\it Proof. }{#1}}

\begin{document}

\title{Regular Substitution Sets: A Means of Controlling
E-Unification
}

\author{Jochen Burghardt}

\institute{GMD Berlin,
\\
jochen@first.gmd.de
\\
http://www.first.gmd.de/persons/Burghardt.Jochen.html
}

\maketitle

\vfill

{\sf

\begin{tabular}[t]{@{}ll@{\hspace*{0.5cm}}l@{}}
Technical Report	\\
Arbeitspapiere der GMD 926	\\
July 1995	\\
ISSN 0723--0508 \\
\end{tabular}
\hfill
\begin{tabular}[t]{@{}ll@{\hspace*{0.5cm}}l@{}}
\mca{2}{GMD -- Forschungszentrum}	\\
\mca{2}{Informationstechnik GmbH}	\\
\mca{2}{D--53754 Sankt Augustin}	\\
Tel. & *49--2241--14--0 \\
Fax & *49--2241--14--2618	\\
Telex & 889469 gmd d	\\
\mca{2}{http://www.gmd.de}	\\
\end{tabular}

}

\newpage

\setcounter{page}{3}

\begin{abstract}
A method for selecting solution constructors in narrowing is
presented. 
The method is based on a sort discipline that describes
regular sets of ground constructor terms as sorts. 
It is extended to cope with regular sets of ground
substitutions, thus allowing different
sorts to be computed for terms with different variable bindings. 
An algorithm for {\em computing} signatures of equationally
defined functions is given that allows potentially
infinite overloading. 
Applications to formal program development are sketched.
\end{abstract}

\newpage

$\;$

\newpage

\section{Motivation}
\label{Motivation}

Solving equations by narrowing has important applications,
e.g.\ in the area of formal software development.
However, the usual narrowing strategies are only able to restrict the
set of  application positions\footnote{%
	Cf.\ e.g.\ the mathematical definition of the notion of
	strategy in \cit{Echahed 1988}{Echahed.1988}.}. 
Ordered paramodulation \cit{Bachmair 1990}{Bachmair.Ganzinger.1990a}
is able to provide a
succession in which the defining equations have to be selected,
but it cannot guarantee that an appropriate one is selected first.
Bockmayr \cit{1991}{Bockmayr.1991}
has shown that, under certain general conditions,
narrowing strategies essentially enumerate the whole term universe
rather than specifically
selecting the appropriate equations of a defined
function to narrow with or the appropriate constructor to insert
into the solution.
In this paper, we present an approach for
restricting the set of applicable
defining equations in a narrowing step that is based on the dynamic
computation of function signatures, rather than their declaration by a
user.

The main idea is as follows\footnote{%
	Notations and naming conventions are consistent with
	Def.~\eqr{1} below.
	}:
As e.g.\ in \cit{Fribourg 1984}{Fribourg.1984},
we distinguish between constructors and
equationally defined functions; each well-defined ground term can be
reduced to a ground constructor term, viz.\ its unique normal form.
For a term $v$, let $V$ be the set of all possible values of $v$, i.e.,
the set of all normal forms of admitted
ground constructor instances of $v$.
Then, a goal equation $v_1 = v_2$ cannot be solved if $V_1 \cap V_2 =
\{\}$; in this case, it can be pruned from the search space of
narrowing.
Unfortunately, $V_1$ and $V_2$ are undecidable in general;
to overcome this problem, we will define computable upper approximations
$\ovl{V}_1 \supset V_1$ and $\ovl{V}_2 \supset V_2$, respectively, and
base the pruning decision on the consideration of $\ovl{V}_1 \cap
\ovl{V}_2$.

To this end, we
provide a framework of ``extended sorts'' to describe infinite sets
of ground constructor terms like $\ovl{V}$ in a closed form, which is
based on regular tree grammars
(e.g.\ \cit{Thatcher 1968}{Thatcher.Wright.1968}).
It is essential that extended sorts are closed wrt.\ intersection and
that their inhabitance can be decided in order to conduct the above
disjointness test.
Moreover, set equality and subsort property can be decided, and
$\ovl{V} = V$ always holds if $v$ is a constructor term. 

An algorithm for computing the extended sort $\ovl{V}$ from a term
$v$ is presented.
In terms of conventional order-sorted rewriting, we thereby
achieve potentially
infinite overloading, since for an arbitrary input sort $S$ we can {\em
compute} a signature \mbox{$f:S \ra \ovl{f[S]}$} rather than being
restricted to a few user-defined signatures which are generally
too coarse
for the disjointness test to be successfully applied.
It is clear that the impact of this test on search-space
reduction depends on the expressiveness of the sort framework and on the
quality of signature approximation.

Consider, for example, the theory comprising equations
{\it a.}\ to {\it i.} in
Fig.~\ref{Sort and function definitions for synthesis of binary
arithmetic algorithms}.
When trying to solve a goal equation like $val(x) = s^5(0)$ wrt.\ this
theory, conventional strategies are unable to decide which of the
equations {\it g.}, {\it h.}, {\it i.}
is to be used for a first narrowing step.
Narrowing (at root position) with equation 
{\it g.}, {\it h.}, and {\it i.}
results in the new goal equations
$0 = s^5(0)$,
$dup(val(x')) = s^5(0)$,
and $s(dup(val(x'))) = s^5(0)$,
respectively.
While the first one is obviously false,
the unsatisfiability of the second one can be detected as our algorithm
computes the sort of
its left-hand side as $Even$ and recognizes that this
is disjoint from
its right-hand side's sort, $\{s^5(0)\}$;
similarly,
the third one is considered to be ``possibly satisfiable'' by the
disjointness test.
Hence, narrowing only makes sense with equation {\it i.},
and any solution to the above goal equation must take the form 
$x = x' \.{::} i$.
In Sect.~\ref{Application in Formal Program Development}
and App.~\ref{Case Study ``Comb Vector Construction''}, examples of
the pruning of infinite search-tree branches are given.
Note that if a user were to declare the signatures
$+\.:Nat \.\times Nat \.\ra Nat$, $dup\.:Nat \.\ra Nat$, 
and $val\.:Bin \.\ra Nat$,
the disjointness test
would allow narrowing with equations {\it h.} and {\it i}.
In more complicated applications, a user cannot know in advance which
signatures might become essential to disjointness tests in the
course of the narrowing proof.

This example also shows that it is important to consider variable
bindings during the computation of a term's sort in order to get good
approximations.
For example, when computing a signature for $dup$,
the term $x+x$ should be assigned the sort $Even$, whereas
$x+y$ can only be assigned $Nat$, assuming that $x$ and $y$
range over $Nat$. 
In conventional order-sorted
approaches, the mapping from a term to its sort is
usually a homomorphic extension of the sort assignment of variables,
thus necessarily ignoring variable bindings, e.g.:
	\display{
	\begin{tabular}[t]{@{}l@{$\;$}ll@{}}
	& $sortof(x+x)$	\\
	$=$ & $get\_range\_from\_signatures(+,sortof(x),sortof(x))$ \\
	$=$ & $get\_range\_from\_signatures(+,sortof(x),sortof(y))$ \\
	$=$ & $sortof(x+y)$.	\\
	\end{tabular}
	}
Instead, we use infinite sets of ground substitutions to denote
sorts of variables, 
e.g.\ $\{[x \.{:=} s^i(0),y \.{:=} s^j(0)] \mid i,j \in \N\}$ to
indicate that $x$ and $y$ range over $Nat$.
The mapping from a term to its set of possible values
can then be achieved by
applying each element of the substitution set, e.g.:
	\display{
	\begin{tabular}[t]{@{}l@{$\;$}ll@{}}
	& $\{[x \.{:=} s^i(0)] \mid i \in \N\} \;\;\; (x+x)$	\\
	$=$ & $\{[x \.{:=} s^i(0)] \; (x+x) \;\mid\; i \in \N\}$ \\
	$=$ & $\{s^i(0)+s^i(0) \mid i \in \N\}$.	\\
	\end{tabular}
	}
Similarly,
$\{[x \.{:=} s^i(0),y \.{:=} s^j(0)] \mid i,j \in \N\} \; (x+y)
= \{s^i(0)+s^j(0) \mid i,j \in \N\}$.
Both sets are different, hence the chance of finding different
approximations for them within our extended sort framework is
not forfeited\footnote{%
	Schmidt-Schau\3 \cit{1988}{Schmidt.1988}
	admits ``term declarations'', allowing the user to declare
	different sorts for terms with different bindings. In our
	approach, however, the sorts are to be computed automatically.
	}.
In Fig.~\ref{Range sort computation for $x+x$},
we show that $Even$ can in fact be obtained
as the sort of $x+x$; obtaining $Nat$ for $x+y$ is similar.

In order to have finite descriptions of such ground substitution sets,
we express ground substitutions as ground constructor terms
(``t-substitutions'') in a lifted algebra, allowing 
sets of them to be treated as tree languages (``t-sets''), and, in
particular, to be described by regular tree grammars (``regular
t-sets'').
Regular 
t-sets can also express simple relations between distinct variables,
allowing e.g.\ the representation of certain conditional equations
by unconditional ones.

We provide a new
class of tree languages, called ``extended sorts'', which
can be described by applying substitutions from a regular t-set
$\sigma$ to an arbitrary constructor term $u$ with $vars(u) \subset
dom(\sigma)$.
In this way, the set of ground-constructor instances of an arbitrary
constructor term can be expressed as an extended sort.

Regular string languages have been used e.g.\ by 
Mishra \cit{1984}{Mishra.1984}
as a basis for sort inference on Horn clauses.
Owing to the restriction to {\em string}
languages describing admissible paths in term trees,
he is only able to express infinite sets that are closed wrt.\ all
constructors;
e.g.\ the set of all lists of naturals containing
at least one $0$ cannot be modeled.
Comon \cit{1990}{Comon.1990}
uses regular tree languages to describe sets of ground
constructor terms as
sorts, and the corresponding automaton constructions to implement sort
operations.
He provides a transformation system to
decide first-order formulas with
equality and sort membership as the only predicates.
He shows the decision of inductive reducibility as an application.
However, he does not consider equationally defined
functions, e.g.\
$(\forall x,y \;\; x\.+y = y\.+x) \ra 0\.+1 = 1\.+0$
reduces to
$(\forall x,y \;\; x\.+y = y\.+x) \ra false$
in his calculus.

Uribe \cit{1992}{Uribe.1992} provides a unification algorithm for
order-sorted terms in the presence of semilinear term 
declarations.
The set of all ground constructor instances of a constructor term can
then be described by a regular tree automaton with equality tests for
direct subterms;
allowing equality tests for arbitrary subterms makes the disjointness of
two tree languages undecidable \cit{Tommasi 1991}{Tommasi.1991}.
In our approach, arbitrary equality constraints may be imposed on
subterms up to a fixed finite depth, whereas below that depth no
equality constraints are allowed at all.
Antimirov \cit{1995}{Antimirov.1995} suggested allowing regular t-sets
with equality tests in extended sorts, thus extending the class of
describable tree languages.
This approach still remains to be investigated.

This paper is organized as follows.
After a short introduction on regular
sorts in Sect.~\ref{Regular Sorts},
regular substitution sets and extended sorts
are presented in
Sect.~\ref{T-Substitutions} -- \ref{Extended Sorts}.
In Sect.~\ref{Equational Theories}, the algorithm for computing
signatures of equationally defined functions is given.
It is shown that an unsorted root-narrowing calculus from
\cit{H\"olldobler 1989}{Holldobler.1989}
remains complete if extended by appropriate sort restrictions.
Section \ref{Application in Formal Program Development} sketches the
application of narrowing to synthesize programs from formal
specifications.
Appendices \ref{Case Study ``Binary Arithmetic''} and \ref{Case Study
``Comb Vector Construction''} contain two case studies in program
synthesis.
For a full version including all proofs, 
see \cit{Burg\-hardt 1993}{Burghardt.1993}.

\section{Regular Sorts}
\label{Regular Sorts}

\DEFINITION{
\eqd{1}

Let \eqi{${\cal V}$} be a countable set of \eqi{variables},
\eqi{${\cal CR}$} a finite set of term \eqi{constructor symbols}, each
with fixed arity, 
\eqi{${\cal F}$} a finite set of symbols for 
\eqi{non-constructor functions}, and 
\eqi{${\cal S}$} a countable set of \eqi{sort names}.
Let \eqi{$ar(g)$} denote the \eqi{arity} of a function symbol $g$.
For a set\footnote{%
	``\eqi{$\subset$}'' denotes subset or equality,
	``\eqi{$\psubset$}'' denotes proper subset. }
of symbols
$X \subset {\cal V} \cup {\cal CR} \cup {\cal F} \cup {\cal S}$,
let \eqi{${\cal T}_X$} be the set of terms formed of symbols from $X$;
we abbreviate ${\cal T}_{X \cup Y}$ to \eqi{${\cal T}_{X,Y}$}.
For example,
the elements of \eqi{${\cal T}_{\cal CR}$}, 
\eqi{${\cal T}_{{\cal CR},{\cal V}}$},
and \eqi{${\cal T}_{{\cal CR},{\cal F},{\cal V}}$}
are called \eqi{ground constructor terms}, \eqi{constructor terms},
and \eqi{terms}, respectively;
the set \eqi{${\cal T}_{{\cal CR},{\cal F},{\cal V},{\cal S}}$}
is introduced in Sect.~\ref{Equational Theories}
for technical reasons. 
Let identifiers like $\eqi{u},u',u_i,\ldots$ always
denote members of ${\cal T}_{{\cal CR},{\cal V}}$;
similarly, $\eqi{v} \in {\cal T}_{{\cal CR},{\cal F},{\cal V}}$,
$\eqi{w} \in {\cal T}_{{\cal CR},{\cal F},{\cal V},{\cal S}}$,
$\eqi{x},\eqi{y},\eqi{z} \in {\cal V}$,
$\eqi{f} \in {\cal F}$,
$\eqi{g} \in {\cal F} \cup {\cal CR}$,
$\eqi{cr} \in {\cal CR}$,
and $\eqi{S} \in {\cal S}$.
}

\DEFINITION{
\eqd{2}
\eqi{$\tpl{v_1,\ldots,v_n}$} denotes an $n$-\eqi{tuple},
\eqi{$\tpl{v_i \mid p(v_i), \; i=1,\ldots,n}$} 
denotes a tuple containing each
$v_i$ such that $p(v_i)$ holds.
We assume the existence of at least one nullary (e.g.\ $nil$) and one
binary constructor (e.g.\ $cons$), so we can model arbitrary tuples as
constructor terms.
% In Sect.~\ref{Regular Substitution Sets}, we will model partial
% operations as operations on sets with at most one element ($\{\}$
% $\corresponds$ undefined\footnote{%
% 	We use ``$\corresponds$'' to denote an informal
% 	correspondence.}).
To improve readability, we sometimes write the application of a unary
function $f$ to its argument $x$ as \eqi{$f \* x$};
\eqi{$x\ty{S}$}
stands, in the following, for the variable or constant $x$
of sort $S$.
We define the
\eqi{elementwise extension}
of a function $f:A \ra B$ to a set $A' \subset A$
by $\eqi{f[A']} := \{ f(a) \mid a \in A'\}$.
\eqi{$A \times B$} denotes the Cartesian product of sets $A$ and $B$.
For a finite set $A$, we denote its cardinality by $\#A$.
We tacitly extend notations like $\bigcup_{i=1}^n A_i$ to several 
binary operators defined in this paper, 
e.g.\ $\eqi{\bigmid_{i=1}^n S_i} := S_1 \mid \ldots \mid S_n$.
}

\DEFINITION{
\eqd{3}
Let \eqi{$vars(v_1,\ldots,v_n)$} denote the set of variables occurring
in any of the terms $v_i$.
A term is called {\eqi{linear}} if it contains no multiple
occurrences of the same variable;
it is called \eqi{pseudolinear}, if any two occurrences of the same
variable are at the same depth;
it is called \eqi{semilinear} if, for any two occurrences of the
same variable, the lists of function symbols on each path from
the root to an occurrence are equal.
We write $\eqi{v_1 \psubterm v_2}$ to express 
that $v_1$ is a proper subterm of $v_2$;
we write \eqi{$v_1 \subterm v_2$}
for $v_1 \psubterm v_2 \;\vee\; v_1 = v_2$.
The \eqi{depth} of a position in a term is its distance from the root.
We distinguish between ``ordinary'' substitutions, defined as usual
(denoted by \eqi{$\beta}, \eqi{\gamma}, \ldots$), and
``t-substitutions'', defined as constructor terms in 
Sect.~\ref{T-Substitutions}, and denoted by $\sigma', \tau', \ldots$.
Application of a substitution $\beta$ to a term $v$ is written in
prefix form, i.e.\ \eqi{$\beta v$}.
For an \eqi{ordinary substitution} $\beta$,
let $\eqi{dom(\beta)} := \{x \in {\cal V} \mid \beta x \neq x\}$,
and $\eqi{ran(\beta)} := \bigcup_{x \in dom(\beta)} vars(\beta x)$.
\eqi{$[x_1 \.{:=} v_1,\ldots,x_n \.{:=} v_n]$} 
denotes the \eqi{substitution} that maps each $x_i$ to $v_i$.
\eqi{$\beta \rs V$} denotes the domain \eqi{restriction} of $\beta$
to a set $V$ of variables.
We assume all substitutions to be idempotent.
If $\beta_1$ and $\beta_2$ agree on the intersection of their domains,
\eqi{$\beta_1 \pscomp \beta_2$} denotes a 
``\eqi{parallel composition}'' of them,
i.e.\
\display{
	$(\beta_1 \pscomp \beta_2) \; (x) := 
	\left\{		%}
	\begin{tabular}{@{}l@{\hspace*{0.5cm}}l@{}}
	$\beta_1 x$ & if $x \in dom(\beta_1)$	\\
	$\beta_2 x$ & if $x \in dom(\beta_2)$	\\
	\end{tabular}
	\right.$
}

$\beta_1 \pscomp \beta_2$ is undefined if $\beta_1$ and $\beta_2$ do
not agree on $dom(\beta_1) \cap dom(\beta_2)$.
A substitution $\beta$ is called {\eqi{linear}} if the term 
$\tpl{\beta x_1,\ldots,\beta x_n}$ is linear, 
where $\{x_1,\ldots,x_n\} = dom(\beta)$;
similarly, $\beta$ is called \eqi{pseudolinear} 
if $\tpl{\beta x_1,\ldots,\beta x_n}$ is pseudolinear.
We use the common notions of \eqi{renaming substitution}
and \eqi{most general unifier} $\beta = \eqi{mgu(v_1,v_2)}$,
however,
we will additionally assume that $v_1$ and $v_2$ have disjoint
variables and write $\beta$ as $\beta_1 \pscomp \beta_2$ with
$dom(\beta_1) = vars(v_1)$ and $dom(\beta_2) = vars(v_2)$.
$mgu$ is tacitly extended to finite sets of terms.
}

We follow the approach of \cit{Burghardt 1993}{Burghardt.1993} in
describing regular sets of ground constructor terms
as fixed points of sort equations, which is
equivalent to the approach using finite tree automata 
\cit{Comon 1990}{Comon.1990},
but provides a unique methodology for algorithms and
proofs.

\DEFINITION{
%[(Sort System)] 
\eqd{4}
We allow \eqi{sort definitions} of the following syntax:
	\display{
	\begin{tabular}{@{}l@{\hspace*{1em}}l@{}}
	$\mathrm{\eqi{SortName} \eqi{\sortdef} 
		SortName \eqi{\mid}\ldots\mid SortName}$,	\\
	$\mathrm{SortName \sortdef Constructor(SortName,
			\ldots,SortName)}$ \\
	\end{tabular}
	}
Let \eqi{$\lg$} be the transitive closure of the relation
$S_i \eqi{\stackrel{\mbox{\Large\bf ..}}{<}} S
:\Lra S \sortdef S_1 \mid \ldots \mid S_n$.
We admit finite systems of sort definitions such that $\lg$ is an
irreflexive partial, hence well-founded, order.
For example,
the sort system consisting of $A \sortdef B$ and $B \sortdef A$
is forbidden.
Each occurring sort name has to be defined.
In examples, we generally
use arbitrary {\eqi{sort expressions}} built from
sort names, constructors, and ``$\mid$'' on the right-hand side
of a sort definition.
Any such \eqi{sort system}
can be transformed to meet the above requirements
while maintaining the least-fixed-point semantics given below.

For example, consider the sort definition
$\eqi{Bin} \sortdef nil \mid Bin \.{::} o \mid Bin \.{::} i$ from
Fig.~\ref{Sort and function definitions for synthesis of binary
arithmetic algorithms} on page~\pageref{Sort and function definitions
for synthesis of binary arithmetic algorithms},
which denotes the lists of binary digits,
where ``$\eqi{o}$'' denotes zero,
``$\eqi{i}$'' denotes one, and $\eqi{::}$ is an infix-$\eqi{snoc}$,
i.e.\ reversed cons.
The sort definition can be transformed into the corresponding
definitions shown in Fig.~\ref{Examples of sort definitions}, which
obey Def.~\eqr{4}, by introducing new auxiliary sort names $Nil$,
$Bino$, $Bini$, $O$, and $I$.

Let $X$ be an arbitrary
mapping from sort names $S$ to subsets \eqi{$S^X$}
of ${\cal T}_{\cal CR}$.
$X$ is extended to sort expressions as follows:
	\display{
	\begin{tabular}[t]{@{}r@{$\;$}l@{}}
	$(S_1 \mid S_1)^X$ 
		& $= S_1^X \cup S_2^X$	\\
	$cr(S_1,\ldots,S_n)^X$
		& $= cr[S_1^X \times \ldots \times S_n^X]$	\\
	$cr^X$ & $= \{cr\}$	\\
	\end{tabular}
	}

We say that $X_1 \subset X_2$ if $S^{X_1} \subset S^{X_2}$ for all sort
names $S$.
According to Thm.~\eqR{5} below, for each admitted
system of sort definition there
exists exactly one mapping \eqi{$M$}, such that $S^M = S'^M$ for
each sort definition $S \sortdef S'$.
The \eqi{semantics} of a sort expression $S$ is then defined as
\eqi{$S^M$}.
}

\THEOREM{
\eqd{5}
Each admitted system of sort definitions has exactly one fixed point.
}
\PROOF{
If $M$ and $M'$ are fixed points of the sort definitions,
use induction on the the lexicographic combination of
$\subterm$ and $\lg$
to show $\forall u \; \forall S \;\;\; u \in S^M \Ra u \in S^{M'}$.
}

\DEFINITION{
\eqd{6}
For a sort name $S$, let \eqi{$use(S)$}
denote the set of all sort names that
occur directly or indirectly in the definition of $S$.
For example, $use(Bin) = \{O, I, Bin, Bino, Bini\}$, cf.\ 
Fig.~\ref{Examples of sort definitions} 
and \ref{Induction principle for sort $Bin$}.
}

A subset $T \subset {\cal T}_{\cal CR}$ is called
{\eqi{regular}} if a 
system of sort definitions exists, such that $T = S^M$ for some sort
expression $S$.
Note that $u^M = \{u\}$ for all $u \in {\cal T}_{\cal CR}$,
e.g., $s(0)^M = \{s(0)\}$.
The empty sort is denoted by $\bot$; it can be defined e.g.\ by
$\bot \sortdef s(\bot)$.
The uniqueness of fixed points validates
the following induction principle,
which is used in almost all correctness proofs of sort
algorithms, cf.\ Alg.~\eqR{10},
\eqR{11},
\eqR{37},
\eqR{40},
\eqR{41},
\eqR{42}, and
\eqR{47}.

\THEOREM{
%[(\eqi{Induction Principle})] 
\eqd{7}
%$\;$\\
Let $p$ be a family of unary predicates, indexed over the set of all
defined sort names.
Show for each defined sort name $S$:
	\display{
	\begin{tabular}[t]{@{}l@{$\;$}ll@{}}
	$\forall u \.\in {\cal T}_{\cal CR} \;\; p_S(u) \lra$
		& $p_{S_1}(u) \vee \ldots \vee p_{S_n}(u)$
		& if $S \sortdef S_1 \mid \ldots \mid S_n$ \\
	$\forall u \.\in {\cal T}_{\cal CR} \;\; p_S(u) \lra$
		& {$\exists u_1,\ldots,u_n
		\in {\cal T}_{\cal CR} \;\;
		u \.= cr(u_1,\ldots,u_n)$}	\\
		& $\wedge p_{S_1}(u_1) \wedge \ldots \wedge
		p_{S_n}(u_n)$ \tab
		& if $S \sortdef cr(S_1,\ldots,S_n)$	\\
	$\forall u \.\in {\cal T}_{\cal CR} \;\; p_S(u) \lra$
		& $u \.= cr$
		& if $S \sortdef cr$	\\
	\end{tabular}
	}
Then, $\forall u \.\in {\cal T}_{\cal CR} \;\; u \in S^M \lra p_S(u)$
holds for each defined sort name $S$.
}
\PROOF{
The mapping $S \mapsto \{u \in {\cal T}_{\cal CR} \mid p_S(u) \}$
is a fixed point of the sort definitions, hence the only one
by Thm.~\eqr{5}.
}

\THEOREM{
\eqd{8}
Let $p$ be a family of unary predicates, indexed over the set of all
defined sort names.
Show for each defined sort name $S$:
	\display{
	\begin{tabular}[t]{@{}lll@{\hspace*{0.5cm}}l@{}}
	$\forall u \.\in {\cal T}_{\cal CR}$
		& $p_S(u)$
		& $\la p_{S_1}(u) \vee \ldots \vee p_{S_n}(u)$
		& if $S \sortdef S_1 \mid \ldots \mid S_n$ \\
	$\forall u_1,\ldots,u_n \.\in {\cal T}_{\cal CR} \;\;$
		& $p_S(cr(u_1,\ldots,u_n))$
		& $\la p_{S_1}(u_1) \wedge \ldots \wedge p_{S_n}(u_n)$
		& if $S \sortdef cr(S_1,\ldots,S_n)$	\\
	$\forall u \.\in {\cal T}_{\cal CR}$
		& $p_S(cr)$
		&& if $S \sortdef cr$	\\
	\end{tabular}
	}
Then, $\forall u \in S^M \;\;\; p_S(u)$
holds for each defined sort name $S$.
}
\PROOF{
Use Scott's fixed-point induction.
The Thm.\ remains valid even if $\lg$ is not irreflexive.
}

\COROLLARY{
\eqd{9}
For each sort name $S$, we provide the following structural induction
principle:
show for each sort definition
$S' \sortdef cr(S'_1,\ldots,S'_n)$
such that $S' \in use(S)$,
and $S'^M \subset S^M$:
	\display{
	$\ds \forall u_1,\ldots,u_n \in {\cal T}_{\cal CR} \;\;\;
	(\bigwedge_{\shortstack{$\scriptstyle i=1 \ldots n$	\\
				$\scriptstyle S'^M_i \subset S^M$}}
	u_i \in S'^M_i \wedge p(u_i))
	\raa p(cr(u_1,\ldots,u_n))$
	}
Then, $\forall u \in S^M \;\; p(u)$ holds.
$S'^M \subset S^M$ can be decided using Alg.~\eqR{12} below.
}
\PROOF{
Use Thm.~\eqr{8} with
$p_{S'}(u) :\Lra \left\{	%}
\begin{tabular}{l@{\hspace{0.5cm}}l}
$p(u)$ & if $S' \in use(S)$ and $S'^M \subset S^M$	\\
$true$ & else	\\
\end{tabular}
\right.$ .
}

\begin{figure}
\begin{center}
\begin{tabular}{@{}|l@{$\;$}l@{\hspace{0.5cm}}l@{\hspace{0.5cm}}l|@{}}
\hline
$Bin$ & $\sortdef Nil \mid Bino \mid Bini$
	&& binary numbers, i.e.\ lists of binary digits	\\
$Bin^{0}$ & $\sortdef Nil \mid Bino^{0}$
	&& binary numbers that contain no ones	\\
$Bin^{n+1}$ & $\sortdef Nil \mid Bino^{n+1} \mid Bini^{n}$ 
	& for $n \geq 0$
	& binary numbers that contain at most $n+1$ ones	\\
$Bin_{0}$ & $\sortdef Bin$
	&& binary numbers that contain at least $0$ ones	\\
$Bin_{n+1}$ & $\sortdef Bino_{n+1} \mid Bini_{n}$
	& for $n \geq 0$
	& binary numbers that contain at least $n+1$ ones	\\
$Bino$ & $\sortdef Bin \.{::} O$
	&& binary numbers with a trailing zero	\\
$Bini$ & $\sortdef Bin \.{::} I$
	&& binary numbers with a trailing one	\\
$Bino^n$ & $\sortdef Bin^n \.{::} O$ &&	\\
$Bini^n$ & $\sortdef Bin^n \.{::} I$ &&	\\
$Bino_n$ & $\sortdef Bin_n \.{::} O$ &&	\\
$Bini_n$ & $\sortdef Bin_n \.{::} I$ &&	\\
$Nil$ & $\sortdef nil$ && empty list	\\
$O$ & $\sortdef o$ && zero-digit	\\
$I$ & $\sortdef i$ && one-digit	\\
\hline
\end{tabular}
\end{center}

\caption{Examples of sort definitions}
\label{Examples of sort definitions}
\end{figure}

\begin{figure}
\begin{center}
\begin{tabular}{@{}|l@{}l@{}l@{}r@{}c@{}r@{}c|@{}}
\hline
&& \multicolumn{5}{r|@{}}
	{\small $S'^M \.\subset Bin^M$} \\
&& \multicolumn{3}{r@{}}
	{\small $S' \.\sortdef cr( \ldots )$} && \vline \\
&& \small $S' \.\in use(Bin)$
	&& \small $\downarrow$ && \small $\downarrow$	\\
\cline{3-3} \cline{5-5} \cline{7-7}
& \tab & \small $O$ & \hspace*{0.1cm} & \small $+$
	& \hspace*{0.1cm} & \small $-$ \\
&& \small $I$ && \small $+$ && \small $-$ \\
&& \small $Bin$ && \small $-$ && \small $+$ \\
$p(nil)$ && \small $Nil$ && \small $+$ && \small $+$ \\
$\forall u_1 \;\; u_1 \in Bin^M \wedge p(u_1) \ra
	p(u_1\.{::}o)$
	&& \small $Bino$
	&& \small $+$ && \small $+$ \\
$\forall u_1 \;\; u_1 \in Bin^M \wedge p(u_1) \ra
	p(u_1\.{::}i)$
	&& \small $Bini$
	&& \small $+$ && \small $+$ \\
\cline{1-1}
$\forall u \.\in Bin^M \;\; p(u)$ &&&&&&	\\
\hline
\end{tabular}
\end{center}

\caption{Induction principle for sort $Bin$}
\label{Induction principle for sort $Bin$}
\end{figure}

Figure~\ref{Induction principle for sort $Bin$} shows an induction
principle for sort $Bin$ from Fig.~\ref{Examples of sort definitions},
using Cor.~\eqr{9}.
Algorithms for computing the \eqi{intersection} and the
\eqi{relative complement} of two regular sorts, as well as for deciding
the \eqi{inhabitance} of a sort, and thus of the \eqi{subsort}
and \eqi{sort equivalence} property, are given below.
They consist essentially of
\eqi{distributivity rules}, \eqi{constructor-matching rules},
and \eqi{loop-checking rules}. The latter stop the algorithm,
when it calls itself recursively with the same arguments, and
generate a corresponding new recursive sort definition.

\begin{figure}
\begin{center}
\begin{tabular}[t]{@{}|r@{$\;$}l@{$\;$}ll|@{}}
\hline
$inf(Bin^{2},Bin_{1})$ & $= Sort_\eqda{1}$ 
	& $\sortdef inf(Nil,Bin_{1})
	\mid inf(Bino^{2},Bin_{1})
	\mid inf(Bini^{1},Bin_{1})$ & by 2.	\\
$inf(Nil,Bin_{1})$ & $= Sort_\eqda{2}$
	& $\sortdef inf(Nil,Bino_{1}) \mid inf(Nil,Bini_{0})$ & by 3. \\
$inf(Bino^{2},Bin_{1})$
	& $= Sort_\eqda{3}$
	& $\sortdef inf(Bino^{2},Bino_{1}) \mid inf(Bino^{2},Bini_{0})$
	& by 3.	\\
$inf(Bini^{1},Bin_{1})$
	& $= Sort_\eqda{4}$
	& $\sortdef inf(Bini^{1},Bino_{1}) \mid inf(Bini^{1},Bini_{0})$
	& by 3.	\\
$inf(Nil,Bino_{1})$ & $= Sort_\eqda{5}$ & $\sortdef \bot$ & by 5. \\
$inf(Nil,Bini_{0})$ & $= Sort_\eqda{6}$ & $\sortdef \bot$ & by 5. \\
$inf(Bino^{2},Bino_{1})$ 
	& $= Sort_\eqda{7}$
	& $\sortdef inf(Bin^{2},Bin_{1}) \.{::} inf(O,O)$
	& by 4.	\\
$inf(Bino^{2},Bini_{0})$
	& $= Sort_\eqda{8}$
	& $\sortdef inf(Bin^{2},Bin_{0}) \.{::} inf(O,I)$
	& by 4.	\\
$inf(Bini^{1},Bino_{1})$
	& $= Sort_\eqda{9}$
	& $\sortdef inf(Bin^{1},Bin_{1}) \.{::} inf(I,O)$
	& by 4.	\\
$inf(Bini^{1},Bini_{0})$
	& $= Sort_\eqda{10}$
	& $\sortdef inf(Bin^{1},Bin_{0}) \.{::} inf(I,I)$
	& by 4.	\\
$inf(Bin^2,Bin_1)$ & $= Sort_\eqra{1}$ && by 1.	\\
$inf(O,O)$ & $= Sort_\eqda{11}$ & $\sortdef O$ & by 4.	\\
$inf(Bin^2,Bin_0)$ & $= Sort_\eqda{12}$ & $\sortdef Bin^2$ 
	by similar computations &	\\
$inf(O,I)$ & $= Sort_\eqda{13}$ & $\sortdef \bot$ & by 5.	\\
$inf(Bin^1,Bin_1)$ & $= Sort_\eqda{14}$ & $\sortdef \ldots$
	by similar computations &	\\
$inf(I,O)$ & $= Sort_\eqda{15}$ & $\sortdef \bot$ & by 5.	\\
$inf(Bin^{1},Bin_{0})$ & $= Sort_\eqda{16}$
	& $\sortdef inf(Bin^{1},Bin)$ & by 3.	\\
$inf(I,I)$ & $= Sort_\eqda{17}$ & $\sortdef I$ & by 4.	\\
$inf(Bin^{1},Bin)$ & $=Bin^{1}$
	& {by similar computations} &	\\
[0.2cm]
Hence,&&&	\\
$Sort_\eqra{1}$ & \mca{2}{$\sortdef Sort_\eqra{2} \mid Sort_\eqra{3}
	\mid Sort_\eqra{4}$,} &	\\
$Sort_\eqra{2}$ 
	& \mca{2}{$\sortdef Sort_\eqra{5} \mid Sort_\eqra{6}$,} & \\
$Sort_\eqra{3}$ & \mca{2}{$\sortdef Sort_\eqra{7}
	\mid Sort_\eqra{8}$,} &	\\
$Sort_\eqra{4}$ & \mca{2}{$\sortdef Sort_\eqra{9} 
	\mid Sort_\eqra{10}$,} &	\\
$Sort_\eqra{5}$ & \mca{2}{$\sortdef \bot$} &	\\
$Sort_\eqra{6}$ & \mca{2}{$\sortdef \bot$} &	\\
$Sort_\eqra{7}$ 
	& \mca{2}{$\sortdef Sort_\eqra{1} \.{::} Sort_\eqra{11}$,} & \\
$Sort_\eqra{8}$ 
	& \mca{2}{$\sortdef Sort_\eqra{12} \.{::} Sort_\eqra{13}$,}&\\
$Sort_\eqra{9}$ 
	& \mca{2}{$\sortdef Sort_\eqra{14} \.{::} Sort_\eqra{15}$,}&\\
$Sort_\eqra{10}$ 
	& \mca{2}{$\sortdef Sort_\eqra{16} \.{::} Sort_\eqra{17}$,} & \\
$Sort_\eqra{16}$ & \mca{2}{$\sortdef Bin^{1}$} &	\\
$Sort_\eqra{11}$ & \mca{2}{$\sortdef O$} & \\
$Sort_\eqra{12}$ & \mca{2}{$= Bin^2$} & \\
$Sort_\eqra{13}$ & \mca{2}{$\sortdef \bot$} & \\
$Sort_\eqra{14}$ & \mca{2}{$\sortdef \ldots$} & \\
$Sort_\eqra{15}$ & \mca{2}{$\sortdef \bot$} & \\
$Sort_\eqra{17}$ & \mca{2}{$\sortdef I$} & \\
\hline
\end{tabular}
\end{center}

\caption{Example computation of sort infimum}
\label{Example computation of sort infimum}
\end{figure}

\ALGORITHM{
\eqd{10}
The following algorithm computes the intersection of two regular
sorts.
Let $S_1$ and $S_2$ be sort names, let $S$ be a new sort name.
Define $\eqi{inf(S_1,S_2)} = S$,
where a new sort definition is introduced for $S$:
\be
\item If $inf(S_1,S_2)$ has already been called earlier,
	$S$ is already defined (loop check).
\item Else, if $S_1 \sortdef S_{11} \mid \ldots \mid S_{1n}$,
	define
	$S \sortdef inf(S_{11},S_2) \mid \ldots \mid inf(S_{1n},S_2)$
\item Else, if $S_2 \sortdef S_{21} \mid \ldots \mid S_{2n}$,
	define
	$S \sortdef inf(S_1,S_{21}) \mid \ldots \mid inf(S_1,S_{2n})$
\item Else, if $S_1 \sortdef cr(S_{11},\ldots,S_{1n})$
	and $S_2 \sortdef cr(S_{21},\ldots,S_{2n})$,	\\
	define
	$S \sortdef cr(inf(S_{11},S_{21}),\ldots,inf(S_{1n},S_{2n}))$
\item Else, define $S \sortdef \bot$
\ee
Using Thm.~\eqr{7}
with
	$p_S(u) :\Lra \left\{	%}
	\begin{tabular}{@{}l@{\hspace*{0.3cm}}l@{}}
	$u \in S_1^M \cap S_2^M$ & if $S = inf(S_1,S_2)$ 	\\
	$u \in S^M$ & else	\\
	\end{tabular}
	\right.$
\\
it can be shown that $inf(S_1,S_2)^M = S_1^M \cap S_2^M$.
The $else$-case in the definition of $p_S(u)$ causes only trivial proof
obligations; in later applications of Thm.~\eqr{7}, it will be
tacitly omitted for the sake of brevity.
The algorithm obviously needs at most $\#use(S_1) * \#use(S_2)$
recursive calls to compute $inf(S_1,S_2)$.
}

\begin{figure}
\begin{center}
\begin{tabular}[t]{@{}|l@{$\;$}lllll|@{}}
\hline
$diff(Bin,Bin_1)$ & $=$ & $Sort_\eqda{18}$
	& $\sortdef$ & $diff(Nil,Bin_1) \mid
	diff(Bino,Bin_1)$
	& by 2.	\\
    &&& $\mid$ & $diff(Bini,Bin_1)$ &	\\
$diff(Nil,Bin_1)$ & $=$ & $Sort_\eqda{19}$
	& $\sortdef$ & $diff(Nil,Bino_1 \mid Bini_0)$
	& by 3.	\\
$diff(Bino,Bin_1)$ & $=$ & $Sort_\eqda{20}$
	& $\sortdef$
	& $diff(Bino,Bino_1 \mid Bini_0)$
	& by 3.	\\
$diff(Bini,Bin_1)$ & $=$ & $Sort_\eqda{21}$
	& $\sortdef$
	& $diff(Bini,Bino_1 \mid Bini_0)$
	& by 3.	\\
$diff(Nil,Bino_1 \mid Bini_1)$ & $=$ & $Sort_\eqda{22}$
	& $\sortdef$ & $diff(Nil,Bino_1)$
	& by 6.	\\
$diff(Nil,Bino_1)$ & $=$ & $Sort_\eqda{23}$
	& $\sortdef$ & $Nil$ & by 7.	\\
$diff(Bino,Bino_1 \mid Bini_0)$
	 & $=$ & $Sort_\eqda{24}$
	 & $\sortdef$ & $Sort_\eqRa{25} \mid Sort_\eqRa{26}
	 \mid Sort_\eqRa{27} \mid Sort_\eqRa{28}$
	 & by 4.	\\
&&
	\begin{tabular}[t]{@{}l@{}}
	$Sort_\eqda{25}$	\\
	$Sort_\eqda{26}$	\\
	$Sort_\eqda{27}$	\\
	$Sort_\eqda{28}$	\\
	\end{tabular}
&
	\begin{tabular}[t]{@{}l@{}}
	$\sortdef$	\\
	$\sortdef$	\\
	$\sortdef$	\\
	$\sortdef$	\\
	\end{tabular}
&
	\begin{tabular}[t]{@{}lllllllll@{}}
	$diff(Bin,$ & $Bin_1$ & $\mid$ & $Bin_0$ 
		& $) \.{::} diff(O,$
		& \multicolumn{3}{@{}c@{}}{$\bot$} & $)$	\\
	$diff(Bin,$ & $Bin_1$ &&
		& $) \.{::} diff(O,$
		&&& $I$ & $)$	\\
	$diff(Bin,$ &&& $Bin_0$ 
		& $) \.{::} diff(O,$
		& $O$ &&& $)$	\\
	$diff(Bin,$ & \multicolumn{3}{@{}c@{}}{$\bot$}
		& $) \.{::} diff(O,$
		& $O$ & $\mid$ & $I$ & $)$	\\
	\end{tabular}
&
	\begin{tabular}[t]{@{}l@{}l@{}}
	\small 1, & \small 1	\\
	\small 1, & \small 2	\\
	\small 2, & \small 1	\\
	\small 2, & \small 2	\\
	\end{tabular}
\\
$diff(Bini,Bino_1 \mid Bini_0)$
	 & $=$ & $Sort_\eqda{29}$
	 & $\sortdef$ & $Sort_\eqRa{30} \mid Sort_\eqRa{31}
	 \mid Sort_\eqRa{32} \mid Sort_\eqRa{33}$
	 & by 4.	\\
&&
	\begin{tabular}[t]{@{}l@{}}
	$Sort_\eqda{30}$	\\
	$Sort_\eqda{31}$	\\
	$Sort_\eqda{32}$	\\
	$Sort_\eqda{33}$	\\
	\end{tabular}
&
	\begin{tabular}[t]{@{}l@{}}
	$\sortdef$	\\
	$\sortdef$	\\
	$\sortdef$	\\
	$\sortdef$	\\
	\end{tabular}
&
	\begin{tabular}[t]{@{}lllllllll@{\hspace*{0.4cm}}l@{}l@{}}
	$diff(Bin,$ & $Bin_1$ & $\mid$ & $Bin_0$ 
		& $) \.{::} diff(I,$
		& \multicolumn{3}{@{}c@{}}{$\bot$} & $)$	\\
	$diff(Bin,$ & $Bin_1$ &&
		& $) \.{::} diff(I,$
		&&& $I$ & $)$	\\
	$diff(Bin,$ &&& $Bin_0$ 
		& $) \.{::} diff(I,$
		& $O$ &&& $)$	\\
	$diff(Bin,$ & \multicolumn{3}{@{}c@{}}{$\bot$}
		& $) \.{::} diff(I,$
		& $O$ & $\mid$ & $I$ & $)$	\\
	\end{tabular}
&
	\begin{tabular}[t]{@{}l@{}l@{}}
	\small 1, & \small 1	\\
	\small 1, & \small 2	\\
	\small 2, & \small 1	\\
	\small 2, & \small 2	\\
	\end{tabular}
\\
$diff(Bin,Bin_1 \mid Bin_0)$ & $=$ & $\bot$
	&& {by similar computations} &	\\
$diff(Bin,Bin_1)$ & $=$ & $Sort_\eqra{18}$ &&& by 1.	\\
$diff(Bin,Bin_0)$ & $=$ & $\bot$
	&& {by similar computations} &	\\
$diff(Bin,\bot)$ & $=$ & $Bin$ &&& by 8.	\\
$diff(O,\bot)$ & $=$ & $O$ &&& by 8.	\\
$diff(O,I)$ & $=$ & $O$ &&& by 7.	\\
$diff(O,O)$ & $=$ & $\bot$ &&& by 5.	\\
$diff(O,O \mid I)$ & $=$ & $\bot$ &&& by 6.,5.	\\
$diff(I,\bot)$ & $=$ & $I$ &&& by 8.	\\
$diff(I,I)$ & $=$ & $\bot$ &&& by 5.	\\
$diff(I,O)$ & $=$ & $I$ &&& by 7.	\\
$diff(I,O \mid I)$ & $=$ & $\bot$ &&& by 6.,5.	\\
[0.2cm]
Hence, &&&&&	\\
$Sort_\eqra{18}$ 
	& $\sortdef$ & \mca{3}{$Sort_\eqra{19} \mid Sort_\eqra{20} 
	\mid Sort_\eqra{21}$} &	\\
$Sort_\eqra{19}$ & $\sortdef$ & \mca{3}{$Sort_\eqra{22}$} &	\\
$Sort_\eqra{20}$ & $\sortdef$ & \mca{3}{$Sort_\eqra{24}$} &	\\
$Sort_\eqra{21}$ & $\sortdef$ & \mca{3}{$Sort_\eqra{29}$} &	\\
$Sort_\eqra{22}$ & $\sortdef$ & \mca{3}{$Sort_\eqra{23}$} &	\\
$Sort_\eqra{23}$ & $\sortdef$ & \mca{3}{$Nil$} &	\\
$Sort_\eqra{24}$ & $\sortdef$ & \mca{3}{$Sort_\eqra{25} 
	\mid Sort_\eqra{26} 
	\mid Sort_\eqra{27} \mid Sort_\eqra{28}$} &	\\
$Sort_\eqra{25}$ & $\sortdef$ & \mca{3}{$\bot \.{::} O$} &	\\
$Sort_\eqra{26}$ & $\sortdef$ & \mca{3}{$Sort_\eqra{18} \.{::} O$}&\\
$Sort_\eqra{27}$ & $\sortdef$ & \mca{3}{$\bot \.{::} \bot$} &	\\
$Sort_\eqra{28}$ & $\sortdef$ & \mca{3}{$Bin \.{::} \bot$} &	\\
$Sort_\eqra{29}$ & $\sortdef$ & \mca{3}{$Sort_\eqra{30} 
	\mid Sort_\eqra{31}
	\mid Sort_\eqra{32} \mid Sort_\eqra{33}$} &	\\
$Sort_\eqra{30}$ & $\sortdef$ & \mca{3}{$\bot \.{::} I$} &	\\
$Sort_\eqra{31}$ & $\sortdef$ & \mca{3}{$Sort_\eqra{18} \.{::} \bot$}&\\
$Sort_\eqra{32}$ & $\sortdef$ & \mca{3}{$\bot \.{::} I$} &	\\
$Sort_\eqra{33}$ & $\sortdef$ & \mca{3}{$Bin \.{::} \bot$} &	\\
\hline
\end{tabular}
\end{center}

\caption{Example computation of sort difference}
\label{Example computation of sort difference}
\end{figure}

\ALGORITHM{
\eqd{11}
The following algorithm computes the relative complement of two regular
sorts.
For technical reasons, the second argument may be an arbitrary union of
sort names.
Let $S_1, \ldots, S_m$ be sort names, let $S$ be a new sort name.
Define $\eqi{diff(S_1,S_2 \mid \ldots \mid S_m)} = S$,
where a new sort definition is introduced for $S$:
\be
\item If $diff(S_1,S_2\.\mid \ldots \.\mid S_m)$ has already been called
	earlier, $S$ is already defined (loop check).
\item If $S_1 \sortdef S_{11} \mid \ldots \mid S_{1n}$,
	define
	$S \sortdef diff(S_{11},S_2\.\mid \ldots \.\mid S_m) \mid
	 \ldots \mid diff(S_{1n},S_2\.\mid \ldots \.\mid S_m)$
\item If $S_i \sortdef S_{i1} \mid \ldots \mid S_{in}$ 
	for $2 \leq i \leq m$,
	define
	$S \sortdef diff(S_1,S_2 \.\mid \ldots \.\mid S_{i\.-1}
	\.\mid S_{i\.+1} \.\mid \ldots  \.\mid S_m \;\mid\;
	S_{i1} \.\mid \ldots \.\mid S_{in})$
\item If $S_1 \sortdef cr(S_{11},\ldots,S_{1n})$ 
	\ldots,
	$S_m \sortdef cr(S_{m1},\ldots,S_{mn})$,
	with $n > 0$,
	\\
	let
	$S_{l_1,\ldots,l_m}$ be a new sort name for each
	$l_1,\ldots,l_m \in \{1,\ldots,n\}$,
	\\
	define
	$\ds S \sortdef \bigmid_{l_1\.=1}^n \ldots
		\bigmid_{l_m\.=1}^n \; S_{l_1,\ldots,l_m}$ \\
	and
	$\ds S_{l_1 \ldots l_m} \sortdef cr(diff(S_{11},
		\bigmid_{j \geq 2,\; l_j\.=1} S_{j1}),
		\ldots,diff(S_{1n},
		\bigmid_{j \geq 2,\; l_j\.=n} S_{jn}))$.
\item If $S_1 \sortdef cr$,
	$S_2 \sortdef cr$ and $m=2$,
	define
	$S \sortdef \bot$
\item If $S_1 \sortdef cr( \ldots )$
	and $S_m \sortdef cr'( \ldots )$
	with $cr \neq cr'$ and $m > 2$,	\\
	define
	$S \sortdef diff(S_1,S_2 \.\mid \ldots \.\mid S_{m-1})$
\item If $S_1 \sortdef cr( \ldots )$
	and $S_m \sortdef cr'( \ldots )$
	with $cr \neq cr'$ and $m = 2$,	\\
	define
	$S \sortdef S_1$
\item If $S_2 \sortdef \bot$ and $m=2$,
	define $S \sortdef S_1$.
\ee

Using Thm.~\eqr{7}
with
\display{
	\begin{tabular}[t]{@{}l@{$\;$}r@{$\;$}l@{\hspace*{0.5cm}}l@{}}
	$p_S(u)$ & $:\Lra$ 
		& $u \in S_1^M \setminus (S_2^M \cup \ldots \cup S_m^M)$
		& if $S = diff(S_1,S_2)$ and	\\
	$p_{S_{l_1,\ldots,l_m}}(u)$ & $:\Lra$
		& $\exists u_1, \ldots, u_n \;\;
		u=cr(u_1,\ldots,u_n)$	\\
	    & $\wedge$ & $\bigwedge_{i=1}^n u_i \in (S_{1i}^M
		\setminus \bigcup_{j \geq 2,\; l_j=i} S_{ji}^M)$
		& if $S_{l_1,\ldots,l_m}$ was defined in rule 4., \\
	\end{tabular}
}
it can be shown that 
$diff(S_1,S_2 \mid \ldots \mid S_m)^M = S_1^M \setminus (S_2^M \cup
\ldots \cup S_m^M)$.
\\
The algorithm needs at most $\#use(S_1) * 2^{\#use(S_2)}$
recursive calls
to compute $diff(S_1,S_2)$.
}

\ALGORITHM{
\eqd{12}
Let $S$ be a sort name,
define $\eqi{inh(S,Occ)} = \tpl{A,B,C,D}$
where $A$ is a finite set of ground constructor terms,
$B \in \{true,false\}$,
$C$, $D$, and $Occ$ are finite sets of sort names,
as follows:
\be
\item If $S \in Occ$,
	define $inh(S,Occ) = \tpl{\{\},false,\{S\},\{S\}}$
\item Else, if $S \sortdef S_1 \mid \ldots \mid S_n$,
	define $inh(S,Occ) 
	=\tpl{A_1 \cup\ldots\cup A_n,B,C,D}$
\item Else, if $S \sortdef cr(S_1,\ldots,S_n)$,
	define $inh(S,Occ) 
	= \tpl{cr[A_1 \times \ldots \times A_n],B,C,D}$
\item Else, if $S \sortdef cr$,
	define $inh(S,Occ) = \tpl{\{cr\},false,\{S\},\{\}}$
\ee
where $\tpl{A_i,B_i,C_i,D_i} := inh(S_i,Occ \cup \{S\})$
for $i=1,\ldots,n$,
$B := B_1 \vee \ldots \vee B_n \vee
S \in C_1 \.\cup \ldots \.\cup C_n$,
$C := C_1 \cup \ldots \cup C_n \cup \{S\}$,
and
$D := (D_1 \cup \ldots \cup D_n) \setminus \{S\}$.
$A$ is used to decide $S^M \neq \{\}$,
$B$ is $true$ if a loop occurs in the definition of $S$,
$C$ is used to compute $B$,
and $D$ is used only for proof technical reasons and need not be
computed in a practical implementation.
Let $inh(S,\{\}) = \tpl{A,B,C,D}$.
Then, $S^M \neq \{\}$ iff $A \neq \{\}$;
$S^M$ finite iff $B \Lra false$,
and in this case $A = S^M$.
The algorithm needs at most $\#use(S) * 2^{\#use(S)}$
recursive calls to compute $inh(S,\{\})$.
Define $\eqi{single(S)} :\Lra inh(S,\{\}) = \tpl{\{u\},false,C,D}$
for some $u$, $C$, $D$.
}
\PROOF{	$\;$
\be
\item Use $Occ_1 < Occ_2 :\Lra Occ_2 \psubset Occ_1$
	to show termination and complexity;
	\\
	note that $Occ_1,Occ_2$ is bounded from above by the finite set
	$use(S)$.
\item Let $inh(S,Occ) = \tpl{A,B,C,D}$,
	and $E := \{ u \mid \exists S' \in D, u' \in S'^M \;\;\; u'
	\subterm u\}$;
	\\
	all following statements are
	proven by induction on the computation tree of $inh(S,Occ)$.
\item Show $D \subset Occ \cap C$, hence $E = \{\}$ if $Occ = \{\}$.
\item If $B = false$, show $S^M \subset A \cup E$.
\item Show $A \subset S^M$.
\item If $A = \{\}$, show $S^M \subset E$
	by induction on the computation tree,
	and (nested) induction on $u \in S^M$.
\item If $Occ = \{\}$,
	from 5.\ (``$\La$''), 3.\ and 6.\ (``$\Ra$'') follows
	$S^M \neq \{\}$ iff $A \neq \{\}$.
\item Show $B \Lra false$ iff $S$ contains no loops (iff $S^M$ is
	finite by the pumping lemma).
\item If $Occ = \{\}$,
	from 5.\ (``$\subset$''), 3.\ and 4.\ (``$\supset$'')
	follows $A = S^M$ if $B \Lra false$.
\ee
}

Using the sort definitions from Fig.~\ref{Examples of sort definitions},
Fig.~\ref{Example computation of sort infimum} shows the computation of
the intersection of $Bin^2$ and $Bin_1$ by Alg.~\eqr{10}; the result
may be simplified to $Sort_\eqra{1} \sortdef Sort_\eqra{1} \.{::} o \mid
Bin^{1} \.{::} i$, which uses the sloppy notation for sort definitions
mentioned in Def.~\eqr{4}, and
intuitively denotes all binary lists with one
or two $i$-digits. Figure~\ref{Example computation of sort difference}
shows the computation of the complement of $Bin_1$ relative to $Bin$ by
Alg.~\eqr{11}; the result may be simplified to $Sort_\eqra{18}
\sortdef nil \mid Sort_\eqra{18} \.{::} o$ which is equivalent to
$Bin^0$.

In \cit{Burghardt 1993}{Burghardt.1993}, sort definitions may include
``\eqi{constraint} formulas'' which are not to be considered by the
sort algorithms, but rather collected and passed to an external prover
in which the sort algorithms are meant to be embedded.
A sort definition (cf.\ Def.~\eqr{4}) may also have the form
\display{
	$SortName \sortdef Constructor(Id:SortName,\ldots,Id:SortName) 
	\lhd Constraint(Id,\ldots,Id)$,
}
with the semantics
\display{
	$(cr(i_1:S_1,\ldots,i_n:S_n) \lhd p(i_1,\ldots,i_n))^M
	= \{cr(u_1,\ldots,u_n) \mid
	u_1 \in S_1^M, \ldots, u_n \in S_n^M,
	p(u_1,\ldots,u_n)\}$;
}
and e.g.\ rule 4.\ of Alg.~\eqr{10} has then the following form
\display{
	If $S_1 \sortdef cr(S_{11},\ldots,S_{1n}) \lhd ct_1$,
	and $S_2 \sortdef cr(S_{21},\ldots,S_{2n}) \lhd ct_2$,
	\\
	define 
	$S \sortdef cr(inf(S_{11},S_{21}),\ldots,inf(S_{1n},S_{2n}))
	\lhd ct_1 \wedge ct_2$.
}
The same applies to Alg.~\eqr{11}. Algorithm
\eqr{12} may yield a proper
predicate $B \not\in \{true,false\}$ if a nontrivial constraint formula
occurs above a loop in the sort definition.
Constraint formulas are mentioned here only because they appear in
App.~\ref{Case Study ``Comb Vector Construction''}.

\section{T-Substitutions}
\label{T-Substitutions}

In this section, we apply the formalism from Sect.~\ref{Regular Sorts}
to define possibly infinite regular sets of ground substitutions.
We define suitable free constructors from which
ground substitutions can be built as terms of a lifted algebra
$\TT_{\pfunc{\cal V}{\cal CR}}$.
We call such terms t-substitutions.
Note that the classical approach, constructing substitutions by
functional composition from simple substitutions, cannot be used, since
functional composition is not free but obeys e.g.\ the associativity
law.
We provide the necessary notions and properties of t-substitutions and
of sets of them, called t-sets.
All results in this section hold for arbitrary t-sets.
\vspace{0.1cm}

\noindent
\parbox[c]{6.9cm}{
\hspace*{\outerparindent}%
We first define suitable free constructors from which
ground substitutions can be built as terms of a lifted algebra.
Expressed informally, to build a substitution term corresponding to
$[x_1 \.{:=} u_1,\ldots,x_n \.{:=} u_n]$ with $u_i$ ground,
we ``overlay'' the $u_i$ 
to obtain the substitution term; on the right, an example is shown for
$[x \.{:=} cons(0,nil), y \.{:=} s(s(0))]$.
}
\hfill
\fbox{
$
\begin{array}[c]{@{}l*{8}{c@{}}c@{}}
x := & cons       &( &0       &  &    &  &, &nil   &)\\
y := &        s   &( &    s   &( &0   &) &  &      &)\\
\cline{2-10}
     & cons_x s_y &( &0_x s_y &( &0_y &) &, &nil_x &)\\
\end{array}
$
}

\DEFINITION{
\eqd{13}
Given a set $V \subset {\cal V}$ of variables,
define the \eqi{constructors for t-substitutions} with domain $V$
as the set \eqi{$\func{V}{\cal CR}$} of all
total mappings from $V$ to $\cal CR$.
T-substitution constructors are denoted 
by $\eqi{\vec{cr}}, \vec{cr}', \ldots$ ,
the empty mapping by \eqi{$\varepsilon$}.
Function \eqi{application} is written as \eqi{$\vec{cr}_x$},
the arity is defined as
$\eqi{ar(\vec{cr})} := \max_{x \in dom(\vec{cr})} ar(\vec{cr}_x)$.
For ${V'} \subset {\cal V}$,
let \eqi{$\vec{cr} \rs {V'}$}
denote the \eqi{restriction} of $\vec{cr}$ to the
variables in ${V'}$,
i.e.\ $dom(\vec{cr} \rs {V'}) = {V'}$
and $(\vec{cr} \rs {V'})_x = \vec{cr}_x$ for all $x \in {V'}$.
For $\vec{cr}$ and $\vec{cr}'$ 
such that $\vec{cr}_x = \vec{cr}'_x$ 
for all $x \in dom(\vec{cr}) \cap dom(\vec{cr}')$,
let \eqi{$\vec{cr} \pecomp \vec{cr}'$} denote 
the ``\eqi{parallel composition}'' of $\vec{cr}$ and $\vec{cr}'$,
i.e.\
$dom(\vec{cr} \pecomp \vec{cr}') = dom(\vec{cr}) \cup dom(\vec{cr}')$,
and 
\display{
	$(\vec{cr} \pecomp \vec{cr}')_x 
	= \left\{ 	%}
	\begin{tabular}{l@{\hspace*{0.5cm}}l}
	$\vec{cr}_x$ & if $x \in dom(\vec{cr})$	\\
	$\vec{cr}'_x$ & if $x \in dom(\vec{cr}')$	\\
	\end{tabular}
	\right.$ .
}
Note that $\vec{cr} \pecomp \vec{cr}'$ is undefined if $\vec{cr}$ and
$\vec{cr}'$ do not agree on their domain intersection.
}

\EXAMPLE{
\eqd{14}
In examples, we write e.g.\ \eqi{$0_x s_y$}
to denote the mapping $(x \.\mapsto 0, y \.\mapsto s)$.
We have 
\begin{tabular}[t]{@{}r@{$\;$}l@{}}
$0_x s_y$ & $\in \func{\{x,y\}}{\cal CR}$,	\\
$ar(0_x s_y)$ & $= max(0,1) = 1$,	\\
$(0_x s_y) \rs {\{y\}}$ & $= s_y$, and	\\
$(0_x s_y) \pecomp (s_y cons_z)$ & $= 0_x s_y cons_z$.	\\
\end{tabular}
}

\DEFINITION{
\eqd{15}
Once we have defined t-substitution constructors, we inherit the 
initial
term algebra \eqi{${\cal T}_{\func{V}{\cal CR}}$} over them.
However, we have to exclude some nonsense terms.
Define the subset
\eqi{$\TT_{\func{V}{\cal CR}}} \subset {\cal T}_{\func{V}{\cal CR}}$
of {\eqi{admissible t-substitutions}} with domain $V$
as the least set such that
\display{
	\begin{tabular}[t]{@{}l@{$\;$}ll@{}}
	$\vec{cr}(\sigma'_1,\ldots,\sigma'_{ar(\vec{cr})})
		\in \TT_{\func{V}{\cal CR}}$ \\
	if $\vec{cr} \in \func{V}{\cal CR}$
		and $\sigma'_i \in
		\TT_{\func{\{x \in V \mid ar(\vec{cr}_x)
		\geq i\}} {\cal CR}}$
		for $i = 1,\ldots,ar(\vec{cr})$.	\\
	\end{tabular}
	}
We denote \eqi{t-substitutions} by 
$\eqi{\sigma'}, \eqi{\tau'}, \eqi{\mu'}, \ldots$ . 
Sets of t-substitutions are called {\eqi{t-sets}}
and are denoted by
$\eqi{\sigma},\eqi{\tau},\eqi{\mu},\ldots$ .
}

\begin{figure}
\begin{center}
\begin{tabular}[t]{@{}|l@{\tab}l|@{}}
\hline
\multicolumn{2}{@{}|l|@{}}{T-substitutions,
	built as constructor terms:}	\\
$s_x(s_x(0_x))$ & $\corresponds [x \.{:=} s(s(0))]$	\\
$s_x s_y(0_x 0_y)$ & $\corresponds [x \.{:=} s(0), y \.{:=} s(0)]$ \\
$s_x 0_y(0_x)$ & $\corresponds [x \.{:=} s(0), y \.{:=} 0]$	\\
\hline
\end{tabular}
\end{center}

\caption{Examples of t-substitutions}
\label{Examples of t-substitutions}
\end{figure}

\EXAMPLE{
\eqd{16}
Definition \eqr{15}
implies that $\vec{cr} \in \TT_{\func{V}{\cal CR}}$
if $\vec{cr} \in \func{V}{\cal CR}$ is
a nullary t-substitution constructor.
For example,
we have
$0_y \in \TT_{\func{\{y\}}{\cal CR}}$,
and hence $0_x s_y(0_y) \in \TT_{\func{\{x,y\}}{\cal CR}}$,
but neither $0_y \in \TT_{\func{\{x,y\}}{\cal CR}}$,
nor $0_x s_y(0_x 0_y) \in \TT_{\func{V}{\cal CR}}$ for any $V$.
Figure~\ref{Examples of t-substitutions} 
shows some more t-substitutions together with 
their intended semantics.
}

\DEFINITION{
\eqd{17}
Since
$\TT_{\func{V_1}{\cal CR}} \cap \TT_{\func{V_2}{\cal CR}}
= \{\}$ for $V_1 \neq V_2$,
we may define $\eqi{dom(\sigma')} := V$
iff $\sigma' \in \TT_{\func{V}{\cal CR}}$.
Let \eqi{$\pfunc{V}{\cal CR}$}
be the set of all \eqi{partial mappings} from $V$ to ${\cal CR}$;
define the set of admissible t-substitutions with a subset of $V$ as
domain by
	%\display
	{
	$\eqi{\TT_{\pfunc{V}{\cal CR}}}
	:= \bigcup_{V' \subset V, V' \mbox{\scriptsize\ finite}}
	\TT_{\func{V'}{\cal CR}}$.
	}
}

\DEFINITION{
%[(T-Substitution Application)]
\eqd{18}
%$\;$\\
Define the t-substitution \eqi{application} \eqi{$\sigma' u$} by
\\
\begin{tabular}[t]{@{}l@{$\;$}l@{\hspace*{1em}}l@{}}
$\sigma' (cr(u_1,\ldots,u_k))$
	& $:= cr[\sigma' u_1 \times \ldots \times \sigma' u_k]$ \\
$\sigma' (cr)$ & $:= \{cr\}$ \\
$(\vec{cr}(\sigma'_1,\ldots,\sigma'_n)) (x)$
	& $:= \vec{cr}_x[\sigma'_1 x \times\ldots\times
	\sigma'_{ar(\vec{cr}_x)} x]$
	& if $x \in dom(\vec{cr})$, $n > 0$	\\
$(\vec{cr}) (x)$ & $:= \{\vec{cr}_x \}$ 
	& if $x \in dom(\vec{cr})$, $n = 0$	\\
$(\vec{cr}(\sigma'_1,\ldots,\sigma'_n)) (x)$
	& $:= \{\}$
	& if $x \not\in dom(\vec{cr})$	\\
\end{tabular}
\\
$\sigma' u$ yields a set with at most one ground
constructor term.
Application is extended elementwise to t-sets by
\eqi{$\sigma u} := \bigcup_{\sigma' \in \sigma} \sigma' u$.
}

Note that, in contrast to an ordinary substitution $\beta$,
a t-substitution $\sigma'$ is undefined outside its domain, i.e.\
it returns the empty set.
We have $\varepsilon u = \{\}$ if $u$ contains variables,
$\varepsilon u = u$ if $u$ is ground,
and always $\bot u = \{\}$.

\LEMMA{
%[(T-Substitution Equality)] 
\eqd{19}
%$\;$\\
$\sigma' = \tau'$ iff 
$\sigma' x = \tau' x$ for all $x \in {\cal V}$,
\\
where ``$=$'' on the left-hand side denotes the syntactic equality in
$\TT_{\pfunc{\cal V}{\cal CR}}$.
}

Although constructors may be written in different ways,
e.g.\ $0_x s_y = s_y 0_x$, 
the initiality condition
$\vec{cr}(u_1,\ldots,u_n) = \vec{cr}'(u'_1,\ldots,u'_{n'})
\Ra \vec{cr} = \vec{cr}' \wedge n = n' \wedge
u_1 = u'_1 \wedge \ldots \wedge u_n = u'_n$
is satisfied in $\TT_{\pfunc{\cal V}{\cal CR}}$.
The desired equivalence of term equality and
function equality from Lemma~\eqr{19}
is the reason for restricting t-substitutions 
to a subset $\TT_{\pfunc{\cal V}{\cal CR}}$
of the initial algebra ${\cal T}_{\pfunc{\cal V}{\cal CR}}$,
excluding nonsense terms like e.g.\
$0_x s_y(0_x 0_y)$ and
$0_x s_y(0_x 1_y)$
which would contradict the initiality requirement.

\LEMMA{
%[(T-Substitutions as Ground Substitutions)] 
\eqd{20}
%$\;$\\
$\TT_{\pfunc{\cal V}{\cal CR}}$ corresponds to the set of all
ordinary ground substitutions in the following sense:
For each $\sigma'$
there exists a $\beta$,
such that $\sigma' u = \{\beta u\}$
whenever $vars(u) \subset dom(\sigma')$.
%\\
Conversely, for each $\beta$ there
exists a $\sigma'$ with the respective
property;
cf.\ Fig.~\ref{Examples of t-substitutions} 
which shows some example correspondences.
}
\PROOF{
Induction on $\sigma'$
with
$\beta_{\vec{cr}(\sigma'_1,\ldots,\sigma'_n)} (x)
:= \vec{cr}_x(\beta_{\sigma'_1} x,\ldots,\beta_{\sigma'_n}x)$.
\\
Conversely:
define
$\{\sigma'_\beta\} := \bigpecomp_{x \in dom(\beta)} [x \.{:=} \beta x]$.
Then,
$\sigma'_\beta u = \{\beta u\}$
whenever $vars(u) \subset dom(\beta)$.
}

\DEFINITION{
%[(T-Substitution Restriction)]
\eqd{21}
Define the t-substitution \eqi{restriction} \eqi{$\sigma' \rs V$} by
\\
\begin{tabular}[t]{@{}l@{$\;$}l@{\hspace*{1em}}l@{}}
$\vec{cr} \rs V$ & as defined in Def.~\eqr{13}
	& if $ar(\vec{cr}) = 0$	\\
$(\vec{cr}(\sigma'_1,\ldots,\sigma'_n)) \rs V$
	& $:= (\vec{cr} \rs V) \; (\sigma'_1 \rs V,\ldots,
	\sigma'_m \rs V)$
	& if $ar(\vec{cr}) > 0$ and $ar(\vec{cr} \rs V) = m$	\\
\end{tabular}
\\
Restriction is extended elementwise to t-sets by
\eqi{$\sigma \rs V} := \{ \sigma' \rs V \;\mid\; \sigma' \in \sigma \}$.
}

\DEFINITION{
%[(T-Substitution Composition)]
\eqd{22}
%$\;$\\
Define the \eqi{parallel composition}
of t-substitutions \eqi{$\sigma' \pecomp \tau'$} by
\\
$\vec{cr}(\sigma'_1,\ldots,\sigma'_n) \pecomp
\vec{cr}'(\tau'_1,\ldots,\tau'_m) :=$
\\
$\;$
\hspace*{\fill}
$\left\{	%}
\begin{tabular}{@{}l@{\hspace*{1em}}l@{}}
$(\vec{cr} \pecomp \vec{cr}') \;
	[(\sigma'_1 \pecomp \tau'_1) \times\ldots\times
	(\sigma'_n \pecomp \tau'_n) \times \{\tau'_{n+1}\} 
	\times \ldots \times \{\tau'_m\}]$
	& if $\vec{cr} \pecomp \vec{cr}'$ is defined, and $n \leq m$ \\
$(\vec{cr} \pecomp \vec{cr}') \;
	[(\sigma'_1 \pecomp \tau'_1) \times\ldots\times
	(\sigma'_m \pecomp \tau'_m) \times \{\sigma'_{m+1}\}
	\times \ldots \times \{\sigma'_n\}]$
	& if $\vec{cr} \pecomp \vec{cr}'$ is defined, and $m \leq n$ \\
$\{\}$ & if $\vec{cr} \pecomp \vec{cr}'$ is undefined	\\
\end{tabular}
\right.$
\\
$\sigma' \pecomp \tau'$ yields a set with at most one t-substitution.
Parallel composition is extended elementwise to t-sets by
\eqi{$\sigma \pecomp \tau}
:= \bigcup_{\sigma' \in \sigma, \tau' \in \tau} \sigma' \pecomp \tau'$.
Note that $\sigma' \pecomp \tau' = \{\}$ if $\sigma'$ and $\tau'$ do
not
agree on $dom(\sigma') \cap dom(\tau')$.
}

\DEFINITION{
\eqd{23}
Define the \eqi{lifting} of a ground constructor term $u$ to a
t-substitution \eqi{$[x \.{:=} u]$},
using the notation from Def.~\eqr{13}, 
by
\\
\begin{tabular}[t]{@{}l@{$\;$}ll@{}}
$[x \.{:=} cr]$ & $:= (x \.\mapsto cr)$ & if $ar(cr) = 0$	\\
$[x \.{:=} cr(u_1,\ldots,u_n)]$
	& $:= (x \.\mapsto cr) \; 
	([x \.{:=} u_1],\ldots,[x \.{:=} u_n])$
	& if $ar(cr) = n > 0$	\\
\end{tabular}
\\
Lifting is extended elementwise to sets of ground constructor terms by
$\eqi{[x \.{:=} S]} := \{ [x \.{:=} u] \;\mid\; u \in S \}$.
}

\begin{figure}
\begin{center}
\begin{tabular}[t]{@{}|r@{$\;$}l@{$\;$}l|@{}}
\hline
\multicolumn{2}{@{}|l@{}}{$(0_x s_y cons_z(0_y 0_z,nil_z)) \; (x)$} 
	& $= \{0\}$ \\
\multicolumn{2}{@{}|l@{}}{$(0_x s_y cons_z(0_y 0_z,nil_z)) \; (y)$} 
	& $= \{s(0)\}$ \\
\multicolumn{2}{@{}|l@{}}{$(0_x s_y cons_z(0_y 0_z,nil_z)) \; (z)$} 
	& $= \{cons(0,nil)\}$ \\
\multicolumn{2}{@{}|l@{}}{$(0_x s_y cons_z(0_y 0_z,nil_z)) \; 
	(cons(x,cons(y,nil)))$}
	& $= \{cons(0,cons(s(0),nil))\}$ \\
\multicolumn{2}{@{}|l@{}}{$(0_x s_y cons_z(0_y 0_z,nil_z)) \; (x')$} 
	& $= \{\}$ \\
[0.2cm]
\multicolumn{2}{@{}|l@{}}{$(0_x s_y cons_z(0_y 0_z,nil_z)) \rs {\{x\}}$}
	& $= 0_x$ \\
\multicolumn{2}{@{}|l@{}}{$(0_x s_y cons_z(0_y 0_z,nil_z)) \rs {\{y\}}$}
	& $= s_y(0_y)$ \\
\multicolumn{2}{@{}|l@{}}{$(0_x s_y cons_z(0_y 0_z,nil_z)) \rs {\{z\}}$}
	& $= cons_z(0_z,nil_z)$ \\
\multicolumn{2}{@{}|l@{}}{$(0_x s_y cons_z(0_y 0_z,nil_z))
	\rs {\{x,y\}}$}
	& $= 0_x s_y(0_y)$	\\
\multicolumn{2}{@{}|l@{}}{$(0_x s_y cons_z(0_y 0_z,nil_z)) 
	\rs {\{x,z\}}$}
	& $= 0_x cons_z(0_z,nil_z)$	\\
\multicolumn{2}{@{}|l@{}}{$(0_x s_y cons_z(0_y 0_z,nil_z)) 
	\rs {\{y,z\}}$}
	& $= s_y cons_z(0_y 0_z,nil_z)$	\\
[0.2cm]
$(0_x)$ & $\pecomp (s_y(0_y))$
	& $= \{0_x s_y(0_y)\}$	\\
$(0_x s_y(0_y))$ & $\pecomp (0_x cons_z(0_z,nil_z))$
	& $= \{0_x s_y cons_z(0_y 0_z,nil_z)\}$	\\
$(0_x cons_z(0_z,nil_z))$ & $\pecomp (s_y cons_z(0_y 0_z,nil_z))$
	& $= \{0_x s_y cons_z(0_y 0_z,nil_z)\}$	\\
$(0_x s_y(0_y))$ & $\pecomp (s_y cons_z(0_y 0_z,nil_z))$
	& $= \{0_x s_y cons_z(0_y 0_z,nil_z)\}$	\\
$(0_x s_y(0_y))$ & $\pecomp (0_y cons_z(0_z,nil_z))$ & $= \{\}$	\\
[0.2cm]
\multicolumn{2}{@{}|l@{}}{$[x \.{:=} 0]$} & $= 0_x$	\\
\multicolumn{2}{@{}|l@{}}{$[y \.{:=} s(0)]$} & $= s_y(0_y)$	\\
\multicolumn{2}{@{}|l@{}}{$[z \.{:=} cons(0,nil)]$} 
	& $= cons_z(0_z,nil_z)$	\\
\hline
\end{tabular}
\end{center}

\caption{Some example computations according to 
	Defs.~\protect\eqr{18} -- \protect\eqr{23}}
\end{figure}

\DEFINITION{
\eqd{24}
Let $\beta$
be an ordinary idempotent substitution with $n \geq 1$,
let $\sigma'$ be a t-substitution
with $ran(\beta) \subset dom(\sigma')$
and $dom(\beta) \cap dom(\sigma') = \{\}$,
define 
$\eqi{\sigma' \circ \beta} 
:= \bigpecomp_{x \in dom(\beta)} [x \.{:=} \sigma' \beta x]$.
We always have $dom(\sigma' \circ \beta) = dom(\beta)$,
$(\sigma' \circ \beta) v = \sigma' (\beta v)$
for all $v$ with $vars(v) \subset dom(\beta)$,
and $(\sigma' \circ \beta) \by \beta = \sigma'$.

For a t-set $\sigma$ with the same domain as $\sigma'$,
define
$\eqi{\sigma \circ \beta}
:= \bigcup_{\sigma' \in \sigma} \sigma' \circ \beta$.
We have $dom(\sigma \circ \beta) = dom(\beta)$,
$(\sigma \circ \beta) v = \sigma (\beta v)$
for all $v$ with $vars(v) \subset dom(\beta)$,
and $(\sigma \circ \beta) \by \beta = \sigma$.
}

\LEMMA{
\eqd{25}
Some properties of application, restriction, parallel composition, and
abstraction are:
\bi
\item $\sigma' \rs V u = \sigma' u$ if $vars(u) \subset V$; 
	\tab
	$\sigma' \rs V u = \{\}$, else
\item $dom(\sigma' \rs V) = dom(\sigma') \cap V$
\item $(\sigma' \rs {V_1}) \rs {V_2} = \sigma' \rs {V_1 \cap V_2}$
\item $\sigma \subset \tau \Ra \sigma \rs V \subset \tau \rs V$
\item $\sigma \rs V u = \sigma u$ if $vars(u) \subset V$
\item $(\sigma \cap \tau) \rs V = \sigma \rs V \cap \tau \rs V$
\item $\pecomp$ is associative
\item $\sigma \pecomp \tau =
	(\sigma \pecomp \tau \rs {T \setminus S}) \cap
	(\sigma \rs {S \setminus T} \pecomp \tau)$
	where $S=dom(\sigma)$, $T=dom(\tau)$
\item $\sigma' u = \tau' u \Lra
	\sigma' \rs {vars(u)} = \tau' \rs {vars(u)}$.
\item $\sigma' \tpl{x_1,\ldots,x_n}
	= \tpl{\sigma' x_1,\ldots,\sigma' x_n}$
\item $\sigma \tpl{x_1,\ldots,x_n}
	\subset \tpl{\sigma x_1,\ldots,\sigma x_n}$.
\ei
}

\DEFINITION{
%[(T-Substitution Factorization)] 
\eqd{26}
%$\;$\\
Define the \eqi{factorization} \eqi{$\sigma' \by \beta$} of
a t-substitution
$\sigma'$ wrt.\ 
to an ordinary substitution
$\beta$ with $dom(\beta) \subset dom(\sigma')$
as follows, let $k := ar(\vec{cr}_x)$:
\\
\begin{tabular}[t]{@{}l@{\hspace*{1em}}l@{$\;$}l@{\hspace*{1em}}l@{}}
1. & $\sigma' \by {[x_1 {:=} u_1,\ldots,x_n {:=} u_n]}$
	& $:= \sigma' \by {[x_1 {:=} u_1]} \pecomp\ldots\pecomp
	\sigma' \by {[x_n {:=} u_n]}$
	& if $n > 1$	\\
2. & $\vec{cr}(\sigma'_1,\ldots,\sigma'_n) \by
	{[x {:=} \vec{cr}_x(u_1,\ldots,u_k)]}$
	& $:= \sigma'_1 \by {[x {:=} u_1]} \pecomp\ldots\pecomp
	\sigma'_k \by {[x {:=} u_k]}$
	& if $k > 0$	\\
3. & $\vec{cr}(\sigma'_1,\ldots,\sigma'_n) \by
	{[x {:=} \vec{cr}_x]}$
	& $:= \{\varepsilon\}$
	& if $k = 0$	\\
4. & $\vec{cr}(\sigma'_1,\ldots,\sigma'_n) \by
	{[x {:=} cr'(u_1,\ldots,u_k)]}$
	& $:= \{\}$
	& if $\vec{cr}_x \neq cr'$	\\
5. & $\sigma' \by {[x {:=} y]}$ & $:= [y \.{:=} \sigma' x]$
	& if $x \.\neq y \.\in {\cal V}$	\\
\end{tabular}
\\
$\sigma' \by \beta$ yields a set with at most one t-substitution;
it is extended elementwise to t-sets by
$\eqi{\sigma \by \beta}
:= \bigcup_{\sigma' \in \sigma} \sigma' \by \beta$.
Note that $[y \.{:=} \sigma' x]$ is a singleton or empty set by Defs.\
\eqr{18} and \eqr{23}.
Factorization by the identity substitution is undefined.
We have $dom(\sigma' \by \beta) = ran(\beta)$ 
if $\sigma' \by \beta \neq\{\}$.
}

\LEMMA{
%[(Matching Property)] 
\eqd{27}
(Pattern-Matching Properties)
\\
\begin{tabular}[t]{@{}l@{\hspace*{1em}}l@{\hspace*{1em}}ll@{}}
a. & $\sigma' \by \beta \beta u = \sigma' u$,
	& if $\sigma' \by \beta \neq \{\}$	\\
b. & $\sigma \by \beta \beta u = \sigma u \cap \top \beta u$	\\
\end{tabular}
}
\PROOF{	$\;$	\\
a. Induction on $n = \#dom(\beta)$:
show $n=1$ by induction on $u$, 
show $n \leadsto n+1$ by induction on $u$.
\\
b. follows from a.
}

\EXAMPLE{
\eqd{28}
We have
\\
\begin{tabular}[t]{@{}l@{$\;$}ll@{}}
& $(0_x s_y (0_y)) \by {[y \.{:=} s(z)]}$	\\
$=$ & $0_y \by {[y \.{:=} z]}$ & by Def.\ \eqr{26}.2	\\
$=$ & $[z \.{:=} (0_y) (y)]$ & by Def.\ \eqr{26}.5	\\
$=$ & $[z \.{:=} \{0\}]$ & by Def.\ \eqr{18}	\\
$=$ & $\{0_z\}$ & by Def.\ \eqr{23}	\\
\end{tabular}
\hfill
\raisebox{\baselineskip}[0cm][0cm]{
\begin{tabular}[t]{@{}l@{$\;$}ll@{}}
& but	\\
& $(0_x s_y (0_y)) \by {[y \.{:=} s(s(z))]}$	\\
$=$ & $0_y \by {[y \.{:=} s(z)]}$ & by Def.\ \eqr{26}.2	\\
$=$ & $\{\}$ & by Def.\ \eqr{26}.4	\\
\end{tabular}
}
\\
and
$\{0_z\} ([y \.{:=} s(z)] (y))
= \{0_z\} (s(z))
= \{s(0)\}
= (0_x s_y (0_y)) (y)$
by Def.~\eqr{18}.
}

\LEMMA{
%[(Matching Conditions)] 
\eqd{29}
The following propositions are equivalent:
\\
\begin{tabular}[t]{@{}l@{\hspace*{1em}}l@{\hspace*{1em}}ll@{}}
a. & $\sigma' \by \beta \neq \{\}$	\\
b. & $\sigma' u \cap \top \beta u \neq \{\}$
	& for all & $u$ with $vars(u) \subset dom(\beta)$	\\
c. & $\sigma' u \cap \top \beta u \neq \{\}$
	& for some & $u$ with $vars(u) = dom(\beta)$	\\
\end{tabular}
}
\PROOF{	$\;$	\\
a.\ $\Ra$ b. by induction on $u$;
\\
b.\ $\Ra$ c. trivial;
\\
c.\ $\Ra$ a.
Show $\sigma' u = \tau' \beta u \Ra \sigma' x = \tau' \beta x$
for all $\tau' \in \top$ and $x \in vars(u)$ by induction on $u$.
\\
Show
$\sigma' u \cap \tau' \beta u \neq \{\}
\Ra \sigma' \by {[x_i \.{:=} u_i]} y = \sigma' \by {[x_j \.{:=} u_j]} y$
for all $\tau' \in \top$ and $y \in vars(u_i) \cap vars(u_j)$,
%$i,j \in \{1,\ldots,n\}$,
$x_i, x_j \in vars(u)$.
\\
Show by induction on $\#dom(\beta)$ that both conditions together
imply a.
}

\LEMMA{\eqd{30} % generalized to n
Let $u_1,\ldots,u_n$ have pairwise disjoint variables,
and let $vars(u_i) \subset dom(\sigma'_i)$.
Then,
$\sigma'_1 u_1 = \ldots = \sigma'_n u_n$
iff
$u_1,\ldots,u_n$ are simultaneously unifiable by
$\beta_1 \pscomp \ldots \pscomp \beta_n$
with $dom(\beta_i) = vars(u_i)$
and
$\sigma'_1 \by {\beta_1} = \ldots = \sigma'_n \by {\beta_n} \neq \{\}$.
}
\PROOF{	$\;$	\\
``$\Ra$'':
Unifiability is obvious,
minimality of $\beta_i$ implies
$\sigma'_i u_i \cap \top \beta_i u_i \neq \{\}$,
hence $\sigma'_i \by {\beta_i} \neq \{\}$ by \eqr{29}.
According to \eqr{27},
$\sigma'_i \by {\beta_i} \beta_i u_i
= \sigma'_i u_i
= \sigma'_j u_j
= \sigma'_j \by {\beta_j} \beta_j u_j
= \sigma'_j \by {\beta_j} \beta_i u_i$,
hence $\sigma'_i \by {\beta_i} = \sigma'_j \by {\beta_j}$.
\\
``$\La$'': According to	\eqr{27}, we have
$\sigma'_i u_i
= \sigma'_i \by {\beta_i} \beta_i u_i
= \sigma'_j \by {\beta_j} \beta_j u_j
= \sigma'_j u_j$.
}

Domain conditions as in Lemma \eqr{30}
can always be satisfied by bounded
renaming, factorizing by a renaming substitution,
cf.\ Alg.~\eqR{47} below.
If $u_1$ and $u_2$ cannot be unified, $\sigma_1 u_1$ and $\sigma_2 u_2$
are always disjoint.

\THEOREM{
\eqd{31}
Let $\beta = mgu(u_1,u_2)$,
$dom(\beta) = vars(u_1,u_2)$,
$vars(u_i) \subset dom(\sigma_i)$,
\\
$dom(\sigma_1) \cap dom(\sigma_2) = \{\}$,
and
$u = \beta u_1$;
then
$\sigma_1 u_1 \cap \sigma_2 u_2
= (\sigma_1 \pecomp \sigma_2) \by \beta \; u$.
}
\PROOF{\footnote{
	Remember that e.g.\ $\sigma'_1 u_1$ yields a {\em set} of ground
	constructor terms with at most one element.
	}
$\;$\\
\begin{tabular}{@{}ll@{}}
``$\subset$'':
& Suppose
	$\sigma'_1 u_1 = \sigma'_2 u_2
	\subset \sigma_1 u_1 \cap \sigma_2 u_2$,
	let $\gamma$ be a renaming substitution; \\
& then,
	$(\sigma'_1 \pecomp \sigma'_2) \tpl{u_1,u_2}
	= (\sigma'_1 \pecomp \sigma'_2) \tpl{u_2,u_1}
	\stackrel{*}{=}
	(\sigma'_1 \pecomp \sigma'_2) \by \gamma \gamma \tpl{u_2,u_1}$.
\\
& Applying Lemma~\eqr{30}
	yields $(\sigma'_1 \pecomp \sigma'_2) \by \beta \neq \{\}$,
	since $\beta \circ \gamma^{-1} 
	= mgu(\tpl{u_1,u_2},\tpl{u_2,u_1})$,
\\
& hence
	$\sigma'_1 u_1
	= (\sigma'_1 \pecomp \sigma'_2) \; u_1
	\stackrel{*}{=}
	(\sigma'_1 \pecomp \sigma'_2) \by \beta \beta u_1
	= (\sigma'_1 \pecomp \sigma'_2) \by \beta u
	\subset (\sigma_1 \pecomp \sigma_2) \by \beta u$;
\\
& similarly,
	$\sigma'_2 u_2 \subset (\sigma_1 \pecomp \sigma_2) \by \beta u$.
\\
``$\supset$'':
& Suppose
	$(\sigma'_1 \pecomp \sigma'_2) \by \beta
	\subset (\sigma_1 \pecomp \sigma_2) \by \beta$,
	i.e.\
	$(\sigma'_1 \pecomp \sigma'_2) \by \beta \neq \{\}$
	hence
	$\sigma'_1 \by \beta \neq \{\}$,	\\
& and
	$(\sigma'_1 \pecomp \sigma'_2) \by \beta u
	= \sigma'_1 \by \beta \beta u_1
	\stackrel{*}{=} \sigma'_1 u_1
	\subset \sigma_1 u_1$;	\\
& similarly,
	$(\sigma_1 \pecomp \sigma_2) \by \beta u 
	\subset \sigma_2 u_2$.	\\
\end{tabular}
\\
The equations marked ``$\stackrel{*}{=}$'' hold by
Lemma~\eqr{27}.
}

\THEOREM{
%[(Extended Sort Inhabitance)]
\eqd{32}
%$\;$\\
$\sigma u \neq \{\}$
iff
$\sigma \rs {vars(u)} \cap \TT_{\func{vars(u)}{\cal CR}} \neq \{\}$.
\\
Note that 
$\TT_{\func{\{x_1,\ldots,x_n\}}{\cal CR}}
= compose(\{abstract(x_1,{\cal T}_{\cal CR}),\ldots,
abstract(x_n,{\cal T}_{\cal CR})\})$
is regular since ${\cal T}_{\cal CR}$ is regular.
}

\THEOREM{
\eqd{33}
Let $u, u_1,\ldots,u_n$ have pairwise disjoint variables,
let $\beta_i \pscomp \gamma_i = mgu(u,u_i)$ exist for all $i$.
Then,
$\sigma u \subset \tau_1 u_1 \cup \ldots \cup \tau_n u_n$
iff
$\forall i \;\;
(\sigma \by {\beta_i} \;\setminus\; \tau_i \by {\gamma_i}) \; \beta_i u
\subset
\bigcup_{j=1, \; j \neq i}^n
\tau_j u_j$
and
$\sigma \subset \bigcup_{i=1}^n \top \circ \beta_i$.
Note that we provide no algorithm to decide the latter condition.
}
\PROOF{
``$\Ra$'':
\\
Let $\sigma' \in \sigma$ 
such that $\sigma' \by {\beta_i} \neq \{\}$
and $\sigma' \by {\beta_i} \not\subset \tau_i \by {\gamma_i}$ 
for all
$i$.
\\
By assumption, $j \in \{1,\ldots,n\}$
and $\tau'_j \in \tau_j$ exist
such that $\sigma' u = \tau'_j u_j$.
\\
By \eqr{30},
$\sigma' \by {\beta_j} = \tau'_j \by {\gamma_i} \neq \{\}$,
hence $j \neq i$,
and $\sigma' \by {\beta_i} \beta_i u = \sigma' u = \tau'_j u_j$
by \eqr{27}.
\\
Next, consider an arbitrary $\sigma' \in \sigma$.
By assumption, $i$ and $\tau'_i \in \tau_i$ exist
such that $\sigma' u = \tau'_i u_i$.
\\
By \eqr{30} and \eqr{27},
$\sigma' \by {\beta_i} \beta_i u = \sigma' u$;
hence, 
$\{\sigma'\} 
= \sigma' \by {\beta_i} \circ \beta_i 
\subset \top \circ \beta_i$
\\
``$\La$'':
\\
Let $\sigma' \in \sigma$;
by assumption, an $i$ exists 
such that $\sigma' = \tau' \circ \beta_i$
for some $\tau' \in \top$.
\\
By \eqr{30},
$\sigma' \by {\beta_i} \neq \{\}$.
Case distinction:
\bi
\item $\sigma' \by {\beta_i} \not\subset \tau_i \by {\gamma_i}$,
	then by assumption
	$\sigma' u
	= \sigma' \by {\beta_i} \beta_i u
	= \tau'_j u_j$
	for some $j \neq i$ and $\tau'_j \in \tau_j$.
\item $\sigma' \by {\beta_i} = \tau'_i \by {\gamma_i} \neq \{\}$
	for some $\tau'_i \in \tau_i$,
	then
	$\sigma' u
	= \sigma' \by {\beta_i} \beta_i u
	= \tau'_i \by {\gamma_i} \gamma_i u_i
	= \tau'_i u_i$.
\ei
}

\THEOREM{
\eqd{34}
Let $u, u_1,\ldots,u_n$ have pairwise disjoint variables,
let $u$ be unifiable with each $u_i$.
\\
For $I \subset \{1,\ldots,n\}$
let
$\beta_I \pscomp \bigpscomp_{i \in I} \beta_{I,i}
= mgu(\{u\} \cup \{u_i \mid i \in I\})$
if it exists,
\\
$dom(\beta_I) = vars(u)$,
$dom(\beta_{I,i}) = vars(u_i)$
\footnote{I.e.\ $\beta_{\{\}}$ is a renaming substitution on $u$},
\\
let $J$ be the set of all $I$ with existing mgu.
\\
Let $\sigma_I := \{\sigma' \in \sigma
\mid \sigma' \by {\beta_{\{i\}}} \neq \{\} \lra i \in I\}$.
\\
Then, $\sigma u \subset \tau_1 u_1 \cup \ldots \cup \tau_n u_n$
iff $\sigma_I \by {\beta_I}
\subset \bigcup_{i \in I} \tau_i \by {\beta_{I,i}}$
for all $I \in J$.
The latter condition reads $\sigma_{\{\}} \subset \{\}$ for $I = \{\}$.
Note that the $\sigma_I$ are not regular, in general.
}
\PROOF{
First observe $(*)$:
for $\sigma' \in \sigma_I$,
$\{u\} \cup \{u_i \mid i \in I\}$ is simultaneously unifiable
\\
since 
$\sigma' \by {\beta_{\{i\}}} \beta_{\{i\}} u
= \sigma' u
= \sigma' \by {\beta_{\{i\}}} \beta_{\{i\},i} u_i$
for all $i \in I$.
\\
``$\Ra$'':
\\
Let $\sigma' \in \sigma_I$ for some $I \in J$;
by assumption, $\sigma' u = \tau'_i u_i$
for some $i $ and $\tau'_i \in \tau_i$.
\\
Applying \eqr{30} to $\sigma' u = \tau'_i u_i$
yields $\sigma' \by {\beta_{\{i\}}} \neq \{\}$;
hence $i \in I$.
\\
Applying \eqr{30} to $(*)$
and $\sigma' u = \tau'_i u_i$
yields $\sigma' \by {\beta_I} = \tau'_i \by {\beta_{I,i}} \neq \{\}$.
\\
``$\La$'':
\\
Let $\sigma' \in \sigma$,
and let $I := \{i \mid \sigma' \by {\beta_{\{i\}}} \neq \{\}\}$.
Then, $\sigma' \in \sigma_I$,
and $I \in J$ by $(*)$.
\\
By \eqR{35}.1 below, it follows that $\sigma' \by {\beta_I} \neq \{\}$,
hence $\sigma' \by {\beta_I} = \tau'_i \by {\beta_{I,i}}$
for some $i \in I$ and $\tau'_i \in \tau_i$
by assumption.
\\
Hence,
$\sigma' u
= \sigma' \by {\beta_I} \beta_I u
= \tau'_i \by {\beta_{I,i}} \beta_{I,i} u_i
= \tau'_i u_i$.
}

\LEMMA{
\eqd{35}
Using the notions of \eqr{34},
\\
let $I, I_1, I_2, I_3 \in J$
with $I_1 \subset I_2 \cap I_3$ and $I_2 \neq I_3$,
$\sigma' \in \sigma$,
$\sigma'_2 \in \sigma_{I_2}$,
$\sigma'_3 \in \sigma_{I_3}$,
then:
\be
\item $\sigma' \by {\beta_I} \neq \{\}$ 
	iff 
	$\forall i \in I \;\;\; \sigma' \by {\beta_{\{i\}}} \neq \{\}$;
\item $\sigma'_2 \by {\beta_{I_1}} \neq \sigma'_3 \by {\beta_{I_1}}$;
\item $\sigma_{I_2} \by {\beta_{I_1}} 
	= \sigma \by {\beta_{I_1}}
	\setminus \bigcup_{I_3 \in J, I_2 \neq I_3 \supset I_1} 
	\sigma_{I_3} \by {\beta_{I_1}}$.
\ee
}
\PROOF{	$\;$
\setlength{\leftmargini}{1.5cm}
\be
\item[1. ``$\Ra$'':]
	Let $i \in I$, then
	$\sigma' u 
	= \sigma' \by {\beta_I} \beta_I u 
	= \sigma' \by {\beta_I} \beta_{I,i} u_i$;
	by \eqr{30}, $\sigma' \by {\beta_{\{i\}}} \neq \{\}$.
\item[1. ``$\La$'':]
	Let $I = \{i_1,\ldots,i_m\}$,
	then
	\\
	$\sigma' u 
	\.= \sigma' \by {\beta_{\{i_1\}}} \beta_{\{i_1\}} u
	\.= \sigma' \by {\beta_{\{i_1\}}} \beta_{\{i_1\},i_1} u_{i_1}
	\.= \ldots
	\.= \sigma' \by {\beta_{\{i_m\}}} \beta_{\{i_m\}} u
	\.= \sigma' \by {\beta_{\{i_m\}}} \beta_{\{i_m\},i_m} u_{i_m}$,
	\\
	hence $\sigma' \by {\beta_I} \neq \{\}$ by \eqr{30}.
\item[2.:]
	By 1.,
	we have 
	$\sigma'_2 \by {\beta_{I_1}} 
	\neq \{\} 
	\neq \sigma'_3 \by {\beta_{I_1}}$;
	assume
	$\sigma'_2 \by {\beta_{I_1}} = \sigma'_3 \by {\beta_{I_1}}$.
	\\
	W.l.o.g., let $i \in I_3 \setminus I_2$,
	\\
	then
	$\sigma'_2 u
	= \sigma'_2 \by {\beta_{I_1}} \beta_{I_1} u
	= \sigma'_3 \by {\beta_{I_1}} \beta_{I_1} u
	= \sigma'_3 u
	= \sigma'_3 \by {\beta_{\{i\}}} \beta_{\{i\}} u
	= \sigma'_3 \by {\beta_{\{i\}}} \beta_{\{i\},i} u_i$,
	\\
	hence $\sigma'_2 \by {\beta_{\{i\}}} \neq \{\}$
	contradicting $i \not\in I_2$.
\item[3. ``$\subset$'':]
	Let $\sigma' \in \sigma_{I_2}$,
	then $\sigma' \in \sigma$
	and $\sigma' \by {\beta_{I_1}} \neq \sigma'' \by {\beta_{I_1}}$
	for all $\sigma'' \in \sigma_{I_3}$, $I_2 \neq I_3 \supset I_1$
	by 2.
\item[3. ``$\supset$:]
	Let $\sigma' \in \sigma$
	with $\sigma' \by {\beta_{I_1}} \neq \{\}$;
	define
	$I := \{i \in \{1,\ldots,n\} \mid
		\sigma' \by {\beta_{\{i\}}} \neq \{\}\}$,
	\\
	then $I \in J$, 
	since
	$\sigma' \pecomp \bigpecomp_{i \in I}
		\sigma' \by {\beta_{\{i\}}} \beta_{\{i\}}$
	unifies $\{u\} \cup \{u_i \mid i \in I\}$,
	\\
	and $I_1 \subset I$,
	since $\sigma' \by {\beta_{\{i\}}} \neq \{\}$
	for all $i \in I_1$
	by 1.
	\\
	Case distinction:
	\be
	\item $I \neq I_2$;
		then 
		$\sigma' \by {\beta_{I_1}}
		\subset \sigma_I \by {\beta_{I_1}}$,
		with $I_2 \neq I \supset I_1$,
		hence $I$ is one of the $I_3$,
		\\
		i.e., $\sigma' \by {\beta_{I_1}}$ 
		is not contained in the right hand side,
		and we have nothing to show.
	\item $I = I_2$;
		then, $\sigma' \by {\beta_{I_1}}$
		is contained in the left hand side.
	\ee
\ee
}

\EXAMPLE{
\eqd{36}
Let $J = \{
	\{\}, 
	\{1\},
	\{2\},
	\{3\},
	\{1,2\},
	\{1,3\}
	\}$,
then \eqr{35}.3 yields following equations,
where e.g.\ $\sigma_{\{i,j\}}$ is written as $\sigma_{ij}$;
similarly for $\beta$:
\\
\begin{tabular}[t]{@{}l@{$\;$}l@{$\;$}l@{}}
$\sigma_{12} \by {\beta_{12}}$
	& $= \sigma \by {\beta_{12}}$	\\
$\sigma_{13} \by {\beta_{13}}$
	& $= \sigma \by {\beta_{13}}$	\\
$\sigma_{12} \by {\beta_{1}}$
	& $= \sigma \by {\beta_{1}}$
	& $\setminus (\sigma_{1} \by {\beta_{1}}
	\cup \sigma_{13} \by {\beta_{1}})$	\\
$\sigma_{1} \by {\beta_{1}}$
	& $= \sigma \by {\beta_{1}}$
	& $\setminus (\sigma_{12} \by {\beta_{1}}
	\cup \sigma_{13} \by {\beta_{1}})$	\\
$\sigma_{12} \by {\beta_{2}}$
	& $= \sigma \by {\beta_{2}}$
	& $\setminus \sigma_{2} \by {\beta_{2}}$	\\
$\sigma_{2} \by {\beta_{2}}$
	& $= \sigma \by {\beta_{2}}$
	& $\setminus \sigma_{12} \by {\beta_{2}}$	\\
$\sigma_{13} \by {\beta_{3}}$
	& $= \sigma \by {\beta_{3}}$
	& $\setminus \sigma_{3} \by {\beta_{3}}$	\\
$\sigma_{3} \by {\beta_{3}}$
	& $= \sigma \by {\beta_{3}}$
	& $\setminus \sigma_{13} \by {\beta_{3}}$	\\
\end{tabular}
}

\section{Regular T-Sets and Algorithms}
\label{Regular T-Sets and Algorithms}

In this section,
we introduce the notion of a regular t-set and provide algorithms to
compute with them.
We obtain a decidability result for a class of Horn clauses that is
isomorphic to regular t-sets (Cor.~\eqR{43}).
We present some simple relations like $x < y$ that can be expressed by
regular t-sets
(Figs.~\ref{Examples of regular t-sets} and
\ref{Some relations expressible by regular t-sets}),
and operations on relations that can be computed
(Fig.~\ref{Computable operations on relations in t-set form}).

Using the result from Sect.~\ref{T-Substitutions},
we can describe regular sets of ground
substitutions as subsets of the initial term algebra
$\TT_{\pfunc{\cal V}{\cal CR}}$.
We will only consider t-sets with a unique domain
$dom(\sigma') = V_\sigma$ for all $\sigma' \in \sigma$;
define $\eqi{dom(\sigma)} := V_\sigma$.
The empty t-set is again denoted by \eqi{$\bot$}; it will be clear from
the context whether $\bot$ denotes the empty sort or the empty t-set.
For each finite $V$, $\eqi{\top_V} := \TT_{\pfunc{V}{\cal CR}}$
is expressible as a regular set.
We write \eqi{$\top$} for $\top_V$ when $V$ is clear from the context;
note that $\TT_{\pfunc{\cal V}{\cal CR}}$ is
not expressible since infinitely many t-substitution constructors exist.

We immediately inherit the mechanisms and algorithms
given in Sect.~\ref{Regular Sorts}, i.e.\ for intersection, relative
complement, and inhabitance.
In addition, the operations defined in \eqr{18},
\eqr{21}, \eqr{22}, and \eqr{23} can be
computed for regular t-sets.

\begin{figure}
\begin{center}
\begin{tabular}[t]{@{}|l@{$\;$}l@{\hspace*{2.0em}}l|@{}}
\hline
\multicolumn{3}{@{}|l|@{}}{T-sets, described as regular sets:}	\\
\eqi{$Nat_x$} & $\sortdef 0_x \mid s_x(Nat_x)$
	& $\corresponds \{[x \.{:=} s^i(0)] \mid i \in \N\}$	\\
\eqi{$Nat_y$} & $\sortdef 0_y \mid s_y(Nat_y)$
	& $\corresponds \{[y \.{:=} s^i(0)] \mid i \in \N\}$	\\
\eqi{$Nat_{x\.=y}$} & $\sortdef 0_x 0_y \mid s_x s_y(Nat_{x\.=y})$
	& $\corresponds \{[x \.{:=} s^i(0), y \.{:=} s^i(0)] \mid
	i \in \N\}$	\\
\eqi{$Nat_{x,y}$} & $\sortdef 0_x 0_y \mid s_x s_y(Nat_{x,y}) \mid$ & \\
	& $\;\;\;\; 0_x s_y(Nat_y) \mid s_x 0_y(Nat_x)$
	& $\corresponds \{[x \.{:=} s^i(0), y \.{:=} s^j(0)] \mid
	i, j \in \N\}$	\\
\eqi{$Nat_{x\.<y}$}
	& $\sortdef 0_x s_y(Nat_y) \mid s_x s_y(Nat_{x\.<y})$
	& $\corresponds \{[x \.{:=} s^i(0), y \.{:=} s^j(0)]
	\mid i,j \in \N, i < j\}$	\\
\hline
\end{tabular}
\end{center}

\caption{Examples of regular t-sets}
\label{Examples of regular t-sets}
\end{figure}

\ALGORITHM{
%[(T-Set Application)]
\eqd{37}
%$\;$\\
The following algorithm computes the elementwise application of a
regular t-set to a variable.
Let $\sigma$ be the name of a regular t-set,
and let $S$ be a new sort name.
Define $\eqi{apply(\sigma,x)} = S$,
where the algorithm introduces a new sort definition for $S$:
\be
\item If $apply(\sigma,x)$ has already been called earlier,
	$S$ is already defined (loop check).
\item Else, if $\sigma \sortdef \sigma_1 \mid \ldots \mid \sigma_n$,
	define
	$S \sortdef apply(\sigma_1,x) \mid \ldots \mid
	apply(\sigma_n,x)$
\item Else, if $\sigma \sortdef \vec{cr}(\sigma_1,\ldots,\sigma_n)$
	with $x \in dom(\vec{cr})$,
	\\
	define $S \sortdef
	\vec{cr}_x(apply(\sigma_1,x),\ldots,
	apply(\sigma_{ar(\vec{cr}_x)},x))$,
\item Else, define $S \sortdef \bot$.
\ee
Using the t-set version of Thm.~\eqr{7}
with $p_S(u) :\Lra u \in \sigma^M x$ if $S = apply(\sigma,x)$,
it can be shown that $apply(\sigma,x)^M = \sigma^M x$.
The algorithm needs at most $\#use(\sigma)$ recursive calls to compute
$apply(\sigma,x)$.
}

% \EXAMPLE{
% Using the t-set definitions from 
% Fig.~\ref{Examples of t-substitutions and regular t-sets}, 
% Alg.~\eqr{37} yields
% 	\\
% 	\begin{tabular}[t]{@{}r@{$\;$}c@{$\;$}ll@{}}
% 	$apply(Nat_x,x)$ & $= Sort_\eqda{34}$ & $\sortdef apply(0_x,x)
% 		\mid apply(s_x(Nat_x),x)$ & by 2	\\
% 	$apply(0_x,x)$ & $= Sort_\eqda{35}$ & $\sortdef 0$ & by 3 \\
% 	$apply(s_x(Nat_x),x)$ & $= Sort_\eqda{36}$
% 		& $\sortdef s(apply(Nat_x,x))$ & by 3	\\
% 	$apply(Nat_x,x)$ & $= Sort_\eqra{34}$ && by 1	\\
% 	\end{tabular}
% 	\\
% Hence $Sort_\eqra{34} \sortdef Sort_\eqra{35} \mid Sort_\eqra{36}$,
% $Sort_\eqra{35} \sortdef 0$,
% $Sort_\eqra{36} \sortdef s(Sort_\eqra{34})$,
% and the sort equivalence algorithm mentioned in 
% Sect.~\ref{Regular Sorts} will yield that $Sort_\eqra{34}^M = Nat^M$,
% using the sort definitions from 
% Fig.~ \ref{Sort and function definitions for synthesis of binary
% arithmetic algorithms}. 
% }

Although t-substitutions are homomorphic wrt.\ all
constructors in ${\cal CR}$, t-sets are generally not; e.g., using the
definitions from Fig.~\ref{Examples of regular t-sets},
$Nat_{x<y}^M (\tpl{x,y}) 
\psubset \tpl{Nat_{x<y}^M (x),Nat_{x<y}^M (y)}$,
cf.\ Lemma~\eqr{25}.
Such t-sets can express certain relations
between distinct variables, e.g.\ $Nat_{x\.<y}$ always assigns a value
to $x$ that is less than the value assigned to $y$.
Figure~\ref{Some relations expressible by regular t-sets} 
shows some more
nontrivial relations that are expressible by regular t-sets.
Figure \ref{Computable operations on relations in t-set form} shows
operations on relations that can be computed for t-sets.

\DEFINITION{
\eqd{38}
We call a t-set $\sigma$ {\eqi{independent}} if it is homomorphic
on linear terms, i.e.\ if
$\sigma \tpl{x_1,\ldots,x_n} = \tpl{\sigma x_1,\ldots,\sigma x_n}$,
otherwise we call it ``\eqi{dependent}''.
An independent t-set assigns the value of one variable independently of
the value of the others, e.g.\ $Nat_{x,y}$ in
Fig.~\ref{Examples of regular t-sets}.
A finite union $\sigma_1 \cup \ldots \cup \sigma_n$
of independent t-sets $\sigma_i$ is called \eqi{semi-independent}.
The intersection of two (semi-)independent t-sets is again
(semi-)independent;
the union of two semi-independent t-sets is trivially semi-independent.
}

\ALGORITHM{
\eqd{39}
The following algorithm computes the elementwise application of a
regular t-set to a linear constructor term. 
Let $u$ be a linear constructor term,
let $\sigma$ be the name of a regular t-set 
such that $\sigma \rs {vars(u)}$ is independent.
Define
$apply(\sigma,cr(u_1,\ldots,u_n))
\sortdef cr(apply(\sigma,u_1),\ldots,apply(\sigma,u_n))$;
if $u \in {\cal V}$, compute $apply(\sigma,u)$ 
by Alg.~\eqr{37}.
Then, $apply(\sigma,u)^M = \sigma^M u$.
}

\ALGORITHM{
%[(T-Set Restriction)] 
\eqd{40}
%$\;$\\
The following algorithm computes the elementwise restriction of a
regular t-set to a set of variables.
Let $\sigma$ be the name of a regular t-set,
$V \subset {\cal V}$,
and let $\tau$ be a new name for a regular t-set.
Define $\eqi{restrict(\sigma,V)} = \tau$,
where the algorithm introduces a new t-set definition for $\tau$:
\be
\item If $restrict(\sigma,V)$ has already been called earlier,
	$\tau$ is already defined (loop check).
\item Else, if $\sigma \sortdef \sigma_1 \mid \ldots \mid \sigma_n$,
	define
	$\tau \sortdef restrict(\sigma_1,V) \mid \ldots \mid
	restrict(\sigma_n,V)$
\item Else, if $\sigma \sortdef \vec{cr}(\sigma_1,\ldots,\sigma_n)$,
	define $\tau \sortdef
	(\vec{cr} \rs V) \; (restrict(\sigma_1,V),\ldots,
	restrict(\sigma_m,V))$,
	\\
	where $m = ar(\vec{cr} \rs V)$.
\ee
Using Thm.~\eqr{7}
with $p_\tau(\tau') :\Lra \tau' \in \sigma^M \rs V$
if $\tau = restrict(\sigma,V)$,
\\
it can be shown that $restrict(\sigma,V)^M = \sigma^M \rs V$.
The algorithm needs at most $\#use(\sigma)$
recursive calls to compute $restrict(\sigma,V)$.
If $\sigma$ is (semi-)independent, then so is $restrict(\sigma,V)$.
}

\ALGORITHM{
\eqd{41}
The following algorithm computes the elementwise parallel composition
of two regular t-sets $\sigma$ and $\tau$.
Let $\mu$ be a new name for a regular t-set.
Define $\eqi{compose(\sigma,\tau)} = \mu$,
where the algorithm introduces a new t-set definition for $\mu$:
\be
\item If $compose(\sigma,\tau)$ has already been called earlier,
	$\mu$ is already defined (loop check).
\item Else, if $\sigma \sortdef \sigma_1 \mid \ldots \mid \sigma_n$,
	define
	$\mu \sortdef compose(\sigma_1,\tau) \mid \ldots \mid
	compose(\sigma_n,\tau)$.
\item Else, if $\tau \sortdef \tau_1 \mid \ldots \mid \tau_n$,
	define
	$\mu \sortdef compose(\sigma,\tau_1) \mid \ldots \mid
	compose(\sigma,\tau_n)$.
\item Else, if $\sigma \sortdef \vec{cr}(\sigma_1,\ldots,\sigma_n)$,
	\hfill
	$\tau \sortdef \vec{cr}'(\tau_1,\ldots,\tau_m)$,
	\hfill
	$\vec{cr}$ and $\vec{cr}'$ agree on their domain intersection,
	\\
	and w.l.o.g.\ $n \leq m$,
	define
	$\mu \sortdef
	(\vec{cr} \pecomp \vec{cr}') \;
	(compose(\sigma_1,\tau_1),\ldots,
	compose(\sigma_n,\tau_n),\tau_{n+1},\ldots,\tau_m)$.
\item Else, if $\sigma \sortdef \vec{cr}(\sigma_1,\ldots,\sigma_n)$,
	$\tau \sortdef \vec{cr}'(\tau_1,\ldots,\tau_m)$,
	and $\vec{cr}$ and $\vec{cr}'$ do not 
	agree on their domain intersection,
	define
	$\mu \sortdef \bot$.
\ee
Using Thm.~\eqr{7}
with $p_\mu(\mu') :\Lra \mu' \in \sigma^M \pecomp \tau^M$
if $\mu = compose(\sigma,\tau)$,
it can be shown that $compose(\sigma,\tau)^M = \sigma^M \pecomp \tau^M$.
The algorithm needs at most $\#use(\sigma) * \#use(\tau)$
recursive calls to compute $compose(\sigma,\tau)$.
If $\sigma$ and $\tau$ are both (semi-)independent, then so is
$compose(\sigma,\tau)$.
\\
We write $compose(\{\sigma_1,\ldots,\sigma_n\})$
for $compose(\sigma_1,compose(\ldots,compose(\sigma_{n-1},
\sigma_n) \ldots))$.
}

\ALGORITHM{
%[(Abstraction)] 
\eqd{42}
%$\;$\\
The following algorithm computes the elementwise lifting of a
regular sort to a regular t-set.
Let $S$ be the name of a regular sort,
$x \in {\cal V}$,
and let $\sigma$ be a new name for a regular t-set.
Define $\eqi{abstract(S,x)} = \sigma$,
where the algorithm introduces a new t-set definition for $\sigma$:
\be
\item If $abstract(S,x)$ has already been called earlier,
	$\sigma$ is already defined (loop check).
\item Else, if $S \sortdef S_1 \mid \ldots \mid S_n$,
	define
	$\sigma \sortdef abstract(S_1,x) \mid \ldots \mid
	abstract(S_n,x)$.
\item Else, if $S \sortdef cr(S_1,\ldots,S_n)$,
	define $\sigma \sortdef
	(x \.\mapsto cr) \; (abstract(S_1,x),\ldots,
	abstract(S_n,x))$.
\ee
Using Thm.~\eqr{7}
with $p_\sigma(\sigma') :\Lra \sigma' \in [x \.{:=} S^M]$
if $\sigma = abstract(S,x)$,
\\
it can be shown that $abstract(S,x)^M = [x \.{:=} S^M]$.
The algorithm needs at most $\#use(S)$
recursive calls to compute $abstract(x,S)$.
$abstract(x,S)$ always yields an independent t-set.
}

\begin{figure}
\begin{center}
\begin{tabular}[t]{@{}|l@{$\;$}l|@{}}
\hline
\multicolumn{2}{@{}|l|@{}}{Expressible relations e.g.:}	\\
& \\
\multicolumn{2}{@{}|l|@{}}{\sf prefix $x$ of length $y$ 
	of a $snoc$-list $z$ with regular element sort $Elem$}	\\
$\eqi{Pref_{x,y,z}}$ & $\sortdef nil_x 0_y nil_z
	\mid snoc_x s_y snoc_z(Pref_{x,y,z},Elem_{x=z})
	\mid nil_x 0_y snoc_z(List_x,Elem_x)$	\\
$List_x$ & $\sortdef nil_x \mid snoc_x(List_x,Elem_x)$	\\
$Elem_x$ & $= abstract(x,Elem)$	\\
$Elem_{x=z}$ & $= dup(Elem_x,[z \.{:=} x])$	\\
[0.2cm]
\multicolumn{2}{@{}|l|@{}}{\sf lexicographical order on $cons$-lists 
	wrt.\ regular element ordering $Elem_{x<y}$}	\\
$\eqi{Lex_{x<y}}$ & $\sortdef nil_x cons_y(Elem_y,List_y)
	\mid cons_x cons_y(Elem_{x=y},Lex_{x<y})$	\\
    & $\mid cons_x cons_y(Elem_{x<y},List_{x,y})$	\\
$Elem_x$ & $= abstract(x,Elem)$	\\
$Elem_y$ & $= abstract(y,Elem)$	\\
$Elem_{x=y}$ & $= dup(Elem_x,[y \.{:=} x])$	\\
$List_x$ & $\sortdef nil_x \mid cons_x(Elem_x,List_x)$	\\
$List_y$ & $\sortdef nil_y \mid cons_y(Elem_y,List_y)$	\\
$List_{x,y}$ & $= compose(List_x,List_y)$	\\
[0.2cm]
% \multicolumn{2}{@{}|l|@{}}{\sf lifting of element relation to list} \\
% [0.2cm]
\multicolumn{2}{@{}|l|@{}}{\sf matching of tree $x$ 
	at the root of tree $y$ 
	(variable bindings not considered)}	\\
\multicolumn{2}{@{}|l|@{}}{\sf set of functions symbols $F$, 
	only one variable symbol $v$}	\\
$\eqi{Mtch_{x,y}}$ & $\sortdef \bigmid_{f \in F} \;
	f_x f_y(Mtch_{x,y},\ldots,Mtch_{x,y}) 
	\mid v_x f_y(Term_y,\ldots,Term_y)$	\\
$Term$ & $\sortdef v \mid \bigmid_{f \in F} \; f(Term,\ldots,Term)$ \\
$Term_y$ & $= abstract(y,Term)$	\\
[0.2cm]
\multicolumn{2}{@{}|l|@{}}{\sf sum $z$ of binary strings $x$ and $y$ 
	($cons$-lists, least bit first)}	\\
$\eqi{Sum_{x,y,z}}$ & $\sortdef Sum_{0,x,y,z}$	\\
$Sum_{0,x,y,z}$ & $\sortdef nil_x nil_y nil_z$	\\
    & $\mid cons_x nil_y cons_z(i_x i_z \.\mid o_x o_z,Bin_{0,x=z})$ \\
    & $\mid cons_x cons_y cons_z(o_x o_y o_z \.\mid o_x i_y i_z
		\.\mid i_x o_y i_z, Sum_{0,x,y,z})$	\\
    & $\mid cons_x cons_y cons_z(i_x i_y o_z,Sum_{1,x,y,z})$	\\
$Sum_{1,x,y,z}$ & $\sortdef nil_x nil_y cons_z(i_z,nil_z)$	\\
    & $\mid cons_x nil_y cons_z(i_x o_z,Bin_{1,x=z})$	\\
    & $\mid cons_x cons_y cons_z(o_x o_y i_z,Sum_{0,x,y,z})$	\\
    & $\mid cons_x cons_y cons_z(o_x i_y o_z \.\mid i_x o_y o_z
		\.\mid i_x i_y i_z,Sum_{1,x,y,z})$	\\
$Bin'$ & $\sortdef nil \mid cons(o,Bin') \mid cons(i,Bin')$	\\
$Bin_x$ & $=abstract(x,Bin')$	\\
$Bin_{0,x=z}$ & $=dup(Bin_x,[z \.{:=} x])$	\\
$Bin_{1,x=z}$ & $\sortdef nil_x cons_z(i_z,nil_z)
	\mid cons_x cons_z(o_x i_z,Bin_{0,x=z})$	\\
	& $\mid cons_x cons_z(i_x o_z,Bin_{1,x=z})$	\\
% [0.2cm]
% \multicolumn{2}{@{}|l|@{}}{\sf zip $z$ of $cons$-lists $x$ and $y$} \\
% $\eqi{Zip_{x,y,z}}$ & $\sortdef nil_x nil_y nil_z
% 	\mid cons_x cons_y cons_z(Elem_{x\.{::}y=z},Zip_{x,y,z})$\\
% $Elem_{x\.{::}y=z}$ & $fact(dup(abstract(z'',Elem\.{::}Elem),
% 	[z'' \.{:=} z']),[z'' \.{:=} x \.{::} y,z' \.{:=} z])$	\\
\hline
\end{tabular}
\end{center}

\caption{Some relations expressible by regular t-sets} 
\label{Some relations expressible by regular t-sets}
\end{figure}

\begin{figure}
\begin{center}
\begin{tabular}[t]{@{}|l@{\hspace*{0.5cm}}l|@{}}
\hline
\multicolumn{2}{@{}|l|@{}}{Operations on relations  e.g.:}	\\
& \\
relation join & $\sigma \pecomp \tau$	\\
relational image &	\\
$R[x_0] = \{ y \mid x_0 \mathrel{R} y\}$
	& $R[x_0] = apply(compose(R,abstract(x,x_0)),y)$	\\
factorization wrt.\ equivalence &	\\
$x \mathrel{R_E} y \Lra \exists x', y' \;\;
	x \mathrel{E} x' \mathrel{R} y' \mathrel{E} y$
	& $R_E = compose(\{E,R,E\})$	\\
equivalence from mapping  &	\\
$x \mathrel{E_M} y \Lra M(x) = M(y)$
	& $E_M = compose(fact(M,[x \.{:=} x',y \.{:=} y']),
		fact(M,[x \.{:=} y',y \.{:=} x']))$	\\
restriction & $\sigma \rs V$	\\
bounded renaming & $\sigma \by \beta$	\\
conjunction & $\sigma \cap \tau$	\\
disjunction & $\sigma \mid \tau$	\\
negation & $\top \setminus \sigma$	\\
\hline
\end{tabular}
\end{center}
\vspace{0.3ex}

For example, using the definitions
from Fig.~\ref{Some relations expressible by regular t-sets},
$restrict(Pref_{x,y,z},\{x,y\})$
yields the length function on $snoc$-lists,
\\
and $apply(compose(Pref_{x,y,z},abstract(y,s^3(0))),x)$
yields the regular sort of all $snoc$-lists of length 3.

\caption{Computable operations on relations in t-set form} 
\label{Computable operations on relations in t-set form}
\end{figure}

\noindent
\parbox[b]{6.6cm}{%
\hspace*{\outerparindent}%
Regular sorts from Sect.~\ref{Regular Sorts} can be shown to
correspond to Horn clauses with unary predicates and thus decide this
theory class by extending the form of sort expressions allowed on the
right-hand side of a sort definition to include intersections, too.
}
\hfill
\begin{picture}(4.4,2.3)
	%\put(0,0){\makebox(0,0){+}}
\put(1.900,0.213){\line(0,1){1.875}}
\put(2.300,0.213){\line(0,1){1.875}}
\put(1.000,2.275){\makebox(0.000,0.000){$\scriptstyle \vec{y}_1$}}
\put(2.100,2.275){\makebox(0.000,0.000){$\scriptstyle \vec{y}_j$}}
\put(4.000,2.275){\makebox(0.000,0.000){$\scriptstyle \vec{y}_n$}}
\put(0.300,1.713){\makebox(0.000,0.000){$\scriptstyle \vec{x}_1$}}
\put(0.300,0.587){\makebox(0.000,0.000){$\scriptstyle \vec{x}_m$}}
\put(1.000,1.713){\makebox(0.000,0.000){$\scriptstyle x_{11}$}}
\put(2.850,1.713){\makebox(0.000,0.000){$\scriptstyle x_{1a}$}}
\put(1.000,0.587){\makebox(0.000,0.000){$\scriptstyle x_{m1}$}}
\put(3.400,0.587){\makebox(0.000,0.000){$\scriptstyle x_{ma}$}}
\put(1.900,0.450){\grey{4}{3}}
\put(1.900,0.850){\grey{4}{3}}
\put(1.900,1.600){\grey{4}{3}}
\thicklines
\put(0.700,0.400){\line(1,0){3.000}}
\put(0.700,0.775){\line(1,0){3.600}}
\put(0.700,1.150){\line(1,0){3.600}}
\put(0.700,1.525){\line(1,0){2.400}}
\put(0.700,1.900){\line(1,0){2.400}}
\put(0.700,0.400){\line(0,1){1.500}}
\put(3.700,0.400){\line(0,1){0.375}}
\put(4.300,0.775){\line(0,1){0.375}}
\put(1.300,1.150){\line(0,1){0.375}}
\put(3.100,1.525){\line(0,1){0.375}}
	\put(0.720,0.415){\line(1,0){3.000}}
	\put(0.720,0.790){\line(1,0){3.600}}
	\put(0.720,1.165){\line(1,0){3.600}}
	\put(0.720,1.540){\line(1,0){2.400}}
	\put(0.720,1.915){\line(1,0){2.400}}
	\put(0.720,0.415){\line(0,1){1.500}}
	\put(3.720,0.415){\line(0,1){0.375}}
	\put(4.320,0.790){\line(0,1){0.375}}
	\put(1.320,1.165){\line(0,1){0.375}}
	\put(3.120,1.540){\line(0,1){0.375}}
\put(0.680,0.385){\line(1,0){3.000}}
\put(0.680,0.760){\line(1,0){3.600}}
\put(0.680,1.135){\line(1,0){3.600}}
\put(0.680,1.510){\line(1,0){2.400}}
\put(0.680,1.885){\line(1,0){2.400}}
\put(0.680,0.385){\line(0,1){1.500}}
\put(3.680,0.385){\line(0,1){0.375}}
\put(4.280,0.760){\line(0,1){0.375}}
\put(1.280,1.135){\line(0,1){0.375}}
\put(3.080,1.510){\line(0,1){0.375}}
\end{picture}

%\noindent
Similarly,
regular t-sets correspond to Horn clauses of the following form:
\\
$p(cr_1(\vec{x}_1),\ldots,cr_m(\vec{x}_m))
\la p_1(\vec{y}_1) \wedge \ldots \wedge p_1(\vec{y}_n)$
where
$\vec{x}_i := \tpl{x_{i1},\ldots,x_{ia_i}}$
for $i = 1,\ldots,m$,
and
$\vec{y}_j := \tpl{x_{ij} \mid i=1,\ldots,m, j \leq ar(cr_i)}$
for $j = 1,\ldots,n$ 
with $n := max_{i=1,\ldots,m} ar(cr_i)$.
The relation between $\vec{x}_i$ and $\vec{y}_j$ is shown in the
above diagram.
If all term constructors $cr_i$ have the same arity $n$, 
the $\vec{y}_j$ are the column vectors of an $m \times n$
matrix built from the
$\vec{x}_i$ as line vectors.
For example, the definition
$Lgth_{x,y} \sortdef 0_x nil_y \mid s_x snoc_y(Lgth_{x,y},Nat_y)$
corresponds to the Horn clauses
	$lgth(0,nil)$ and
	$lgth(s(x),snoc(y_1,y_2)) \la lgth(x,y_1) \wedge nat(y_2)$.
We thus have the following

\COROLLARY{
\eqd{43}
The satisfiability of any
predicate defined by Horn clauses of the above form can be decided.
The set of such predicates is closed wrt.\ conjunction,
disjunction, and negation.
}

\ALGORITHM{
\eqd{44}
The following algorithm ``duplicates'' each t-substitution $\sigma'$
in $\sigma$, i.e., it composes $\sigma'$ with a renamed copy of itself.
Let $\sigma$ be the name of a regular t-set, 
let $\beta$ be an ordinary idempotent substitution
with $\beta x \in {\cal V}$ for all $x \in dom(\beta)$,
$\beta$ need not be linear.
Let $\tau$ be a new name for a regular t-set.
Define $\eqi{dup(\sigma,\beta)} = \tau$,
where the algorithm introduces a new t-set definition for $\tau$:
\be
\item If $dup(\sigma,\beta)$ has already been called earlier, $\tau$ is
	already defined (loop check).
\item Else, if $\sigma \sortdef \sigma_1 \mid \ldots \mid \sigma_n$,
	define
	$\tau \sortdef dup(\sigma_1,\beta) \mid \ldots \mid
	dup(\sigma_n,\beta)$.
\item Else, if $\sigma \sortdef \vec{cr}(\sigma_1,\ldots,\sigma_n)$
	and $cr_{x_i} = cr_{x_j}$ whenever $\beta x_i = \beta x_j$, \\
	let $\vec{cr}'$ be a t-substitution constructor 
	such that $\vec{cr}'_{\beta x} = \vec{cr}_x$ 
	for all $x \in dom(\beta)$,
	\\
	define
	$\tau \sortdef
	(\vec{cr} \pecomp \vec{cr}') \;
	(dup(\sigma_1,\beta),\ldots,dup(\sigma_n,\beta))$.
\item Else, define $\theta \sortdef \bot$.
\ee
Using Thm.~\eqr{7} with 
$p_\tau(\tau') :\Lra \tau' \in \bigcup_{\sigma' \in \sigma}
\sigma' \pecomp (\sigma' \by \beta)$,
it can be shown that 
$dup(\sigma,\beta)^M 
= \bigcup_{\sigma' \in \sigma} \sigma' \pecomp (\sigma' \by \beta)$.
The algorithm needs at most $\#use(\sigma)$
recursive calls to compute $dup(\sigma,\beta)$.
}

\EXAMPLE{
\eqd{45}
Using the definitions in Fig.~\ref{Examples of
regular t-sets}, we get $dup(Nat_x,[y \.{:=} x]) = Nat_{x\.=y}$.
}

\ALGORITHM{
\eqd{46}
If $\sigma$ is regular and $\beta x \in {\cal V}$ 
for all $x \in dom(\beta)$,
$\sigma \circ \beta$ is again regular;
in general, it is not.
In the former case, the following algorithm computes a regular t-set
definition for $\sigma \circ \beta$:
\be
\item If $\beta = \beta_1 \pscomp \beta_2$ such that $\beta_1$ and
	$\beta_2$ are each injective, i.e.\ renamings,
	let $\gamma$ be a renaming on $ran(\beta_2)$,
	then
	$\sigma \circ \beta
	= fact(dup(\sigma,\gamma),
	\beta_1^{-1} \pscomp (\beta_2^{-1} \circ \gamma^{-1}))$.
\item Any other $\beta$ can be represented 
	as $\beta_1 \circ \ldots \circ \beta_n$
	such that 
	$1 \leq \#\{x \mid \beta_i x = y\} \leq 2$ for all $y$
	and for all $i$,
	i.e.\ each $\beta_i$ has the form required by 1.;
	then
	$\sigma \circ \beta
	= ( \ldots (\sigma \circ \beta_1) \circ \ldots ) \circ \beta_n$.
\ee
}

\ALGORITHM{\eqd{47}
The following algorithm computes $fact(\sigma,\beta)$ 
if $\beta x \in {\cal V}$ for all $x$.
Let $\sigma$ be a regular t-set,
let $\mu$ be a new name for a regular t-set;
define $\eqi{fact(\sigma,\beta)} = \mu$,
where a new t-set definition is introduced for $\mu$:
\be
\item If $fact(\sigma,\beta)$ has been called earlier,
	$\mu$ is already defined (loop checking).
\item Else, if $\sigma \sortdef \sigma_1 \mid \ldots \mid \sigma_n$,
	define
	$\mu \sortdef fact(\sigma_1,\beta) \mid \ldots \mid
	fact(\sigma_n,\beta)$.
\item Else, if $\sigma \sortdef \vec{cr}(\sigma_1,\ldots,\sigma_n)$
	with $dom(\beta) \subset dom(\vec{cr})$,	\\
	and $\vec{cr}_x = \vec{cr}_y$ whenever $\beta x = \beta y$,
	define $\vec{cr}': ran(\beta) \ra {\cal CR}$
	by $\vec{cr}'_{\beta x} := \vec{cr}_x$, \\
	define $\mu \sortdef
	\vec{cr}'(fact(\sigma_1,\beta),\ldots,
fact(\sigma_{ar(\vec{cr}')},\beta))$.
	\item Else, define $\mu \sortdef \bot$.
\ee
Using the induction principle from Thm.~\eqr{7}, lifted to t-sets,
with $p_\mu(\sigma') :\Lra \sigma' \in \sigma^M \by \beta$
if $\mu = fact(\sigma,\beta)$,
it can be shown that $fact(\sigma,\beta)^M = \sigma^M \by \beta$.
The algorithm needs at most $\#use(\sigma)$
recursive calls to compute $fact(\sigma,\beta)$.
}

\DEFINITION{\eqd{48}
$\beta$ is called \eqi{homogeneous}
if all variables in the range of $\beta$ occur at the same depth,
i.e.,
if $\beta x \in {\cal V}$ for each $x \in dom(\beta)$, 
or if $\beta x = cr_x(u_{x1},\ldots,u_{x \; ar(cr_x)})$
for all $x \in dom(\beta)$ and appropriate $u_{xi}$,
and 
$[x \.{:=} u_{xi} \mid x \in dom(\beta),
ar(cr_x) \geq i]$
is again homogeneous for each $i$.
}

% \EXAMPLE{
% \ eqd{3120}
% Intuitively, a substitution $\beta$ is homogeneous if all its range
% terms have the same structure up to different arities.
% Let $c$ and $d$ be a binary and a unary constructor,
% respectively;
% $[x \.{:=} c(x_1,x_2), y \.{:=} y_1]$
% is not homogeneous,
% but both
% $[x \.{:=} c(x_1,x_2), y \.{:=} c(y_2,y_3)]$
% and $[x \.{:=} c(x_1,x_2), y \.{:=} d(y_2)]$
% are.
% }

\ALGORITHM{\eqd{49}
The following algorithm computes $fact(\sigma,\beta)$
if $\beta$ is homogeneous.
Let $\sigma$ be a regular t-set,
define $\eqi{fact(\sigma,\beta)}$ by:
\be
\item If $\beta = [\;]$,
	define $fact(\sigma,\beta) := \{\varepsilon\}$.
\item Else, if $\beta x \in {\cal V}$ for all $x$,
	compute $fact(\sigma,\beta)$ by Alg.\ \eqr{47}.
\item Else, if $\sigma \sortdef \sigma_1 \mid \ldots \mid \sigma_n$,
	define
	$fact(\sigma,\beta) :=
	fact(\sigma_1,\beta) \mid \ldots \mid fact(\sigma_n,\beta)$.
\item Else, if $\sigma \sortdef
	\vec{cr}(\sigma_1,\ldots,\sigma_n)$,
	$dom(\beta) \subset dom(\vec{cr})$,
	and $\beta x = \vec{cr}_x(u_{x,1},\ldots,u_{x,ar(\vec{cr}_x)})$
	for all $x \in dom(\beta)$,
	\\
	define
	\begin{tabular}[t]{@{}l@{$\;$}ll@{}}
	$fact(\sigma,\beta) := compose(\{$
		& $fact(\sigma_1,[x \.{:=} u_{x,1}
		\mid x \in dom(\beta), ar(\vec{cr}_x) \geq 1]), 
		\ldots,$	\\
	& $fact(\sigma_n,[x \.{:=} u_{x,n}
		\mid x \in dom(\beta), ar(\vec{cr}_x) \geq n])\})$. \\
	\end{tabular}
\item Else, define $fact(\sigma,\beta) := \bot$.
\ee
Using the lexicographic combination of
the size of range terms of $\beta$ and $\lg$,
it can be shown that
the algorithm always terminates and
yields $fact(\sigma,\beta)^M = \sigma^M \by \beta$
for $\beta \neq [\;]$.
The algorithm needs at most $depth(\beta)$
recursive calls to compute $fact(\sigma,\beta)$.
If $\sigma$ is semi-independent,
then so is $fact(\sigma,\beta)$.
}

\ALGORITHM{\eqd{50}
Let $\beta$ be pseudolinear,
$dom(\beta) \subset dom(\sigma)$,
$V := \{ x \in dom(\beta) \mid \beta x \in {\cal V}\}$.
Define the finite set $hom(\sigma,\beta)$ of ordinary homogenizing
substitutions for $\beta$ wrt.\ $\sigma$:
\be
\item If $\sigma \sortdef \sigma_1 \mid \ldots \mid \sigma_n$,
	define $hom(\sigma,\beta) :=
	hom(\sigma_1,\beta) \cup \ldots \cup hom(\sigma_n,\beta)$.
\item Else, if $\sigma \sortdef \vec{cr}(\sigma_1,\ldots,\sigma_n)$,
	let $\gamma_0 :=
	[\beta x \.{:=} \vec{cr}_x(y_{x,1},\ldots,y_{x,ar(\vec{cr}_x)})
	\mid x \in V]$
	where the $y_{x,i}$ are new variables,
	define $hom(\sigma,\beta) :=
	(hom(\sigma,\gamma_0 \circ \beta) \circ \gamma_0) \rs
	{ran(\beta)}$.
\item Else, if $\sigma \sortdef \vec{cr}(\sigma_1,\ldots,\sigma_n)$,
	$\beta$ not homogeneous,
	$V = \{\}$,
	\\
	and 
	$\beta x = \vec{cr}_x(u_{x1},\ldots,u_{x \; ar(\vec{cr}_x)})$
	for all $x \in dom(\beta)$,
	\\
	define $hom(\sigma,\beta) :=
	\bigpscomp_{i=1}^n hom(\sigma_i,[x \.{:=} u_{xi} \mid
	x \in dom(\beta), ar(\vec{cr}_x) \geq i])$.
\item Else, if $\beta$ homogeneous, define
	$hom(\sigma,\beta) := \{[x \.{:=} y_x \mid x \in ran(\beta)]\}$,
	\\
	where each $y_x$ is a new variable.
\item Else, define $hom(\sigma,\beta) := \{\}$.
\ee
Then, for each $\gamma \in hom(\sigma,\beta)$:
\be
\item $\gamma \circ \beta$ is homogeneous,
\item $dom(\gamma) = ran(\beta)$,
\item for each $\sigma' \in \sigma$ with $\sigma' \by \beta \neq \{\}$
	there exists a $\gamma \in hom(\sigma,\beta)$ such that
	$\sigma' \by {\gamma \circ \beta} \neq \{\}$, and
\item for each $\sigma' \in \sigma$ with $\sigma' \by \beta \neq \{\}$
	there exists at most one $\gamma \in hom(\sigma,\beta)$ 
	such that $\sigma' \by {\gamma \circ \beta} \neq \{\}$.
\ee
}
\PROOF{
Use the lexicographic combination of
the size of range terms of $\beta$ and $\lg$
as termination ordering and as induction ordering for 1.\ to 4.
The algorithm needs at most $\#use(\sigma) * depth(\beta)$
recursive calls to compute $hom(\sigma,\beta)$.
}

\THEOREM{\eqd{51}
If $\sigma$ regular and $\beta$ pseudolinear,
then
\mbox{$\sigma \by \beta u
= \bigcup_{\gamma \in hom(\sigma,\beta)}
\sigma \by {\gamma \circ \beta} \gamma u$}
for all $u$ with $vars(u) \subset ran(\beta)$,
where the $\sigma \by {\gamma \circ \beta}$ are regular.
}
\PROOF{
Regularity follows from \eqr{50} and \eqr{49}.
Since for each $\{\} \neq \sigma' \by \beta \in \sigma \by \beta$
there exists exactly one $\gamma$ with
$\sigma' \by {\gamma \circ \beta} \neq \{\}$
by \eqr{50},
the claimed equality follows from \eqr{34}.
}

\EXAMPLE{\eqd{52}
Consider the definition of $nat_{x\.<y}$
in Fig.~\ref{Examples of regular t-sets};
let $\beta := [x \.{:=} x',y \.{:=} s(y')]$.
We first homogenize $\beta$ wrt.\ $nat_{x\.<y}$
by \eqr{50},
yielding
$hom(nat_{x\.<y},\beta)
= \{[x' \.{:=} 0,y' \.{:=} y''], [x' \.{:=} s(x''),y' \.{:=} y'']\}$.
Then, using \eqr{49},
we factorize $nat_{x\.<y}$ wrt.\ the homogeneous substitutions
$[x' \.{:=} 0,y' \.{:=} y''] \circ \beta$
and $[x' \.{:=} s(x''),y' \.{:=} y''] \circ \beta$,
yielding
$fact(nat_{x\.<y},[x \.{:=} 0,y \.{:=} s(y'')])
= nat_{y''}$
and
$fact(nat_{x\.<y},[x \.{:=} s(x''),y \.{:=} s(y'')])
= nat_{x''\.<y''}$,
respectively.
Using \eqr{51}, we can thus compute
$nat_{x\.<y} \by \beta \tpl{x',y'}$
as
$nat_{y''} \tpl{0,y''} \mid nat_{x''\.<y''} \tpl{s(x''),y''}$.
}

\EXAMPLE{\eqd{53}
Let
$\sigma \sortdef 0_x cr_y(0_y,0_y) \mid cr_x cr_y(\sigma,\sigma)$,
and $\beta = [x \.{:=} x',y \.{:=} cr(y',x')]$.
Then, $\sigma^M \by \beta$ is the infinite set of
complete binary trees $A$ that is minimal with
$0_{x'} 0_{y'} \in A$
and $cr_{x'} cr_{y'}(\sigma',\sigma') \in A$
for $\sigma' \in A$.
$\sigma \by \beta x'$ is a similar set of complete binary trees
which cannot be written 
as $\tau_1 u_1 \cup \ldots \cup \tau_n u_n$ with regular $\tau_i$.
}
\PROOF{
\be
\item[]
\item Show $\sigma' \in A \Ra \sigma' \in \sigma^M \by \beta$
	by induction on $\sigma'$.
\item Show 
	$\sigma^M \by \beta
	= \{0_{x'} 0_{y'}\} 
	\cup (\sigma^M \by {[x \.{:=} x_1,y \.{:=} y']}
	\pecomp \sigma^M \by \beta \by {[x' \.{:=} x_2,y' \.{:=} x_1]})
	\circ [x' \.{:=} cr(x_1,x_2)]$
	by direct computation.
\item Show
	$\sigma^M \by {[x \.{:=} x_1,y \.{:=} y']}
	\pecomp \sigma' \by {[x' \.{:=} x_2,y' \.{:=} x_1]}
	= \sigma' \by {[x' \.{:=} x_2,y' \.{:=} x_1]}
	\pecomp cr_{y'}(\sigma' \rs {y'},\sigma' \rs {y'})$
	by induction on $\sigma'$ using 2.
\item Show $\sigma^M \by \beta \subset A$
	by induction on the order
	$\sigma'_1 < \sigma'_2
	:\Lra \sigma'_1 x' \psubterm \sigma'_2 x'$
	using 3.\ and 4.
\item Show that no infinite set of complete binary trees can be
        represented as $\tau u$ with regular t-set $\tau$ and
        constructor term $u$ by induction on $u$, using in the base
	case \eqr{37} and a pumping lemma.
\ee

}

\THEOREM{
%[(Independent T-Set Factorization)] 
\eqd{54}
%$\;$\\
If $\sigma$ is independent,
then
$\sigma \by \beta
= \bigpecomp_{x \in dom(\beta)} \sigma \by {[x {:=} \beta x]}$.
\\
The right-hand side can be algorithmically computed
using \eqr{49},
since $[x \.{:=} \beta x]$ is always homogeneous.
}

\ALGORITHM{
\eqd{55}
Let $\sigma$ be the name of a regular t-set.
The following algorithm decides 
whether $\sigma \subset \top \circ \beta$,
\\
i.e.\ whether $\sigma' \by \beta \neq \{\}$
for all $\sigma' \in \sigma$.
\\
Define
\begin{tabular}[t]{@{}l@{$\;$}r@{$\;$}l@{}}
$\eqi{div(\sigma,\beta)}$
	& $:\Lra$ & $\bigwedge_{x \in dom(\beta)}
	div(\sigma,[x \.{:=} \beta x])$	\\
& $\wedge$ & $\bigwedge_{x,x' \in dom(\beta), x \neq x'}
	\bigwedge_{y \in vars(\beta x) \cap vars(\beta x')}$	\\
&& $single(apply(fact(\sigma,[x \.{:=} \beta x]),y) \mid
	apply(fact(\sigma,[x' \.{:=} \beta x']),y))$;	\\
\end{tabular}
\\
where $div(\sigma,[x \.{:=} u])$ is computed as follows:
\be
\item If $\sigma \sortdef \sigma_1 \mid \ldots \mid \sigma_n$,
	define
	$div(\sigma,[x \.{:=} u]) :\Lra
	div(\sigma_1,[x \.{:=} u]) \wedge \ldots \wedge
	div(\sigma_n,[x \.{:=} u])$.
\item Else, if $\sigma \sortdef \vec{cr}(\sigma_1,\ldots,\sigma_n)$ 
	and $u = \vec{cr}_x(u_1,\ldots,u_k)$,
	\\
	define
	\begin{tabular}[t]{@{}l@{$\;$}r@{$\;$}l@{}}
	$div(\sigma,[x \.{:=} u])$ & $:\Lra$
		& $\bigwedge_{i=1}^k div(\sigma_i,[x \.{:=} u_i])$ \\
	& $\wedge$ & $\bigwedge_{1 \leq i < j \leq k}
		\bigwedge_{y \in vars(u_i) \cap vars(u_j)}$	\\
	&& $single(apply(fact(\sigma_i,[x \.{:=} u_i]),y) \mid
		apply(fact(\sigma_j,[x \.{:=} u_j]),y))$.
	\end{tabular}
\item Else, if $\sigma \sortdef \vec{cr}(\sigma_1,\ldots,\sigma_n)$
	and $u = cr'(u_1,\ldots,u_k)$ 
	with $cr' \neq \vec{cr}_x$,
	define
	$div(\sigma,[x \.{:=} u]) :\Lra false$.
\item Else, if $u \in {\cal V}$ (also $u=x$),
	define
	$div(\sigma,[x \.{:=} u]) :\Lra x \in dom(\sigma)$.
\ee
Using the lexicographical combination of $\subterm$ and $\lg$,
the correctness and termination of the computation of 
$div(\sigma,[x \.{:=} u])$ can be shown 
by induction on $\tpl{\sigma,u}$.
The proof, as well as the correctness proof for $div(\sigma,\beta)$,
uses the fact
that $\sigma y \cup \tau y$ is a singleton set 
for all $y \in dom(\sigma) \cap dom(\tau)$
iff $\sigma' \pecomp \tau' \neq \{\}$ 
for all $\sigma' \in \sigma$, $\tau' \in \tau$.
The algorithm needs at most $\#use(\sigma)$
recursive calls to decide $div(\sigma,[x \.{:=} u])$.
}

\section{Extended Sorts}
\label{Extended Sorts}

In this section,
we discuss several possible ways of defining a class of sorts that can
express more subsets
of ${\cal T}_{\cal CR}$ than regular tree languages.
In particular, we define a class called
``extended sorts'' that can express
for arbitrary $u \in {\cal T}_{{\cal CR},{\cal V}}$
the set of all possible values $U$ of $u$, mentioned in
Sect.~\ref{Motivation}, as an extended sort.
We define the notion of an ``annotated term'' $\vs \sigma v$
where the t-set $\sigma$ indicates the set of admitted ground
constructor instances of $v$'s variables, i.e.\ their sort.

The results obtained in Sect.~\ref{Regular T-Sets and Algorithms}
allow us to define three different language classes which are all proper
extensions of regular tree languages.
In each class, a language of ground constructor terms is described by
applying regular t-sets $\sigma$
to constructor terms $u$, the classes differing in the
form that is allowed for $\sigma$ and $u$:
\be
\item $\sigma u$ with $\sigma$ semi-independent, $u$ arbitrary.
	\\
	The intersection can be computed using Thms.~\eqr{31} and
	\eqr{54}.
	The subset property $\sigma u \subset \tau_1 u_1$
	(hence equivalence and inhabitance)
	can be decided using Thm.~\eqr{34} and Lemma~\eqr{35},
	since we have $J = \{ \{\}, \{1\} \}$, \
	$\sigma_{\{\}} \subset \{\} \Lra div(\sigma,\beta_{\{1\}})$,
	and 
	$\sigma_{\{1\}} \by {\beta_{\{1\}}}
	= \sigma \by {\beta_{\{1\}}}$.
	However, this class is not closed wrt.\ union.
        Any regular sort $S^M$ from Sect.~\ref{Regular Sorts} can be
        expressed as $[x \.{:=} S^M] \; x$, but
        the converse is false, e.g.\ $Nat_x^M \; \tpl{x,x}$ is not a
        regular sort, as can be shown using a pumping lemma
        \cit{Burghardt 1993}{Burghardt.1993}. 
\item $\sigma_1 u_1 \cup \ldots \cup \sigma_n u_n$
	with $\sigma_i$ independent, $u_i$ arbitrary.
	\\
	This is a proper superclass of the class given in 1.
	The intersection can be computed using Thms.~\eqr{31} and
	\eqr{54};
	union is trivial;
	inhabitance can be decided using Thm.~\eqr{32}.
	However, we do not provide an
	algorithm to decide the subset property
	in general.
	Again, any regular sort $S^M$ can be expressed as 
	$[x \.{:=} S^M] \; x$.
\item $\sigma_1 u_1 \cup \ldots \cup \sigma_n u_n$
	with $\sigma_i$ arbitrary, $u_i \in T$
	for some set $T$
	such that for any $u,u' \in T$,
	$\beta = mgu(u,u')$ is always
	pseudolinear if it exists, and again $\beta u \in T$.
	\\
	The intersection can be computed using Thms.~\eqr{31},
	and \eqr{51}.
	Inhabitance can be decided as in class 2.,
	but again we do not provide an algorithm to decide subsort in
	general.
	If we take $T$ to be the set of all constructor terms in which
	variables occur only at a fixed unique depth $n$, the
	requirements to $T$ are fulfilled, and each regular sort can be
	expressed.

        As shown in section \ref{Regular T-Sets and Algorithms},
        dependent regular t-sets can express certain relations between
        distinct variables, e.g.\ the conditional equation
        \mbox{$x\ty{Nat} < y\ty{Nat} \ra f(x,y) = g(x,y)$} can be
        expressed unconditionally by $f(x,y) = g(x,y)$, where
	the value combinations of
	$x$ and $y$ are restricted by $Nat_{x<y}$ from
	Fig.~\ref{Examples of regular t-sets}.
        Since we have the problem that Thms.~\eqr{31} and
        \eqr{34} use factorization $\sigma \by \beta$ which is not
        always a regular t-set, cf.\ Ex.~\eqr{53}, we have to
	restrict $T$ as above.
        It is an unsolved
        problem whether a superclass of 2.\ exists that
        allows dependent t-sets but is still closed wrt.\ the required
        operations, especially intersection.

% 	To obtain a set $T$, we can proceed as follows:
% 	Define ${\cal T}'$ as the least set of sets of constructor
% 	terms such that
% 	%\display{
% 		%\begin{tabular}[t]{@{}l@{$\;$}ll@{}}
% 		$\{x\} \in {\cal T}'$, and
% 		$\{[x \.{:=} cr(x_1,\ldots,x_n)] \; t 
% 			\mid cr \in {\cal CR},
% 			x_1,\ldots,x_n \in {\cal V} \mbox{ new},
% 			t \in T'\}
% 			\in {\cal T}'$
% 			if $T' \in {\cal T}'$ and $x \in vars(t)$
% 			for some $t \in T'$.
% 		%\end{tabular}
% 	%}
% 	Choose any $T' \in {\cal T}'$, and define 
% 	$T := \{ \beta t \mid \beta \mbox{ pseudolinear}, t \in T',
% 		dom(\beta) \supset vars(t) \}$.
% 	Show by induction on the structure of ${\cal T}'$ that any
% 	regular language $S^M$ can be represented:
% 	for $\{x\} \in {\cal T}'$, we have $S^M = [x \.{:=} S^M] \; x$;
% 	for the recursive case, we have
% 	\\
% 	\begin{tabular}[t]{@{}l@{$\;$}l@{\hspace*{1.0cm}}l@{}}
% 	& $S^M$	\\
% 	$=$ & $\bigcup_{i=1}^n \sigma_i u_i$ & by I.H.	\\
% 	$=$ & $\bigcup_{i=1}^n \bigcup_{cr \in {\cal CR}}
% 		(\sigma_i \cap \top \circ [x \.{:=} cr(x_1,\ldots,x_n)])
% 		\; u_i$
% 		& Def.\ $\top$	\\
% 	$=$ & $\bigcup_{i=1}^n (\bigcup_{cr \in {\cal CR}}
% 		\sigma \by {[x \.{:=} cr(x_1,\ldots,x_n)]} )
% 		\; [x \.{:=} cr(x_1,\ldots,x_n)] u_i$
% 		& by \eqr{27}.b	\\
% 	\end{tabular}

\ee

All classes follow the philosophy of allowing
arbitrary nonlinearities up
to a finite depth and forbidding any below.
Since class 1.\ is sufficient to represent the set of all possible
values $U$ of an arbitrary constructor term $u$,
we will use this class in the rest of this paper.
In classical order-sorted
approaches, each variable in a term is assigned a sort,
e.g.\
$x\ty{Nat}$ + x\ty{Nat}. We will, instead,
use a semi-independent t-set to specify the set
of possible ground instances, written e.g.\ $\vs {(Nat_x)} \; (x+x)$,
with the informal meaning that each (``admissible'')
substitution instantiating $x+x$ must be
extendable to a ground substitution contained in $Nat_x^M$.
This approach still allows
variable bindings in a term to be reflected by its sort, as sketched
in Sect.~\ref{Motivation};
what we lose is the
possibility of expressing nontrivial relations between variables.

\DEFINITION{
\eqd{56}
We define an {\eqi{annotated term}} as a pair of a semi-independent
regular t-set
$\sigma$ and an (unsorted) term $v$; it is written as 
\eqi{$\vs \sigma v$}.
The t-set
$\sigma$ denotes the admitted instances of $v$, cf.\ the use of $\vs
\sigma v$ in Defs.~\eqR{61} and~\eqR{74} below.
}

\DEFINITION{
\eqd{57}
We call an expression of the form
$\sigma u$
with semi-independent regular $\sigma$
and $u \in {\cal T}_{{\cal CR},{\cal V}}$
an {\eqi{extended sort}}.
The set of all possible values $U$ of an annotated term $\vs \sigma u$
is always an extended sort, viz.\ $U = \sigma u$.
Sets of the form $\sigma v$, where $v$
contains non-constructor functions,
will be approximated by extended sorts later, cf.\
Sect.~\ref{Equational Theories}.
}

% Thus, regular sorts are a proper subclass of extended
% sorts, and we need new algorithms to compute with the
% latter. Theorems~\eqr{31} and \eqr{34} reduce the
% computation of intersection and the decision of subset
% property of extended sorts to the respective problems of
% regular t-sets, which can be solved by ``lifted''
% versions of the algorithms mentioned in
% Sect.~\ref{Regular Sorts}. Thus, both Theorems provide
% the basis for the necessary algorithms to conduct the
% disjointness test from Sect.~\ref{Motivation} on
% extended sorts. The notion of factorization from
% Def.~\eqr{26} plays a central role here, in particular
% Lemma \eqr{30}, from which we get the following 

\COROLLARY{\eqd{58}
A sorted equation $\vs \sigma u_1 = \vs \sigma u_2$
between annotated constructor terms is solvable
iff an (unsorted) mgu $\beta$ of $u_1$ and $u_2$ exists
and $\sigma \by \beta \neq \{\}$.
The latter set contains the admissible ground instances of variables in
$ran(\beta)$.
An equation system
$\vs {\sigma_1} \, u_1 = \vs {\tau_1} \, u'_1 \wedge \ldots \wedge
\vs {\sigma_n} \, u_n = \vs {\tau_n} \, u'_n$
can be reduced to a single equation
$\vs \sigma {\tpl{u_1,\ldots,u_n}}
= \vs \sigma {\tpl{u'_1,\ldots,u'_n}}$
by defining $\sigma := \bigpecomp_{i=1}^n \sigma_i \pecomp \tau_i$.
}

\EXAMPLE{
\eqd{59}
Using the definitions from \eqr{53},
$\sigma \tpl{x,y} \cap \top_{\{x'',y''\}} \tpl{x'',cr(y'',x'')}$
cannot be represented as an extended sort.
}
\PROOF{
Observe that 
	$\beta'
	= \beta \pscomp [x'' \.{:=} x', y'' \.{:=} y']
	= mgu(\tpl{x,y},\tpl{x'',cr(y'',x'')})$,
	\\
	and 
	$(\sigma \pecomp \top_{\{x'',y''\}}) \by {\beta'}
	= \sigma \by \beta$;
	\\
hence, by \eqr{31},
	$\sigma \tpl{x,y} \cap \top_{\{x'',y''\}} \tpl{x'',cr(y'',x'')}
	= (\sigma \pecomp \top_{\{x'',y''\}}) \by {\beta'}
	\tpl{x',cr(y',x')}
	= \sigma \by \beta \tpl{x',cr(y',x')}$,
	\\
	the latter
	cannot be written as $\tau_1 u_1 \cup \ldots \cup \tau_n u_n$
	with $\tau_i$ regular,
	by an argument similar to \eqr{53}.5.
}

\section{Equational Theories}
\label{Equational Theories}

In this section, we extend the previous formalism to allow equationally
defined functions $f$.
We allow \eqi{defining equations} of the
form given in Def.~\eqR{60}, thus ensuring the
``executability'' of $f$.
Signatures of such a function are computed
from its defining equations by the \eqi{$rg$} algorithm
presented below in Alg.~\eqR{73},
which will play a central role in pruning the
search space of narrowing. 
The algorithm takes a regular t-set and a term
with non-constructor functions and computes an upper approximation by
an extended sort, e.g.\
$rg([x,y \.{:=} Nat],x+y) = [z \.{:=} Nat] \; z$.
In terms of Sect.~\ref{Motivation}, we have $rg(\sigma,v) = \ovl{V}$
where $\sigma$ denotes the values over which the variables in $v$ may
range.
The $rg$ algorithm consists of local transformations like
rewriting and some simplification rules (cf.\ Def.~\eqR{64}),
global transformations
looking at a sequence of local transformation steps and recognizing
certain kinds of self-references (cf.\ Lemma \eqR{67}),
and an approximation rule.
Only the main rules can be discussed here;
the complete algorithm is given in \cit{Burghardt 1993}{Burghardt.1993}.

In Theorem~\eqR{75}, a narrowing calculus from
\cit{H\"olldobler 1989}{Holldobler.1989} is equipped with sorts.
In \cit{Burghardt 1993}{Burghardt.1993},
the calculus is shown
to remain complete if the 
applicability of its main rule is restricted by the
disjointness test from Sect.~\ref{Motivation}.

\DEFINITION{
\eqd{60}
In the rest of this section, we assume that $f$ has the following 
\eqi{defining equations}:
% $\eqi{\vs {\mu_i} f(u_{i1},\ldots,u_{in})}$ 
\display{
\begin{tabular}[t]{@{}l@{$\;$}ll@{}}
${\vs {\mu_1} f(u_{11},\ldots,u_{1n})}$ 
	& $= \vs {\mu_1} \, v_1,$	\\
& $\ldots$,	\\
$\vs {\mu_m} f(u_{m1},\ldots,u_{mn})$
	& $= \vs {\mu_m} \, v_m$,	\\
\end{tabular}
}
where $vars(v_i) \subset vars(u_{i1},\ldots,u_{in})$.
We assume that the variables of different defining equations are
disjoint.
Define
$\eqi{dom(f,I)} := \bigcup_{i \in I} \mu_i \tpl{u_{i1},\ldots,u_{in}}$,
and $\eqi{dom(f)} := dom(f,\{1,\ldots,m\})$.
}

\DEFINITION{
%[(\eqi{Sorted Rewriting})] 
\eqd{61}
%$\;$\\
Define the \eqi{rewrite relation} induced by the defining equations by:
$\;$
$\eqi{\vs \sigma v_1 \abldi \vs \sigma v_2$}
iff
\be
\item a defining equation
	$\vs \mu f(u_1,\ldots,u_n) = \vs \mu v$,
	a substitution $\beta$, and a term $v'(x)$ linear in $x$ exist
	such that
	$v_1 = v'(\beta f(u_1,\ldots,u_n))$, $v_2 = v'(\beta v)$,
\item and for all $\sigma' \in \sigma$
	there exists $\mu' \in \mu$ such that
	for all $x \in vars(u_1,\ldots,u_n)$
	\\
	$\vs \varepsilon \sigma' \beta x
	\ablds \vs \varepsilon \mu' x$
	if $\vs \varepsilon \sigma' \beta x$
	is well-defined.
\ee
While the former condition is merely rewriting by pattern matching,
the latter is an analogue to the classical
well-sortedness requirement for $\beta$, requiring any well-defined
variable instance to be admitted by the defining equation's sort.
A ground term is called {\eqi{well-defined}} if it is reducible to a
ground constructor term.
\eqi{$\ablds$} and \eqi{$\Ablds$} are defined as usual;
the definition of $\abldi$ is recursive, but well-founded.
We require
confluence and termination of $\abldi$,
ensuring
${\cal T}_{\cal CR} \subset {\cal T}_{{\cal CR},{\cal F}}/{\Ablds}$,
where $\eqi{{\cal T}_{{\cal CR},{\cal F}}/{\Ablds}}$ denotes the set of
equivalence classes of terms in ${\cal T}_{{\cal CR},{\cal F}}$ modulo
$\Ablds$.
In other words,
$\Ablds$ does not identify terms in ${\cal T}_{\cal CR}$,
but new irreducible terms like $nil+nil$ may
arise which we will
regard as ``\eqi{junk terms}'' and exclude from equation
solutions.
For a well-defined ground term $v$,
let $\eqi{nf(v)} \in {\cal T}_{\cal CR}$
denote its unique normal form;
for $A \subset {\cal T}_{{\cal CR},{\cal F}}$,
let $\eqi{nf[A]} := \{ nf(v) \mid v \in A, v\mbox{ well-defined}\}$.
}

\begin{figure}
\begin{center}
\begin{tabular}[t]{@{}|l|l|l|@{}}
\hline
& Classical order-sorted terms & Annotated terms \\
\hline
term & $w$ where $x_1\ty{$s_1$},\ldots,x_m\ty{$s_m$}$
	& $\vs \sigma w$ where $dom(\sigma) = \{x_1,\ldots,x_m\}$ \\
\hline
sort & $sortof(w)$ & $\sigma w$	\\
\hline
def. eq. & $f(l_1,\ldots,l_n) = r$ where
	& $\vs \mu f(l_1,\ldots,l_n) = \vs \mu r$ where	\\
& $y_1\ty{$t_1$},\ldots,y_m\ty{$t_m$}$
	& $dom(\mu) = \{y_1,\ldots,y_m\}$	\\
\hline
rewriting & $v(\beta f(l_1,\ldots,l_n)) \abldi v(\beta r)$ where
	& $\vs \sigma v(\beta f(l_1,\ldots,l_n)) \abldi
	\vs \sigma v(\beta r)$ where	\\
& $\forall y \in V \;\;
	sortof(\beta y) \subset sortof(y)$
	& $\forall \sigma' \.\in \sigma \; \exists \mu' \.\in \mu
	\; \forall y \.\in V \;\;
	\vs \varepsilon \sigma' \beta y \ablds \vs \varepsilon \mu' y$\\
\hline
equation & $w_1 = w_2$ where $x_1\ty{$s_1$},\ldots,x_m\ty{$s_m$}$
	& $\vs \sigma w_1 = \vs \sigma w_2$	\\
\hline
solution & $\gamma w_1 \Ablds \gamma w_2$ where
	& $\gamma w_1 \Ablds \gamma w_2$ and $\tau$ where \\
& $\forall x \in W \;\; sortof(\gamma x) \subset sortof(x)$
	& $nf[\tau \gamma \tpl{x_1,\ldots,x_m}]
	\subset \sigma \tpl{x_1,\ldots,x_m}$	\\
&& and $nf[\tau \beta \tpl{u_1,\ldots,u_n}] \neq \{\}$	\\
\hline
\multicolumn{3}{@{}l@{}}{
	$w, w_1, w_2 \in {\cal T}_{{\cal CR},{\cal F},{\cal V}}$,
	$l_1,\ldots,l_n \in {\cal T}_{{\cal CR},{\cal V}}$,
	$\sigma, \tau, \mu \subset \TT_{\pfunc{\cal V}{\cal CR}}$,} \\
\multicolumn{3}{@{}l@{}}{$W = vars(w) = vars(w_1,w_2)
	= \{x_1,\ldots,x_m\} =dom(\sigma)$,}	\\
\multicolumn{3}{@{}l@{}}{
	$V = vars(l_1,\ldots,l_n) = \{y_1,\ldots,y_m\}$}	\\
\end{tabular}

\begin{tabular}[t]{@{}|l|l|l|@{}}
\mca{3}{Examples:}	\\ \hline
& Classical order-sorted terms & Annotated terms	\\ \hline
term & $x\ty{Nat}+x\ty{Nat}$ & $\vs {[x \.{:=} Nat]\;} x+x$ \\ \hline
sort & $Nat+Nat = Nat$ & $[x \.{:=} Nat] \; (x+x) = Even$ \\ \hline
def. eq.
	& \begin{tabular}[t]{@{}l@{$\;$}lr@{}}
	$a\ty{Nat}+0$ & $= a\ty{Nat}$	\\
	$a\ty{Nat}+s(b\ty{Nat})$ & $= s(a\ty{Nat}+b\ty{Nat})$	\\
	$s(a\ty{Nat})+b\ty{Nat}$ & $= s(a\ty{Nat}+b\ty{Nat})$	\\
	\end{tabular}
	& \begin{tabular}[t]{@{}l@{$\;$}lr@{}}
	$\vs {[a \.{:=} Nat]\;} a+0$ & $= \vs {[a \.{:=} Nat]\;} a$ \\
	$\vs {[a,b \.{:=} Nat]\;} a\.+s(b)$
		& $= \vs {[a,b \.{:=} Nat]\;} s(a\.+b)$ \\
	$\vs {[a,b \.{:=} Nat]\;} s(a)\.+b$
		& $= \vs {[a,b \.{:=} Nat]\;} s(a\.+b)$ \\
	\end{tabular}
	\\ \hline
rewriting &
	\begin{tabular}[t]{@{}l@{$\;$}lr@{}}
	& $s(x\ty{Nat}+x\ty{Nat}+s(y\ty{Nat}))$ \\
	$\abldi$ & $s(s(x\ty{Nat}+x\ty{Nat}+y\ty{Nat}))$	\\
	\mca{2}{where $\beta = [a \.{:=} x+x,b \.{:=} y]$}	\\
	\end{tabular}
	& \begin{tabular}[t]{@{}l@{$\;$}lr@{}}
	& $\vs {[x,y \.{:=} Nat]\;} s(x+x+s(y))$	\\
	$\abldi$ & $\vs {[x,y \.{:=} Nat]\;} s(s(x+x+y))$	\\
	\mca{2}{where $\beta = [a \.{:=} x+x,b \.{:=} y]$}	\\
	\end{tabular}
	\\
& $sortof(\beta a) = Nat\.+Nat = sortof(a)$
	& for $[x \.{:=} s^i(0),y \.{:=} s^j(0)]$	\\
& $sortof(\beta b) = Nat = sortof(b)$
	& choose $[x \.{:=} s^{2 \* i}(0),y \.{:=} s^j(0)]$ \\ \hline
equation & $s(s(x\ty{Nat})) = y\ty{Nat}+y\ty{Nat}$
	& $\vs {[x,y \.{:=} Nat]\;} s(s(x)) = y+y$	\\ \hline
solution & \begin{tabular}[t]{@{}l@{}}
	$\gamma = [x\.{:=}z\ty{Nat}+z\ty{Nat},y\.{:=}s(z\ty{Nat})]$ \\
	$sortof(\gamma x) = Nat\.+Nat = sortof(x)$	\\
	$sortof(\gamma y) = s(Nat) \subset sortof(y)$	\\
	\end{tabular}
	& \begin{tabular}[t]{@{}l@{}}
	$\gamma = [x\.{:=}z+z,y\.{:=}s(z)]$	\\
	and $\tau = [z\.{:=}Nat]$, \hspace*{0.1cm}
	$nf[\tau \gamma \tpl{x,y}]$	\\
	$= \{\tpl{s^{2\*i}(0),s^{i\.+1}(0)} \mid i \in \N\}$	\\
	$\subset \tpl{Nat,Nat}$ \\
	\end{tabular}	\\ \hline
\end{tabular}

\caption{Comparison of classical order-sorted terms and annotated terms}
\label{Comparison of classical order-sorted terms and annotated terms}
\end{center}
\end{figure}

Figure \ref{Comparison of classical order-sorted terms and annotated
terms} shows a comparison of classical order-sorted terms and annotated
terms.
The applicability of $\abldi$ is not decidable in general
owing to the well-sortedness condition~\eqr{61}.2.
It is possible to compute sufficiently large t-sets
$\mu_i$ for the defining equations such that
\eqr{61}.2
becomes trivial, cf.\ Alg.~\eqR{71};
however, if the $\mu_i$ are too large,
well-sorted terms arise that are not well-defined. 
As in any order-sorted term rewriting approach, we cannot
overcome both problems simultaneously.

Range sorts are computed using expressions of the form
$(w_1\.:u_1) \ldots (w_n\.:u_n)$,
which can intuitively be thought of as generalized
equation systems; the semantic is the set of all t-substitutions making
each $u_i$ equal ($\Ablds$) to an element of $w_i$.
For example, $(Nat:x)$ denotes $\{[x \.{:=} s^i(0)] \mid i \in
\N\}$,
and $(x+x:z)$ can be evaluated to $\{[x \.{:=} s^i(0),z \.{:=}
s^{2 * i}(0)] \mid i \in \N\}$.

\DEFINITION{
%[(Generalized Term Semantics)] 
\eqd{62}
%$\;$\\
Define
$\eqi{w^M} \subset {\cal T}_{{\cal CR},{\cal F},{\cal V}}$
by:	\\
\begin{tabular}[t]{@{}r@{$\;$}l@{\hspace*{1em}}l@{}}
$S^M$ & $:= S^M$ as in Def.~\eqr{4}
	& for $S \in {\cal S}$ \\
$g(v_1,\ldots,v_n)^M$ & $:= \{g(v'_1,\ldots,v'_n) \mid
	v'_1 \in v_1^M,\ldots,v'_n \in v_n^M\}$
	& for $g \in {\cal CR} \cup {\cal F}$ \\
$x^M$ & $:= \{x\}$ & for $x \in {\cal V}$	\\
\end{tabular}
\\
For $w \in {\cal T}_{{\cal CR},{\cal S}}$,
$w^M$ agrees with Def.~\eqr{4}.
For $w \in {\cal T}_{{\cal CR},{\cal F},{\cal V}}$
we always have $w^M = \{w\}$.
We tacitly extend the operations of 
Sect.~\ref{T-Substitutions}
to ${\cal T}_{{\cal CR},{\cal F},{\cal V}}$ by treating function
symbols from ${\cal F}$ like constructors from ${\cal CR}$, e.g.\
$(0_x) \; (x+x) = \{0+0\}$.
\\
Let $\eqi{(w:u)^M} := \{\sigma' \mid dom(\sigma') = vars(w,u),
	\exists w' \in w^M, u' \in u^M \;\;
	\sigma' w' \Ablds \sigma' u'\}$
and
$\eqi{((w_1:u_1) \; (w_2:u_2))^M} := (w_1:u_1)^M \pecomp (w_2:u_2)^M$.
We write $\eqi{(\sigma)$}
to denote an expression $(w_1\.:u_1) \ldots (w_n\.:u_n)$
such that $((w_1\.:u_1) \ldots (w_n\.:u_n))^M = \sigma$,
e.g.\ $(Nat_{x,y})$ denotes $(Nat:x)(Nat:y)$,
but
note that $\sigma$ need neither be independent nor even regular.
The terms are unsorted in order to deal with t-sets explicitly;
$(\vs \sigma w:\vs \tau u)$
can be written as $(\sigma) \; (\tau) \; (w:u)$.
Note that the t-substitutions in $(w:u)^M$ always yield ground
constructor terms.
}

\LEMMA{
\eqd{63}
Let $\vs \varepsilon f(w_1,\ldots,w_n) \ablds \vs \varepsilon u$,
for $w_1,\ldots, w_n \in {\cal T}_{{\cal CR},{\cal F}}$
and $u \in {\cal T}_{\cal CR}$;
let $I \subset \{1,\ldots,m\}$
such that 
$nf[\top \tpl{w_1,\ldots,w_n}] \cap dom(f,I)
= nf[\top \tpl{w_1,\ldots,w_n}] \cap dom(f)$;
\\
then, $i \in I$ and $\mu'_i \in \mu_i$
exists such that $w_j \ablds \mu'_i u_{ij}$ for $j=1,\ldots,n$
and $\mu'_i v_i \ablds u$.
}
\PROOF{
Consider the first reduction at root position within the chain
$\vs \varepsilon f(w_1,\ldots,w_n) \Ablds \vs \varepsilon u$.
}

\DEFINITION{
%[(Range Computation -- Local Transformation Rules)]
\eqd{64}
%$\;$\\
The following \eqi{local transformation rules}
for $rg$ are defined (excerpt):
\be
\item	$(f(v'_1,\ldots,v'_n) : u) = \bigmid_{i \in I} (v_i : u) \;
	(v'_1 : u_{i1}) \; \ldots \; (v'_n : u_{in}) \; (\mu_i)$ \\
	if $I \subset \{1,\ldots,m\}$
	arbitrary such that
	$nf[\top \tpl{w_1,\ldots,w_n}] \cap dom(f,I)
	= nf[\top \tpl{w_1,\ldots,w_n}] \cap dom(f)$,
	cf.\ the remarks on page \pageref{4042}.
\item	$(u' : x) \; (v : u)
	= (u' : x) \; ([x \.{:=} u'] \; v : [x \.{:=} u'] \; u)$
	\tab if $x \not\in vars(u')$,
\item	$(S:x) \; (S':x) = (S \cap S':x)$,
\ee
Rules 2.\ and 3.\ also show ``$(\cdot:\cdot)$'' as a
generalization of term equality and sort membership, respectively.
All local rules satisfy the correctness criterion
$lhs^M \rs {vars(lhs)} = rhs^M \rs {vars(rhs)}$.
}
\PROOF{
Use \eqr{63} for correctness of rule 1.;
correctness of 2.\ and 3.\ follows by simple computations.
}

Only one proper \eqi{approximation rule} is needed, viz.\
$(w_1:u_1) \; \ldots \; (w_n:u_n)^M \subset
(w_2:u_2) \; \ldots \; (w_n:u_n)^M$;
all other rules can be made exact by including the left-hand side in
the right-hand side.

Applying local transformations creates a 
{\eqi{computation tree}} with
{\eqi{alternatives}} (separated by ``$\mid$'')
as nodes, each alternative having a unique 
{\eqi{computation path}} from the root, cf.\ Fig.~\ref{Range sort
computation for $x+x$}.
Global
transformations operate on such computation trees. A proof
methodology (``rank induction'')
is provided in Def.~\eqR{65} and Lemma~\eqR{66}
for their verification that also allows
the introduction of new global rules, if necessary, for some class
of applications.

\DEFINITION{
%[(\eqi{Rank of a T-Substitution})]
\eqd{65}
%$\;$\\
Let $\sigma' \in ((w_1\.:u_1) \; \ldots \; (w_n\.:u_n))^M$,
then e.g.\ $\sigma' w_1 \Ablds \sigma' u_1$,
where $\sigma' u_1 \in {\cal T}_{\cal CR}$ is in normal form.
Owing to confluence and termination,
each ``$\abldi$'' chain starting from $\sigma' w_1$
ends after finitely many steps at $\sigma' u_1$.
\\
Define
$\eqi{rank(\sigma',(w_1:u_1))}$ as the length of the longest such chain,
which always exists.
Define
$\eqi{rank(\sigma',(w_1:u_1) \; \ldots \; (w_n:u_n))} :=
\sum_{i=1}^n rank(\sigma',(w_i:u_i))$.
We always have $rank(\sigma',(\sigma)) \in \N$ and
$rank(\sigma',(\tau)) = 0$ for $\tau$ regular t-set.
}

\LEMMA{
%[(Rank Induction)]
\eqd{66}
%$\;$\\
Let $(\sigma) = (\sigma_1) \mid \ldots \mid (\sigma_m)$
be the result of repeated application of the rules from Def.\
\eqr{64},
let $z \in dom(\sigma) \cap
dom(\sigma_1) \cap \ldots \cap dom(\sigma_m)$.
Then for each $\sigma' \in (\sigma)^M$,
an $i \in \{1,\ldots,m\}$
and a $\sigma'_i \in (\sigma_i)^M$
exists such that $\sigma' z = \sigma'_i z$
and $rank(\sigma',(\sigma)) \geq rank(\sigma'_i,(\sigma_i)) + n_i$,
where $n_i$ denotes the number of applications of Def.~\eqr{64}.1 in
the path from $(\sigma)$ to $(\sigma_i)$.
}
\PROOF{
Induction on the number of applications of rules from Def.~\eqr{64}.
}

\LEMMA{
\eqd{67}
%$\;$\\
(Global Transformation: Loop-Checking Rule)
\\
Assume $z \not\in dom(\sigma) \supset vars(v) \not\ni x$ and
a computation tree
of the form
\\[0.3ex]
\begin{tabular}[t]{@{}l@{$\;$}l@{}}
& $(\sigma) \; (v : z) \;\;\; = \ldots$ \\
$=$ & $(\sigma) \; (v:x) \; (u_1(x) : z)
	\mid \ldots \mid (\sigma) \; (v:x) \; (u_n(x) : z)
	\mid (u_{n+1} : z) \mid \ldots \mid (u_m : z)$	\\
\end{tabular}
\\[0.5ex]
where in each alternative's path at least one application of rule
\eqr{64}.1 occurred.
Then, $((\sigma) \; (v : z))^M \subset (S : z)^M$,
where $S$ is a new sort name defined by
$S \sortdef u_1(S) \mid \ldots \mid u_n(S)
\mid u_{n+1} \mid \ldots \mid u_m$.
If all $u_i(x)$ are linear in $x$, we have equality.
}
\PROOF{
Show $\sigma' \in ((\sigma) \; (v:z))^M \;\Ra\; \sigma' z \in s^M$
by induction on $rank(\sigma',(\sigma) \; (v:z))$,
using Lemma \eqr{66}.
% If all $u_i(x)$ are linear in $x$,
% use \eqr{8} with
% $p_{v(S)}(u) :\Lra \exists u' \in v(\sigma^M w) \;\;\; u' \ablds u$
% if $v(S) \in use(S)$
% to show
% $u \in S^M 
% \Ra \exists u' \in (\sigma) \; (v:z) \; ^M (z) \;\;\; u' \ablds u$.
% 
% {\Huge ???}
}

\LEMMA{\eqd{68}
Let $1 \leq k \leq n \leq m$,
\\[0.3ex]
\begin{tabular}[t]{@{}l@{\hspace*{1em}}l@{}}
assume
&
\begin{tabular}[t]{@{}c*{8}{@{$\;$}c}@{}}
& $(\sigma)$ & $(v:z)$	\\
$=$ & $(\sigma)$ & $(v:u_1)$ & $(u'_1:z)$
& $\mid \ldots \mid$ & $(\sigma)$ & $(v:u_n)$ & $(u'_n:z)$ \\
$\mid$ & $(\sigma_{n\.+1})$ & $(v_{n\.+1}\.:u_{n\.+1})$
& $(u'_{n\.+1}\.:z)$
& $\mid \ldots \mid$ & $(\sigma_m)$ & $(v_m:u_m)$
& $(u'_m:z)$ \\
\end{tabular} \\[6.0ex]
\mca{2}{\parbox{\textwidth}{where in each alternative's path at least
one application of Def.~\eqr{64}.1 occurred.}} \\
Then,
&
\begin{tabular}[t]{@{}c*{8}{@{$\;$}c}@{}}
& $(\sigma)$ & $(v:z)$	\\
$=$ & $(\sigma)$ & $(v:u_{k\.+1})$ & $(u'_{k\.+1}:z)$
& $\mid \ldots \mid$
& $(\sigma)$ & $(v:u_n)$ & $(u'_n:z)$	\\
$\mid$ & $(\sigma_{n\.+1})$ & $(v_{n\.+1}\.:u_{n\.+1})$
& $(u'_{n\.+1}\.:z)$
& $\mid \ldots \mid$
& $(\sigma_m)$ & $(v_m:u_m)$ & $(u'_m:z)$	\\
\end{tabular}
\end{tabular} \\[1.0ex]
provided $(u'_i : u_j)^M = \{\}$
for all $i \in \{k+1,\ldots,m\}, j \in \{1,\ldots,k\}$.
\\
Intuitively, constructor terms $u_1,\ldots,u_k$ may be produced as
the value of $z$ or $v$ only in alternatives $1,\ldots,k$,
but this in turn requires a constructor term $u_1,\ldots,u_k$.
Hence, there is no recursion basis,
i.e.\ $v$ may not have a $u_i$ as its value,
i.e.\ the first $k$ alternatives are superfluous.
}
\PROOF{
Show
$\sigma' \in ((\sigma) \; (v:z))^M
\;\;\Raa\;\; \bigwedge_{j=1}^k \forall \tau' \.\in \top \;\;
\sigma' z \neq \tau' u_j$	\\
by induction on
$rank(\sigma',(\sigma) \; (v:z))$, using \eqr{66}.
}

\LEMMA{
\eqd{69}
Let $y, x_1,\ldots,x_k \in {\cal V}$,
$w \in {\cal T}_{{\cal CR},{\cal F},{\cal V}}$,
$u(x_1,\ldots,x_k) \in {\cal T}_{{\cal CR},\{x_1,\ldots,x_k\}}$,
$u_{ij}, v_{ij}, v_i \in {\cal T}_{{\cal CR},{\cal V}}$, \\
$v_i(y) \in {\cal T}_{{\cal CR},\{y\}}$ for $1 \.\leq i \.\leq n_1$,
$v_i \in {\cal T}_{\cal CR}$ for $n_1+1 \.\leq i \.\leq n_2$, \\
We abbreviate $u(y,u_{12},\ldots,u_{1k})$
to $u(y,\vec{u}_1)$.
\\
Assume
\\
\begin{tabular}[t]{@{}r@{$\;$}l@{$\;$}l@{$\;$}l@{$\;$}l@{}}
& $(\sigma)$ & $(\tau)$ & $(w:z)$	\\
$=$ & $(\beta_1 \sigma)$ & $(\tau_1)$ & $(\beta_1 w:u(y,\vec{u}_1))$
	& $(u(v_1(y),\vec{v}_1):z)$	\\
$\mid$ & $ \ldots $	\\
$\mid$ & $(\beta_{n_1} \sigma)$ & $(\tau_{n_1})$
	& $(\beta_{n_1} w:u(y,\vec{u}_{n_1}))$
	& $(u(v_{n_1}(y),\vec{v}_{n_1}):z)$ \\
$\mid$ & $(\beta_{n_1+1} \sigma)$ & $(\tau_{n_1+1})$
	& $(\beta_{n_1+1} w:u_{n_1+1})$
	& $(u(v_{n_1+1},\vec{v}_{n_1+1}):z)$ \\
$\mid$ & $ \ldots $	\\
$\mid$ & $(\beta_{n_2} \sigma)$ & $(\tau_{n_2})$
	& $(\beta_{n_2} w:u_{n_2})$ & $(u(v_{n_2},\vec{v}_{n_2}):z)$ \\
$\mid$ && $(\tau_{n_2+1})$ & $(\beta_{n_2+1} w:u_{n_2+1})$
	& $(v_{n_2+1}:z)$	\\
$\mid$ & $ \ldots $	\\
$\mid$ && $(\tau_{n_3})$ & $(\beta_{n_3} w:u_{n_3})$ & $(v_{n_3}:z)$ \\
\end{tabular}
\\
and $(u_i:u(x_1,\vec{x})) = \bot_s = (v_j:u(x_1,\vec{x}))$
for $n_1+1 \leq i \leq n_2$ and $n_2+1 \leq j \leq n_3$.	\\
Define $S \sortdef v_1(S) \mid \ldots \mid v_{n_1}(S) \mid
	v_{n_1+1} \mid \ldots \mid v_{n_2}$.	\\
Then,
\\
\begin{tabular}[t]{@{}r@{$\;$}l@{$\;$}l@{$\;$}l@{$\;$}l@{$\;$}ll@{}}
& $(\sigma)$ & $(\tau)$ & $(w:z)$	\\
$=$ & $(\beta_1 \sigma)$ & $(\tau_1)$ & $(\beta_1 w:u(y,\vec{u}_1))$
	& $(u(v_1(y),\vec{v}_1):z)$ & $(S:y)$	\\
$\mid$ & $ \ldots $	\\
$\mid$ & $(\beta_{n_1} \sigma)$ & $(\tau_{n_1})$
	& $(\beta_{n_1} w:u(y,\vec{u}_{n_1}))$
	& $(u(v_{n_1}(y),\vec{v}_{n_1}):z)$ & $(S:y)$	\\
$\mid$ & $(\beta_{n_1+1} \sigma)$ & $(\tau_{n_1+1})$
	& $(\beta_{n_1+1} w:u_{n_1+1})$
	& $(u(v_{n_1+1},\vec{v}_{n_1+1}):z)$ \\
$\mid$ & $ \ldots $	\\
$\mid$ & $(\beta_{n_2} \sigma)$ & $(\tau_{n_2})$
	& $(\beta_{n_2} w:u_{n_2})$ & $(u(v_{n_2},\vec{v}_{n_2}):z)$
	&& $\mid $ \\
$\mid$ && $(\tau_{n_2+1})$ & $(\beta_{n_2+1} w:u_{n_2+1})$
	& $(v_{n_2+1}:z)$	\\
$\mid$ & $ \ldots $	\\
$\mid$ && $(\tau_{n_3})$ & $(\beta_{n_3} w:u_{n_3})$ & $(v_{n_3}:z)$ \\
\end{tabular}
\\
Intuitively, a constructor term of the form $u(v,\ldots)$ can occur
only in two places:
\bi
\item in one of the alternatives $1,\ldots,n_1$,
	$v$ having the form $v_i(v')$ where $u(v',\ldots)$ occurred
	earlier; or
\item in one of the alternatives $n_1+1,\ldots,n_2$,
	$v$ having the form $v_i$.
\ei
Thus, it is always true that $v \in S^M$.
}
\PROOF{
Show 
$\exists \sigma' \in (\sigma) (w:z) \; ^M \;\;\;
u(u'_1,\vec{u}') = \sigma' z \Ra u'_1 \in S^M$
by induction along the order
$u(u'_1,\vec{u}') < u(u''_1,\vec{u}'') 
:\Lra u'_1 \subterm u''_1$.
}

\ALGORITHM{
\eqd{70}
The following algorithm provides an initial, coarse approximation
$max_f$ of the range sorts for an equationally defined function $f$.
\\
\begin{tabular}[t]{@{}l@{$\;$}lr@{}}
$\eqi{max_f}$ & $\sortdef max_{\mu_1,v_1} \mid \ldots \mid
	max_{\mu_m,v_m}$	\\
\mca{2}{where $max_{\mu,w}$ for
	$w \in {\cal T}_{{\cal CR},{\cal F},{\cal V}}$
	is defined by:} \\
$max_{\mu,g(w_1,\ldots,w_n)}$ & $:= max_g$ & if $g \in {\cal F}$ \\
$max_{\mu,cr(w_1,\ldots,w_n)}$
	& $\sortdef cr(max_{\mu,w_1},\ldots,max_{\mu,w_n})$
	& if $cr \in {\cal CR}$	\\
$max_{\mu,x}$ & $\sortdef apply(\mu,x)$
	& if $x \in {\cal V}$	\\
\end{tabular}
\\
If $f(w_1,\ldots,w_n) \ablds u \in {\cal T}_{\cal CR}$,
then $u \in max_f^M$,
as can be shown by induction on the length of the $\abldi$ chain.
}

\ALGORITHM{
\eqd{71}
Assume $f$ is defined by the (yet unsorted) equations
\\
\begin{tabular}[t]{@{}l@{$\;$}l@{}}
$f(u_{11},\ldots,u_{1n_1})$ & $= v_1$ \\
\ldots	\\
$f(u_{m1},\ldots,u_{mn_m})$ & $= v_m$ \\
\end{tabular}
\\
Assume that for each $f \in {\cal F}$ a set $F_f \subset {\cal F}$
of admitted function symbols for
arguments of $f$ is given.
The following algorithm finds minimal independent t-sets $\mu_i$ such
that the applicability of $\abldi$ becomes trivial if only subterms
starting with a $g \in F_f$ appear at the argument positions of $f$. 
\\
\begin{tabular}[t]{@{}l@{$\;$}lr@{}}
$\eqi{max'_f}$ 
	& $\sortdef max'_{f,v_1} \mid \ldots \mid max'_{f,v_m}$ \\
\mca{2}{where $max'_{f,w}$ for
	$w \in {\cal T}_{{\cal CR},{\cal F},{\cal V}}$
	is defined by:} \\
$max'_{f,g(w_1,\ldots,w_n)}$ & $:= max'_g$ & if $g \in {\cal F}$ \\
$max'_{f,cr(w_1,\ldots,w_n)}$
	& $\sortdef cr(max'_{f,w_1},\ldots,max'_{f,w_n})$
	& if $cr \in {\cal CR}$ \\
$max'_{f,x}$ & $\sortdef \bigmid_{g \in F_f} max'_g$
	& if $x \in {\cal V}$	\\
\end{tabular}
\\
Define 
$\mu_i = compose(\{ abstract(x,max'_{f,x}) \;\mid\;
	x \in vars(u_{i1},\ldots,u_{in_i}) \})$.
\\
Then, $f(v_1,\ldots,v_n) \abldi v$
iff \eqr{61}.1 is satisfied and
($v_i \in {\cal V}$ 
or $v_i = g(\vec{v}_i)$ with $g \in {\cal CR} \cup {\cal F}_f$).
}

\EXAMPLE{
\eqd{72}
For the functions defined by the unstarred equations of
Fig.~\ref{Sort and function definitions for synthesis of binary
arithmetic algorithms}, allowing arbitrary argument terms
for $+$ and $dup$, but only constructor terms from $Bin$
as arguments for $val$, one gets
$F_+ = F_{dup} = \{+,dup,val\}$,
$F_{val} = \{\}$,
and
\\
\begin{tabular}{@{}l@{$\;$}l@{}}
$max'_+$ & $\sortdef max'_{+,x} \mid s(max'_+)$	\\
$max'_{+,x}$ & $\sortdef max'_+ \mid max'_{dup} \mid max'_{val}$ \\
$max'_{dup}$ & $\sortdef max'_+$	\\
$max'_{val}$ & $\sortdef 0 \mid max'_{dup} \mid s(max'_{dup})$	\\
\end{tabular}
\hfill
or, simplified:
\hfill
\begin{tabular}{@{}l@{$\;$}l@{}}
$max'_+$ & $\sortdef Nat$	\\
$max'_{+,x}$ & $\sortdef Nat$	\\
$max'_{dup}$ & $\sortdef Nat$	\\
$max'_{val}$ & $\sortdef Nat$	\\
\end{tabular}
\\
which corresponds to
the implicit t-set shown in Fig.~\ref{Sort and function
definitions for synthesis of binary arithmetic algorithms}.
}

\ALGORITHM{
\eqd{73}
To compute $\eqi{rg(\sigma,v)}$,
start with the expression $(\sigma) \; (v:z)$, where $z$ is new,
and repeatedly apply rules in the following order:
global rules, approximation rule, simplifying local rules (like
Defs.~\eqr{64}.2 and \eqr{64}.3), and rewriting (Def.~\eqr{64}.1).
Apply approximation only if certain conditions make it necessary;
apply all other rules wherever possible.
By setting certain parameters in the termination criterion,
the trade-off
between computation time and precision of the result can be
controlled.
On termination, an expression
$(\sigma_1) \; (u_1:z) \mid \ldots \mid (\sigma_n) \; (u_n:z)$ with
regular t-sets $\sigma_i$ is obtained.
The final result is then
$rg(\sigma,v) := \sigma_1 u_1 \mid \ldots \mid \sigma_n u_n$,
satisfying $nf[\sigma^M v] \subset rg(\sigma,v)^M$.
}

\begin{figure}
\begin{center}
\begin{tabular}[t]{@{}l@{$\;$}lr@{}}
$x+0$ & $= x$	\\
$x+s(y)$ & $= s(x)+y$	\\
\end{tabular}
\\[0.3cm]
\begin{tabular}[t]{@{}r@{$\;$}lr@{}}
& $(Nat:x) \; (Nat:y) \; (x+y:z)$	\\
$=$ & $(Nat:x) \; (Nat:y) \; (x:x_1) \; (y:0) \; (x_1:z)$
	& \eqr{64}.1	\\
$\mid$ & $(Nat:x) \; (Nat:y) \; (x:x_1) \; (y:s(y_1)) \;
	(s(x_1)+y_1:z)$ \\
$=$ & $(Nat:x) \; (x:z)$
	& simplification	\\
$\mid$ & $(Nat:x_1) \; (Nat:y_1) \; (s(x_1)+y_1:z)$	\\
$=$ & $(Nat:x) \; (x:z)$
	& \eqr{64}.1	\\
$\mid$ & $(Nat:x_2) \; (x_2:z)$
	& + simplification	\\
$\mid$ & $(Nat:x_1) \; (Nat:y_2) \; (s(s(x_1))+y_2:z)$	\\
$=$ & $ \ldots $	\\
$=$ & $(Nat:x) \; (x:z)$
	& \eqr{64}.1	\\
$\mid$ & $(Nat:x) \; (Nat:y) \; (s^i(x)+y:z)$
	& + simplification	\\
$=$ & $ \ldots $	\\
\end{tabular}
\end{center}

\caption{Nonterminating sort rewriting computation}
\label{Nonterminating sort rewriting computation}
\end{figure}

The termination of Alg.~\eqr{73} has to be artificially enforced.
Certainly, the rewrite relation $\abldi$ is Noetherian, i.e.\ each
computation chain starting from a {\em term} will terminate.
However, Alg.~\eqr{73} computes with {\em sorts} that represent
infinitely many terms in general, and the length of their computation
chains may increase unboundedly.
Hence, {\em sort} rewriting need not terminate even though {\em term}
rewriting terminates.
As an example, consider the equational theory
and the computation shown in 
Fig.~\ref{Nonterminating sort rewriting computation}.

In principle, Alg.~\eqr{73} can be stopped after every step, using
the approximation by $max_f$;
in other words,
there is a trade-off between computation time and the precision of
the result.
We suggest the following termination criterion:
applying Rule \eqr{64}.1 to an expression $(f(\ldots):\ldots)$ is
allowed only if less than $\#use(dom(f))$ rewrite steps wrt.\ $f$
have occurred in the current path\footnote{
	For an extended sort $S = \sigma u$,
	we define $use(S) = use(\sigma)$.

}.
The --~heuristic~-- justification considers that $f$ is defined
recursively over the structure of $dom(f)$ and that no more than
$\#use(dom(f))$ rewrite steps are necessary to ``get back to the
starting expression'', thus making e.g.\ Rule \eqr{67}
applicable.
Since all other local and global transformations except \eqr{64}.1
can be applied only a finite number of times,
this criterion ensures termination.

\begin{figure}
\begin{center}
\begin{tabular}[b]{@{}lll@{}}
\\
\mca{3}{Index sets:}	\\
$f_1:$ & $\{1,2\}$ & $\{1,3\}$	\\
$f_2:$ & $\{1,2\}$ & $\{2,3\}$	\\
$f_3:$ & $\{1,2\}$	\\
$f_4:$ & $\{1,2\}$ & $\{2,3\}$	\\
\end{tabular}
\hfill
\begin{picture}(7.7,3.2)
% {
% !2}textree g wd 1.0 lt 0.1 fs 1.2 | tc

% \tree{$\bf a$}
%   {\tree{$\bf b$}
%     {\tree{$e$} }
%     {\tree{$\bf f$}}
%     {\tree{$\bf g$}}
%   }
%   {\tree{$c$}
%     {\tree{$\bf h$}}
%     {\tree{$i$} }
%   }
%   {\tree{$\bf d$}
%     {\tree{$j$} }
%     {\tree{$\bf k$}}
%     {\tree{$\bf l$}}
%   }

\ % text after tree ignored
\put(0.500,0.200){\makebox(0.000,0.000)[t]{$e$}}
\put(1.500,0.200){\makebox(0.000,0.000)[t]{$\bf f$}}
\put(2.500,0.200){\makebox(0.000,0.000)[t]{$\bf g$}}
\put(1.500,1.600){\makebox(0.000,0.000)[t]{$\bf b$}}
\put(1.500,1.300){\line(-1,-1){1.000}}
\put(1.500,1.300){\line(0,-1){1.000}}
\put(1.500,1.300){\line(1,-1){1.000}}
\put(3.500,0.200){\makebox(0.000,0.000)[t]{$\bf h$}}
\put(4.500,0.200){\makebox(0.000,0.000)[t]{$i$}}
\put(4.000,1.600){\makebox(0.000,0.000)[t]{$c$}}
\put(4.000,1.300){\line(-1,-2){0.500}}
\put(4.000,1.300){\line(1,-2){0.500}}
\put(5.500,0.200){\makebox(0.000,0.000)[t]{$j$}}
\put(6.500,0.200){\makebox(0.000,0.000)[t]{$\bf k$}}
\put(7.500,0.200){\makebox(0.000,0.000)[t]{$\bf l$}}
\put(6.500,1.600){\makebox(0.000,0.000)[t]{$\bf d$}}
\put(6.500,1.300){\line(-1,-1){1.000}}
\put(6.500,1.300){\line(0,-1){1.000}}
\put(6.500,1.300){\line(1,-1){1.000}}
\put(4.000,3.000){\makebox(0.000,0.000)[t]{$\bf a$}}
\put(4.000,2.700){\line(-5,-2){2.500}}
\put(4.000,2.700){\line(0,-1){1.000}}
\put(4.000,2.700){\line(5,-2){2.500}}

\put(4.200,2.300){\makebox(0.000,0.000){$f_1$}}
\put(1.700,0.800){\makebox(0.000,0.000){$f_2$}}
\put(4.000,0.800){\makebox(0.000,0.000){$f_3$}}
\put(6.700,0.800){\makebox(0.000,0.000){$f_4$}}
\end{picture}
\end{center}

\caption{Applying a global transformation in an example computation 
	tree}
\label{Applying a global transformation in an example computation tree}
\end{figure}
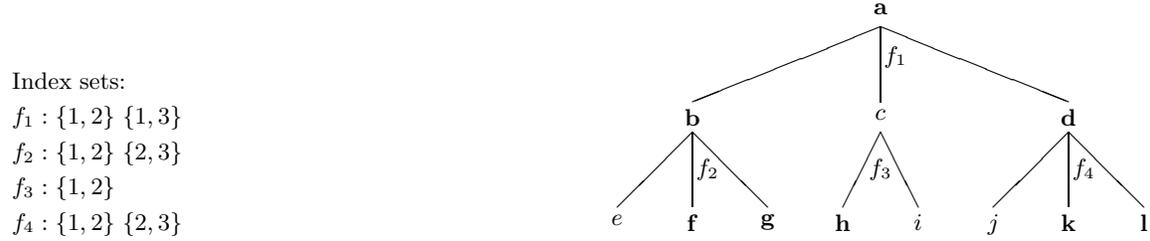

\label{4042}

The defining equations of a function $f$ need not be independent in a
logical / axiomatic sense; arbitrarily many ``derived'' equations may be
added, cf.\ Fig.~\ref{Range sort computation for $x+x$}.
Accordingly, it suffices to use only a subset of the equations for
a rewrite step by \eqr{64}.1, provided $dom(f)$ is still completely
covered (cf.\ the role of the index set $I$ in \eqr{64}.1).
Since it is not possible to select a suitable $I$ at the time the
rewrite step is conducted we proceed the other way round:
we use {\em all} equations for $f$ in each rewrite step, making a global
transformation applicable not only if all alternatives have the
required form but even when a subset of alternatives has the required
form and confinement to these alternatives still leads to index sets
completely covering $dom(f)$ in all relevant rewrite steps.

In this way, supplying additional derived function equations may
result in making ``better'' global transformations applicable,
and hence in enhancing the precision of the computed sort. Thus,
we may get an effect similar to that obtained by term
declarations in \cit{Schmidt-Schau\3 1988}{Schmidt.1988}. 

The test for applicability of a global transformation works as follows:
for each alternative $(\sigma)$ that does not meet the
applicability criterion, delete all complete index sets in the last
rewrite step leading to $(\sigma)$.
If no index set remains, omit this rewrite step, and delete in turn all
complete index sets in the previous rewrite step in the computation
tree.
If no previous rewrite step exists, the global transformation cannot be
made applicable.
If a complete index set still exists in the first, highest-level
rewrite step after all deletions are done, the transformation has been
made applicable.

As an example, 
consider the computation tree shown in Fig.~\ref{Applying a global
transformation in an example computation tree}.
Rewrite steps have been conducted for functions $f_1,f_2,f_3,f_4$,
transforming alternative $a$ into $b \mid c \mid d$, and in turn to 
$e \mid f \mid g \mid h \mid i \mid j \mid k \mid l$ (only applications
of \eqr{64}.1 are shown).
The complete index sets for each function are listed in the table.
Suppose a global transformation were applicable if we could restrict
our attention to
$f \mid g \mid h \mid k \mid l$ (shown in bold face).
The procedure described above results in selection of the index set
$\{2,3\}$ for both, $f_2$ and $f_4$, and $\{1,3\}$ for $f_1$,
deleting alternative $c$, as well.
Hence, the transformation has been made applicable.

\begin{figure}
Taking the definitions in 
Fig.~\ref{Sort and function definitions for synthesis of binary
arithmetic algorithms}, 
we can compute $rg(Nat_x,x+x)$:
\begin{center}
\begin{tabular}[t]{@{}|r*{3}{@{$\;$}l}|@{}}
\hline
& \multicolumn{3}{@{}l|@{}}{$(Nat:x) \; (x+x:z)$}	\\
$=$ & \multicolumn{3}{@{}l|@{}}{$(Nat:x) \; (x_1:z) \; (x:x_1) \; (x:0)
	\;\mid\; (Nat:x) \; (x_1:z) \; (x:0) \; (x:x_1) \;\mid$} \\
    & \multicolumn{3}{@{}l|@{}}{$(Nat\.:x) \; (s(x_1+y_1)\.:z)
	\; (x\.:x_1) \; (x\.:s y_1)
	\mid (Nat\.:x) \; (s(x_1+y_1)\.:z) \; (x\.:s x_1) 
	\; (x\.:y_1)$} \\
$=$ & $(0:z)$
	& $\mid (Nat:y_1) \; (s y_1 +y_1:z_1) \; (s z_1 :z)$
	& $\mid (Nat:x_1) \; (x_1+s x_1 :z_1) \; (s z_1 :z)$	\\
$=$ & $(0:z)$
	& $\mid (Nat:y_2) \; (s s y_2\.+y_2:z_2) \; (s s z_2\.:z)$
	& $\mid (Nat:y_2) \; (y_2+y_2:z_2) \; (s s z_2  :z)$ \\
    && $\mid (Nat:y_2) \; (y_2+y_2:z_2) \; (s s z_2  :z)$
	& $\mid (Nat:x_2) \; (x_2\.+s s x_2:z_2)\; (s s z_2\.:z)$ \\
$=$ & $(0:z)$
	& $\mid (Nat:y_2) \; (y_2+y_2:z_2) \; (s s z_2  :z)$ &	\\
$=$ & \multicolumn{3}{@{}l|@{}}{$(Even:z)$}	\\
\hline
\end{tabular}
\end{center}
where the new sort definition
$Even \sortdef 0 \mid s(s(Even))$ is generated.
The performed steps are:
Rule~\eqr{64}.1 with equations a.-d.;
simplification;
Rule~\eqr{64}.1 with c.-d.\ twice in parallel, including
simplification;
deletion of the 2nd, 4th, and 5th alternative, since they are covered by
the 3rd one;
this makes Lemma \eqr{67} applicable as the final step.

\caption{Range sort computation for $x+x$}
\label{Range sort computation for $x+x$}
\end{figure}

\begin{figure}
\begin{center}
\begin{tabular}{@{}|r@{$\;$}ll|@{}}
\hline
& $(Bin:x) \; (val \* x:z)$ &	\\
$=$ & $(Bin:x) \; (x:nil) \; (0:z)$ & Def.\ $val$	\\
$\mid$ & $(Bin:x) \; (x:x_1\.{::}o) \; (dup \* val \* x_1:z)$ &	\\
$\mid$ & $(Bin:x) \; (x:x_1\.{::}i) \; (s \* dup \* val \* x_1:z)$ & \\
$=$ & $(0:z)$ &	\\
$\mid$ & $(Bin:x_1) \; (dup \* val \* x_1:z)$ &	\\
$\mid$ & $(Bin:x_1) \; (dup \* val \* x_1:z_1) \; (s \* z_1:z)$ & \\
$=$ & $(0:z)$ & Def.\ $dup$	\\
$\mid$ & $(Bin:x_1) \; (val \* x_1:0) \; (0:z)$ & \\
$\mid$ & $(Bin:x_1) \; (val \* x_1:s \* x_2) \;
	(s \* s\* dup \* x_2:z)$ &	\\
$\mid$ & $(Bin:x_1) \; (val \* x_1:0) \; (0:z_1) \; (s \* z_1:z)$ & \\
$\mid$ & $(Bin:x_1) \; (val \* x_1:s \* x_2) \;
	(s \* s\* dup \* x_2:z_1) \; (s \* z_1:z)$ &	\\
$=$ & $(0:z)$ & \\
$\mid$ & $(Bin:x_1) \; (val \* x_1:s \* x_2) \;
	(dup \* x_2:z_2) \; (s \* s\* z_2:z)$ &	\\
$\mid$ & $(Bin:x_1) \; (val \* x_1:0) \; (s \* 0:z)$ & \\
$\mid$ & $(Bin:x_1) \; (val \* x_1:s \* x_2) \;
	(dup \* x_2:z_2) \; (s \* s\* s \* z_2:z)$ &	\\
$\stackrel{(\subset)}{=}$ & $(0:z)$ & $(*)$	\\
$\mid$ & $(Bin:x_1) \; (max_{val}:s \* x_2) \;
	(dup \* x_2:z_2) \; (s \* s\* z_2:z)$ &	\\
$\mid$ & $(Bin:x_1) \; (max_{val}:0) \; (s \* 0:z)$ & \\
$\mid$ & $(Bin:x_1) \; (max_{val}:s \* x_2) \;
	(dup \* x_2:z_2) \; (s \* s\* s \* z_2:z)$ &	\\
$=$ & $(0:z)$ & $(**)$	\\
$\mid$ & $(Nat:s \* x_2) \; (dup \* x_2:z_2) \; (s \* s\* z_2:z)$ & \\
$\mid$ & $(Nat:0) \; (s \* 0:z)$ & \\
$\mid$ & $(Nat:s \* x_2) \;
	(dup \* x_2:z_2) \; (s \* s\* s \* z_2:z)$ &	\\
$=$ & $(0:z)$ & $dup$, see above	\\
$\mid$ & $(s \* s\* Even:z)
	\mid (s \* 0:z) \mid (s \* s\* s \* Even:z)$ &	\\
$=$ & $(Nat:z)$ & \\
\hline
\end{tabular}

\begin{tabular}{@{}rp{11.3cm}@{}}
$(*)$: &
Here, $val \* x_1$ is estimated upwards,
since the original expression $(val \* x_1: \ldots )$ occurs as part of
the actual expression, but the introduction of a new recursive sort
definition is prohibited by the presence of the non-constructor
function $dup$.
We write ``$\stackrel{(\subset)}{=}$'' to indicate that $rg$ here
differs from the real, semantic range sort.	\\
$(**)$: &
One can trivially transform the defining equations of $val$ and $dup$
into sort definitions by replacing function applications with
corresponding sort names. This yields an upper bound for the range
sorts:	\\
&
	\begin{tabular}[t]{@{}l@{$\;$}lr@{}}
	$max_{val}$ & $\sortdef 0 \mid s \* max_{dup} \mid max_{dup}$
		& and \\
	$max_{dup}$ & $\sortdef 0 \mid s \* s\* max_{dup}$	\\
	\mca{3}{i.e.\ $max_{dup}^M = Even$ and $max_{val}^M = Nat^M$} \\
	\end{tabular}
\end{tabular}
\end{center}

\caption{Range sort computation for $val$}
\label{Range sort computation for $val$}

\end{figure}

\DEFINITION{
%[(Sorted Equation Solution)] 
\eqd{74}
%$\;$\\
A substitution $\beta$ is called a \eqi{solution of an equation} 
$\vs \sigma v_1= \vs \sigma v_2$
iff a t-set $\tau$ exists that denotes the sorts of variables in
the $ran(\beta)$ such that
\be
\item $\vs \tau \beta v_1 \Ablds \vs \tau \beta v_2$,
\item $\forall \tau' \in \tau \;\;
	\exists \sigma' \in \sigma \;\;
	\forall x \in vars(v_1,v_2) \;\;\;
	\tau' \beta x \mbox{ well-defined } 
	\Ra \tau' \beta x \Ablds \sigma' x$,
	\\
	or equivalently:
	$nf[\tau \beta \tpl{x_1,\ldots,x_n}]
	\subset \sigma \tpl{x_1,\ldots,x_n}$,
	where $\{x_1,\ldots,x_n\} = vars(v_1,v_2)$,
	\\
	similar to the classical
	well-sortedness requirement for $\beta$, and
\item $nf[\tau \beta v_1] \neq \{\}$,
	i.e.\ the solution has at least
	one well-defined ground instance.
\ee
}

\THEOREM{
%[(\eqi{Sorted Narrowing})] 
\eqd{75}
%$\;$\\
An arbitrary narrowing calculus preserving solution sets remains
complete if restricted appropriately by sorts.
For example, for \eqi{lazy narrowing} 
\cit{H\"olldobler 1989}{Holldobler.1989},
abbreviating $\tau := \sigma \pecomp \top_{vars(u_1,\ldots,u_n)}$,
we get for the main rules:
\vspace{0.3cm}

\noindent
\eqi{(ln)}
\tab
\begin{tabular}{@{}c@{}}
$\vs \tau v_1 = \vs \tau u_1 \wedge \ldots \wedge
	\vs \tau v_n = \vs \tau u_n \wedge
	\vs \tau v = \vs \tau {v'}$\\
\hline
$\vs \sigma f(v_1,\ldots,v_n) = \vs \sigma v$
\end{tabular}
	\tab
\begin{tabular}{@{}l@{}}
$\vs \top f(u_1,\ldots,u_n) = \vs \top {v'}$ defining equation	\\
$inh(inf(rg(\tau,v),rg(\tau,v')))$,	\\
$inh(inf(rg(\tau,v_1),rg(\tau,u_1)))$, \ldots	\\
$inh(inf(rg(\tau,v_n),rg(\tau,u_n)))$,	\\
\end{tabular}
\vspace{0.5cm}

\noindent
\eqi{(d)}
\tab
\begin{tabular}{@{}c@{}}
$\vs \tau u_1 = \vs \tau v_1 \;\wedge\; \ldots \;\wedge\; 
	\vs \tau u_n = \vs \tau v_n$ \\
\hline
$\vs \sigma f(u_1,\ldots,u_n) = \vs \sigma f(v_1,\ldots,v_n)$	\\
\end{tabular}
	\tab
\begin{tabular}{@{}l@{}}
$inh(inf(rg(\tau,u_1),rg(\tau,v_1)))$,\ldots,	\\
$inh(inf(rg(\tau,u_n),rg(\tau,v_n)))$	\\
\end{tabular}
\vspace{0.5cm}

\noindent
In rule (ln), the remaining equations 
$\vs \tau \; v_1 = \vs \tau \; u_1$, \ldots,
$\vs \tau \; v_n = \vs \tau \; u_n$
can often be solved by purely syntactic unification.
In this case, the non-disjointness criteria
$inh(inf(rg(\tau,v_1),rg(\tau,u_1)))$,\ldots,
$inh(inf(rg(\tau,v_n),rg(\tau,u_n)))$
are trivially satisfied and may be omitted in practical
implementations.
Note that the variables in defining equations have to be assigned
the sort $\top$.
Starting from a conditional narrowing calculus,
nontrivially sorted defining equations become possible.
}
\PROOF{
Rules may be restricted using the fact that the solution set of
$\vs {\sigma_1} v_1 = \vs {\sigma_2} v_2$
is inhabited only if
$rg(\sigma_1,v_1)^M \cap rg(\sigma_2,v_2)^M
\supset nf[\sigma_1^M v_1] \cap nf[\sigma_2^M v_2] \neq \{\}$.
\\
To prove the completeness of assigning sorts to variables in goal
equations,
observe that each definition of a regular
t-set $\sigma$ can be transformed
into a  definition of a function $f_\sigma$ admitted by
Def.~\eqr{61} such that $\sigma' \in \sigma^M$
iff $f_\sigma(\sigma' x_1,\ldots,\sigma' x_n) = true$,
where $dom(\sigma) = \{x_1,\ldots,x_n\}$ and $true \in {\cal CR}$.
Hence, a sorted equation $\vs \sigma v_1 = \vs \sigma v_2$
can be simulated
without sorts
by $v_1 = v_2 \wedge f_\sigma(x_1,\ldots,x_n) = true$.
Assigning non-trivial sorts to variables in defining equations is
possible for conditional calculi in a similar way.
}

\LEMMA{
%[(Sorted Equation Solvability)] 
\eqd{76}
%$\;$\\
Let $\sigma$ be independent, and let $x$ be new;
then, the equation
$\vs \sigma v_1 = \vs \sigma v_2$
has a solution iff
$rg(\sigma,\tpl{v_1,v_2})^M \cap
\top \tpl{x,x} \neq \{\}$,
provided the approximation rule was not used in $rg$ computation.
}

Lemma \eqr{76} shows that the amount of search space reduction by the
sorts depends only on the quality of approximations by
$rg$ and the expressiveness of our sort language.
Without the reflection of variable bindings
in sorts, such a result is impossible, even
if no ``occur check'' and no non-constructor functions are involved,
e.g.\
$\tpl{x,y} = \tpl{0,s(0)}$ is solvable, but
$\tpl{x,x} = \tpl{0,s(0)}$ is not.
\vspace{0.3cm}

It is possible to extend the presented framework to cope with unfree
constructors, too.
This allows us, for example, to define a sort $Set$ of sets of natural
numbers, cf.\ App.~\ref{Case Study ``Comb Vector Construction''}.
As we show below, it is sufficient to be able to compute
the closure of a sort wrt.\ the congruence relation induced by the
equations between constructors.

\DEFINITION{
\eqd{77}
Assume we are given certain \eqi{equations between constructors}
in addition
to the equations for defined functions.
As in Def.~\eqr{61}, we define the rewrite relation
$\eqi{\ablci}$ to be induced by the equations between constructors.
We do not require $\ablci$ to be confluent, nor to be
Noetherian.
For the union of both equation sets, we similarly
define $\eqi{\ablcdi}$.
We require the defining equations to be compatible with the constructor
equations, i.e.\
	\display{
	$\bigwedge_{i=1}^n v_i \Ablcs v'_i
	\Raa nf(f(v_1,\ldots,v_n)) \Ablcs nf(f(v'_1,\ldots,v'_n))$
	}
whenever at least one of the two normal forms exists.
Note that the sort algorithms work only on free sorts and hence ignore
the relation $\Ablcs$.
}

\LEMMA{
\eqd{78}
If $v, w \in {\cal T}_{{\cal CR},{\cal F}}$
are well-defined,
we have $v \Ablcds w \;\;\Lra\;\; nf(v) \Ablcs nf(w)$.
}
\PROOF{
``$\La$'' trivial;
``$\Ra$'' by induction on the length of the $\Ablcds$ chain.
}

In each equivalence class wrt.\ $\Ablcs$, we may select an arbitrary
element and declare it to be the normal form, thus defining 
\eqi{$nf_c$}.
We adapt the notion of solution of an equation system from
Def.~\eqr{74} by replacing $\Ablds$ with $\Ablcds$, leaving condition
\eqr{74}.3 unchanged. 
Then, using Lemma~\eqr{78}, we can show that an equation
$\vs \sigma v = \vs \sigma v'$ has a solution only if
$nf_c[rg(\sigma,v)^M] \cap nf_c[rg(\sigma,v')^M] \neq \{\}$.
If we have an algorithm \eqi{$rg_c$} to compute upper approximations for
$nf_c[\cdot]$, similar to $rg$ for $nf[\cdot]$,
we can extend the sorted narrowing rules from Def.~\eqr{75}
to cope with unfree constructors by replacing $rg(\tau,v)$
with $rg_c(rg(\tau,v))$, etc.
However, such an algorithm is not provided here.

\section{Application in Formal Program Development}
\label{Application in Formal Program Development}

\parbox[b]{11.5cm}{
To support formal program development,
we employ the paradigm of implementation proof, starting from an
``abstract'' operation $ao$ on abstract data of sort $as_1$ or
$as_2$ which are to be implemented by a corresponding ``concrete''
operation $co$ on concrete data of sorts $cs_1$ or $cs_2$,
respectively.
The connection 
between
abstract
and
concrete
data
is
established
by
representation 
func-
}
\hfill
\begin{picture}(3.2,1.5)
\put(0.800,0.400){\makebox(0.000,0.000){$cs_1$}}
\put(2.800,0.400){\makebox(0.000,0.000){$cs_2$}}
\put(0.800,1.400){\makebox(0.000,0.000){$as_1$}}
\put(2.800,1.400){\makebox(0.000,0.000){$as_2$}}
\put(1.100,0.400){\vector(1,0){1.400}}
\put(1.100,1.400){\vector(1,0){1.400}}
\put(0.800,0.600){\vector(0,1){0.600}}
\put(2.800,0.600){\vector(0,1){0.600}}
\put(1.800,0.500){\makebox(0.000,0.000)[b]{$co$}}
\put(1.800,1.500){\makebox(0.000,0.000)[b]{$ao$}}
\put(0.700,0.900){\makebox(0.000,0.000)[r]{$r_1$}}
\put(2.900,0.900){\makebox(0.000,0.000)[l]{$r_2$}}
\end{picture}

\noindent
tions
$r_1: cs_1 \ra as_1$ and $r_2: cs_2 \ra as_2$,
representing each concrete data term as an abstract one. Different
concrete terms may represent the same abstract term.
Thus,
it is possible to perform the computation on the concrete level, and
interpret the result on the abstract level.
The correspondence between the concrete and abstract operation imposes
correctness requirements on the concrete operation.

We wish to synthesize the concrete operation $co$ as the
Skolem function for $y$ in the formula
$\forall \vec{x} \; \exists y \;\; ao(r_1(\vec{x})) = r_2(y)$.
A suitable method for the constructive
correctness proof is induction on the form of
a data term $\vec{x} \in cs_1$,
leading to a case distinction according
to (one of) the head constructor(s)
of $\vec{x}$. In each case, we have to solve an
equation $ao(r_1(\vec{x}_i)) = r_2(y)$ wrt.\ $y$.
The synthesized function $co$ is then given by equations $co(\vec{x}_i)
= \beta_i y$,
where $\vec{x}_i$ is a data term starting with the $i^{th}$
constructor, and $\beta_i$ is the solving substitution for this case.
After having solved an equation, one still has to check whether the
solution $\beta_i y$ is of the required sort $cs_2$, if not,
a different solution must be found.

The sort discipline presented here
supports specifically this method. Besides
allowing recursive sort definitions of $cs_1$, $cs_2$, $as_1$, and
$as_2$ as well as recursive function definitions of $ao$, $r_1$, and
$r_2$, the induction principle from Thm.~\eqr{9}
provides the case distinction and proof goals for an induction on $x \in
cs_1^M$. 
The sort discipline is able to cope with the additional problems of
synthesis as
compared with verification, i.e., to direct the
construction of the solution term, to the extent that
disjoint subsorts of a
concrete sort are assigned with disjoint subsorts of the corresponding
abstract sort. In this manner, the sort of an ``abstract'' term
indicates which ``concrete'' terms are representing it.

As an example, consider the formal development of
algorithms for binary numbers,
Consider the sort and function definitions in 
Fig.~\ref{Sort and function definitions for synthesis of binary
arithmetic algorithms}. 
All terms are sorted by the t-set 
$[x\.{:=}Nat] \pecomp [y\.{:=}Nat] \pecomp [z\.{:=}Bin]$,
which is omitted in the equations for the sake of brevity.
Equations marked
by ``$^*$'' are redundant and can be proven by structural
induction. The rightmost column contains the sort computed by
$rg$ for each equation; all sorts happen to be regular.
We have axioms defining the ``representation function''
$val:Bin \raa Nat$, and an auxiliary function $dup$ to duplicate
natural numbers which uses the addition $+$ on natural numbers.

\begin{figure}
\begin{tabular}{@{}l@{$\;$}ll@{}}
$\eqi{Nat}$ & $\sortdef 0 \mid s(Nat)$	\\
$\eqi{Bin}$ & $\sortdef nil \mid Bin\.{::}o \mid Bin\.{::}i$ \\[0.9ex]
\mca{2}{E.g.}	\\
\mca{2}{$val(nil\.{::}i\.{::}o\.{::}i) = s^5(0)$}	\\[0.9ex]
\mca{2}{Prove}	\\
\mca{2}{$\forall c \;\exists z \;\; s(val(c)) = val(z)$} \\[0.9ex]
\end{tabular}
\hfill
\begin{picture}(3.2,1.5)
\put(0.6,0.2){\makebox(0,0){$Bin$}}
\put(2.6,0.2){\makebox(0,0){$Bin$}}
\put(0.6,1.2){\makebox(0,0){$Nat$}}
\put(2.6,1.2){\makebox(0,0){$Nat$}}
\put(0.9,0.2){\vector(1,0){1.4}}
\put(0.9,1.2){\vector(1,0){1.4}}
\put(0.6,0.4){\vector(0,1){0.6}}
\put(2.6,0.4){\vector(0,1){0.6}}
\put(1.6,0.3){\makebox(0,0)[b]{$incr$}}
\put(1.6,1.3){\makebox(0,0)[b]{$s$}}
\put(0.5,0.7){\makebox(0,0)[r]{$val$}}
\put(2.7,0.7){\makebox(0,0)[l]{$val$}}
\end{picture}
\hfill
\begin{tabular}{@{}l@{\hspace*{0.5em}}r@{$\;$}l@{\hspace*{1em}}l@{}}
\it a. & $x + 0$ & $= x$ & $Nat$	\\
\it b.$^*$ & $0 + x$ & $= x$ & $Nat$	\\
\it c. & $x + s(y)$ & $= s(x+y)$ & $s(Nat)$	\\
\it d.$^*$ & $s(x) + y$ & $= s(x+y)$ & $s(Nat)$	\\
\it e. & $\eqi{dup(x)}$ & $= x+x$ & $Even$	\\
\it f.$^*$ & $dup(x+y)$ & $=dup(x) + dup(y)$ & $Even$	\\
\it g. & $\eqi{val}(nil)$ & $= 0$ & $0$	\\
\it h. & $val(z\.{::}o)$ & $= dup(val(z))$ & $Even$	\\
\it i. & $val(z\.{::}i)$ & $= s(dup(val(z)))$ & $s(Even)$	\\
\end{tabular}

\caption{Sort and function definitions for synthesis of binary
	arithmetic algorithms}
\label{Sort and function definitions for synthesis of binary arithmetic
	algorithms}
\end{figure}

The main contribution of the sorts is the computation of
$rg(Nat_x,dup(x))$ $= [z \.{:=} Even] \; z = Even$,
where the sort definition $\eqi{Even} \sortdef 0 \mid s(s(Even))$
is automatically introduced.
Although only independent t-sets are involved in the example,
the variable bindings in $x+x$ are reflected by its sort, viz.\ $Even$.
For this range-sort computation, the redundant equation {\it d.} is
necessary, cf.\ Fig.~\ref{Range sort computation for $x+x$}.

Taking the easiest example, let us synthesize an algorithm $incr$
for incrementing a binary number; the synthesis of algorithms for
addition and multiplication is shown in App.~\ref{Case Study ``Binary
Arithmetic''}.
The goal $\forall c \; \exists z \;\; s(val(c)) = val(z)$ is 
proved
by structural induction on $c$, the appropriate induction
scheme being provided by Thm.~\eqr{9}, cf.\ Fig.~\ref{Induction
principle for sort $Bin$}.
For example,
in case $c = c'\.{::}o$,
we have to solve the equation
$s(val(c'\.{::}o)) = val(z)$ wrt.\ $z$,
and the sorted narrowing rule from Thm.~\eqr{75} is only
applicable to equation {\it i.} since the left-hand side's sort
is computed as $s(Even)$.
Note that, in order to get the full benefit of the sort calculus,
narrowing should be applied only at the root of a term, since then
additional sort information is supplied from the other side of the
equation; this is the reason for using lazy narrowing.
The employed calculus' drawback of admitting
only trivially sorted defining equations is overcome by
subsequently checking the solutions obtained for well-sortedness.

Narrowing with equation
{\it h.} instead of {\it i.} would lead into an
infinite branch\footnote{
	Cf.\ Figs.~\ref{Brute Force Search Space for Equation val z = s
	val c1::o F}
	and~\ref{Search space of $incr$ synthesis using sorts}
	in App.~\ref{Case Study ``Binary Arithmetic''},
	where the search space for this example is shown for
	both unsorted and sorted narrowing.
},
trying to solve an equation $s(dup(\ldots)) = dup(\ldots)$.
Such infinite branches are cut off by the sorts, especially by the
global transformation rules which detect certain kinds of recursion
loops.
This seems to justify the computational overhead of sort computation.
Thanks
to the provided proof methodology based on regular
t-sets, new global rules
for detecting new recursion patterns can easily be added if required.

The control information provided by the sort calculus
acquires particular importance in
``proper'' narrowing steps, i.e., the ones actually contributing
to the solution term. While conventional narrowing procedures
essentially enumerate each element of the constructor term
algebra and test whether it is a solution, the presented sort
calculus approaches the solutions directly,
depending on the precision of computed range sorts.

The sort algorithms, especially $rg$, perform, in fact, simple induction
proofs.
For example, it is easy to prove by induction that $x+x$ always has
sort $Even$, and that sorts $Even$ and $s(Even)$
are disjoint, once these
claims have been guessed or intuitively recognized.
However, while a conventional induction prover would not propose these
claims as auxiliary lemmas during the proof of
$\forall c \; \exists z \;\; s(val(c)) = val(z)$,
they are implicitly generated by the sort algorithms.
The sort calculus allows the ``recognition of new
concepts'', so to speak, although
only within the rather limited framework given by the sort language.
In \cit{Heinz 1994}{Heinz.1994a}, an approach to the
automatic generation of
more complex auxiliary lemmas is presented based on E-generalization
using regular sorts, too.

A prototype support system written in Quintus-Prolog takes a total of
41 seconds user time on a Sparc 1
to automatically conduct the 9 induction proofs, with 135 narrowing
subgoals necessary for
the development of incrementation, addition, and multiplication
algorithms on binary numbers, cf.\ App.~\ref{Case Study ``Binary
Arithmetic''}.
In the form of
a paper case study from the area of compiler construction,
an implementation of sets of lists of natural numbers by ordered
son-brother trees has been proved, cf.\ App.~\ref{Case Study ``Comb
Vector Construction''}.
The algorithm for inserting a new list into a tree
is used to construct comb vectors for parse table compression;
it is specified as an implementation of $(\{\cdot\} \cup \cdot)$.
The use of sorts reduces the search space of the synthesis proof to
that of a verification proof, i.e.\ it uniquely determines all proper
narrowing steps or solution constructors.
The computed signatures are too complex for there to be much likelihood
of their being
declared by a user who does not know the proof in advance.

%\section{References}

%\renewcommand{\literaturename}{}

\bibliographystyle{plain}
\bibliography{lit}

\newpage

\noindent
{\LARGE\bf Appendix}

\appendix

\section{Case Study ``Binary Arithmetic''}
\label{Case Study ``Binary Arithmetic''}

In this appendix, the synthesis of algorithms for incrementation,
addition, and
multiplication of binary numbers is shown.
Figure~\ref{Implementation proofs for binary arithmetic}
gives an overview of the induction proofs conducted, together with the
induction variable, the computation time (in seconds on a Sparc 1 under
Quintus Prolog), and the number of subgoals.
$a$ and $b$ denote (Skolem) constants, while $x$ denotes a variable
with respect to which the equation is to be solved.
Note that in our paradigm, universally quantified variables are
skolemized into constant symbols;
induction is performed over just some of these constants.
Figure~\ref{Synthesized algorithms for binary arithmetic} lists
the synthesized algorithms.

On page~\pageref{synthesis session}, a
protocol of the synthesis session is given.
Pure induction proofs, i.e.\ ones that do not solve
an equation wrt.\ some
variable, are omitted.
The notation in the Prolog implementation differs slightly from the one
used in this paper.
The predicate ``$init\_sort\_system$'' computes a range sort for every
defining equation.
The predicate ``$solve$'' tries to solve the given equation;
every time the actual narrowing step is not uniquely determined by the
sorts, the user is shown a menu and prompted to make a decision.
During the session, the user always had to decide
to start an induction, indicated
by ``{\bf ind}'', optionally followed by a list of induction
variables.
Note that none of the proofs required any further user interaction.
The execution trace shows the actual subgoal on entry into ``$solve$'',
and the solved subgoal together with the solving substitution on exit.
At the end of the session, an example computation ($5*6 = 30$) is
performed (predicate ``$eval\_term$'')
using the newly synthesized algorithms,
and the sort definitions (incomplete), proved laws, and
function-defining equations are listed.
Functions $f_{24}$, $f_{188}$, and $f_{429}$ compute the successor
of a binary number, the sum of two binary numbers, and the product of
two binary numbers, respectively.

Starting on page~\pageref{Brute Force Search Space for Equation val z =
s val nil F}, the search space is shown in particular for the synthesis
of the $incr$ algorithm. Figures~\ref{Brute Force Search Space for
Equation val z = s val nil F} to~\ref{Brute Force Search Space for
Equation val z = s val c2::i F} show the search space in cases where
no sorts are
used to control narrowing; Fig.~\ref{Search space of $incr$
synthesis using sorts} shows the search space where sorts are used.

\newpage

\begin{figure}
\begin{center}
\begin{tabular}[t]{|r@{$\;$}l|r|r|r|}
\hline
\multicolumn{2}{|c|}{formula} & ind. & time  & sub \\
			    && var. & (sec.) & goals \\
\hline \hline
\multicolumn{2}{|c|}{initialization} && 3 &	     \\
\hline
$s(val(a))$ & $= val(x)$ & $a$ & 5 & 10 \\
\hline
$0+a$ & $= a$ & $a$ & 1 & 5	\\
$s(a)+b$ & $= s(a+b)$ & $b$ & 1 & 7	\\
$dup(a)+dup(b)$ & $= dup(a+b)$ & $b$ & 2 & 11	\\
\hline
$val(a)+val(b)$ & $= val(x)$ & $a,b$ & 19 & 47	\\
\hline
$a+b+b$ & $= a+dup(b)$ & $b$ & 2 & 11	\\
$a*dup(b)$ & $= dup(a*b)$ & $b$ & 2 & 13	\\
$dup(val(a))+val(b)$ & $= val(add(a\.{::}o,b))$ & $b$ & 3 & 18	\\
\hline
$val(a)*val(b)$ & $= val(x)$ & $b$ & 6 & 13	\\
\hline
\multicolumn{2}{|c|}{total} && 41 & 135 \\
\hline
\end{tabular}
\end{center}

\caption{Implementation proofs for binary arithmetic}
\label{Implementation proofs for binary arithmetic}
\end{figure}

\begin{figure}
\begin{center}
\begin{tabular}[t]{@{}|l@{$\;$}l|@{}}
\hline
$incr(nil)$ & $= nil\.{::}i$	\\
$incr(x\.{::}o)$ & $= x\.{::}i$ \\
$incr(x\.{::}i)$ & $= incr(x)::o$	\\
\hline
$add(nil,nil)$ & $= nil$	\\
$add(nil,y\.{::}o)$ & $= y\.{::}o$	\\
$add(nil,y\.{::}i)$ & $= y\.{::}i$	\\
$add(x\.{::}o,nil)$ & $= x\.{::}o$	\\
$add(x\.{::}o,y\.{::}o)$ & $= add(x,y)\.{::}o$	\\
$add(x\.{::}o,y\.{::}i)$ & $= add(x,y)\.{::}i$	\\
$add(x\.{::}i,nil)$ & $= x\.{::}i$	\\
$add(x\.{::}i,y\.{::}o)$ & $= add(x,y)\.{::}i$	\\
$add(x\.{::}i,y\.{::}i)$ & $= incr(add(x,y))\.{::}o$	\\
\hline
$mult(x,nil)$ & $= nil$ \\
$mult(x,y\.{::}o)$ & $= mult(x,y)\.{::}o$	\\
$mult(x,y\.{::}i)$ & $= add(mult(x,y)\.{::}o,x)$	\\
\hline
\end{tabular}
\end{center}

\caption{Synthesized algorithms for binary arithmetic}
\label{Synthesized algorithms for binary arithmetic}
\end{figure}

\newpage

\newcommand{\1}{\hspace*{0.3cm}}

\noindent
{\bf Synthesis session protocol}

{\bf
\label{synthesis session}

\renewcommand{\*}{\;}

\begin{tabbing}
MMMMMMMMMMMMMMMMMMMMMMMMMMM\=\kill
\\
?- $init\_sort\_system.$	\\
$val \* nil=0$	\> $0$	\\
$val (x:o)=dup \* val \* x$	\> $sort_{1}$	\\
$val (x:i)=s \* dup \* val \* x$	\> $s \* sort_{1}$	\\
$dup \* 0=0$	\> $0$	\\
$dup \* s \* u=s \* s \* dup \* u$	\> $s \* s \* sort_{1}$	\\
$u+0=u$	\> $nat$	\\
$u+s \* v=s (u+v)$	\> $s \* sort_{4}$	\\
$u*0=0$	\> $0$	\\
$u* (s \* v)=u*v+u$	\> $0 \mid nat \mid s \* nat$	\\
\\
?- $solve(s(val(c)) = val(x),S).$	\\
$s \* val \* c=val \* x$	\\
\\
$1$	$[dup \* val \* x_{19}=s \* val \* c]	\la [x:=x_{19}:o]$	\\
$2$	$[dup \* val \* x_{23}=val \* c]	\la [x:=x_{23}:i]$	\> ind	\\
\\
\1$s \* val \* nil=val \* x$	\\
\1\1$s \* 0=val \* x$	\\
\1\1\1$dup \* val \* x_{27}=0$	\\
\1\1\1\1$0=val \* x_{27}$	\\
\1\1\1\1$0=val \* x_{27}$	\> $\la [x_{27}:=nil]$	\\
\1\1\1$dup \* val \* x_{27}=0$	\> $\la [x_{27}:=nil]$	\\
\1\1$s \* 0=val \* x$	\> $\la [x:=nil:i]$	\\
\1$s \* val \* nil=val \* x$	\> $\la [x:=nil:i]$	\\
\1$s \* val (c_{25}:o)=val \* x$	\\
\1\1$s \* dup \* val \* c_{25}=val \* x$	\\
\1\1$s \* dup \* val \* c_{25}=val \* x$	\> $\la [x:=c_{25}:i]$	\\
\1$s \* val (c_{25}:o)=val \* x$	\> $\la [x:=c_{25}:i]$	\\
\1$s \* val (c_{26}:i)=val \* x$	\\
\1\1$s \* s \* dup \* val \* c_{26}=val \* x$	\\
\1\1\1$dup \* val \* x_{38}=s \* s \* dup \* val \* c_{26}$	\\
\1\1\1\1$val \* f_{24}(c_{26})=val \* x_{38}$	\\
\1\1\1\1$val \* f_{24}(c_{26})=val \* x_{38}$	\> $\la [x_{38}:=f_{24}(c_{26})]$	\\
\1\1\1$dup \* val \* x_{38}=s \* s \* dup \* val \* c_{26}$	\> $\la [x_{38}:=f_{24}(c_{26})]$	\\
\1\1$s \* s \* dup \* val \* c_{26}=val \* x$	\> $\la [x:=f_{24}(c_{26}):o]$	\\
\1$s \* val (c_{26}:i)=val \* x$	\> $\la [x:=f_{24}(c_{26}):o]$	\\
$s \* val \* c=val \* x$	\> $\la [x:=f_{24}(c)]$	\\
\\
$S = [x:=f_{24}(c)] $	\\
\\
?- $solve(0+a = a,S).$	\\
% $0+a=a$	\\
% 
% $1$	$[u_{45}=0,0=a]	\la [u_{45}:=a]$	\\
% $2$	$[s (u_{47}+v_{48})=a,s \* v_{48}=a]	\la [u_{47}:=0]$	\> ind	\\
% 
% \1$0+0=0$	\\
% \1\1$0=0$	\\
% \1\1$0=0$	\> $\la [ \; ]$	\\
% \1$0+0=0$	\> $\la [ \; ]$	\\
% \1$0+s \* a_{49}=s \* a_{49}$	\\
% \1\1$s (0+a_{49})=s \* a_{49}$	\\
% \1\1\1$a_{49}=a_{49}$	\\
% \1\1\1$a_{49}=a_{49}$	\> $\la [ \; ]$	\\
% \1\1$s (0+a_{49})=s \* a_{49}$	\> $\la [ \; ]$	\\
% \1$0+s \* a_{49}=s \* a_{49}$	\> $\la [ \; ]$	\\
% $0+a=a$	\> $\la [ \; ]$	\\
% \\
$S = [ \; ] $	\\
\\
?- $solve(s(a)+b = s(a+b),S).$	\\
% $s \* a+b=s (a+b)$	\\
% 
% $1$	$[a+b=a,0=b]	\la [u_{65}:=s (a+b)]$	\\
% $2$	$[u_{84}=s \* a,s \* v_{85}=b]	\la [u_{84}:=a,v_{85}:=b]$	\> ind[b].	\\
% 
% \1$s \* a+0=s (a+0)$	\\
% \1\1$s \* a=s (a+0)$	\\
% \1\1\1$a=a$	\\
% \1\1\1$a=a$	\> $\la [ \; ]$	\\
% \1\1$s \* a=s (a+0)$	\> $\la [ \; ]$	\\
% \1$s \* a+0=s (a+0)$	\> $\la [ \; ]$	\\
% \1$s \* a+s \* b_{86}=s (a+s \* b_{86})$	\\
% \1\1$s (s \* a+b_{86})=s (a+s \* b_{86})$	\\
% \1\1\1$s \* a+b_{86}=s (a+b_{86})$	\\
% \1\1\1\1$s (a+b_{86})=s (a+b_{86})$	\\
% \1\1\1\1$s (a+b_{86})=s (a+b_{86})$	\> $\la [ \; ]$	\\
% \1\1\1$s \* a+b_{86}=s (a+b_{86})$	\> $\la [ \; ]$	\\
% \1\1$s (s \* a+b_{86})=s (a+s \* b_{86})$	\> $\la [ \; ]$	\\
% \1$s \* a+s \* b_{86}=s (a+s \* b_{86})$	\> $\la [ \; ]$	\\
% $s \* a+b=s (a+b)$	\> $\la [ \; ]$	\\
% \\
$S = [ \; ] $	\\
\\
?- $solve(dup(a)+dup(b) = dup(a+b),S).$	\\
$S = [ \; ] $	\\
\\
?- $solve(val(c)+val(d) = val(x),S).$	\\
$val \* c+val \* d=val \* x$	\\
\\
$1$	$[0=val \* c+val \* d]	\la [x:=nil]$	\\
$2$	$[dup \* val \* x_{141}=val \* c+val \* d]	\la [x:=x_{141}:o]$	\\
$3$	$[s \* dup \* val \* x_{153}=val \* c+val \* d]	\la [x:=x_{153}:i]$	\\
$4$	$[0=val \* d]	\la [u_{162}:=val \* x,x:=c]$	\\
$5$	$[s (u_{186}+v_{187})=val \* x,s \* v_{187}=val \* d]	\la [u_{186}:=val \* c]$	\> ind	\\
\\
\1$val \* nil+val \* nil=val \* x$	\\
\1\1$0+val \* nil=val \* x$	\\
\1\1\1$0+0=val \* x$	\\
\1\1\1\1$0=val \* x$	\\
\1\1\1\1$0=val \* x$	\> $\la [x:=nil]$	\\
\1\1\1$0+0=val \* x$	\> $\la [x:=nil]$	\\
\1\1$0+val \* nil=val \* x$	\> $\la [x:=nil]$	\\
\1$val \* nil+val \* nil=val \* x$	\> $\la [x:=nil]$	\\
\1$val \* nil+val (d_{191}:o)=val \* x$	\\
\1\1$0+val (d_{191}:o)=val \* x$	\\
\1\1\1$0+dup \* val \* d_{191}=val \* x$	\\
\1\1\1\1$dup \* val \* d_{191}=val \* x$	\\
\1\1\1\1$dup \* val \* d_{191}=val \* x$	\> $\la [x:=d_{191}:o]$	\\
\1\1\1$0+dup \* val \* d_{191}=val \* x$	\> $\la [x:=d_{191}:o]$	\\
\1\1$0+val (d_{191}:o)=val \* x$	\> $\la [x:=d_{191}:o]$	\\
\1$val \* nil+val (d_{191}:o)=val \* x$	\> $\la [x:=d_{191}:o]$	\\
\1$val \* nil+val (d_{192}:i)=val \* x$	\\
\1\1$0+val (d_{192}:i)=val \* x$	\\
\1\1\1$0+s \* dup \* val \* d_{192}=val \* x$	\\
\1\1\1\1$s (0+dup \* val \* d_{192})=val \* x$	\\
\1\1\1\1\1$s \* dup \* val \* d_{192}=val \* x$	\\
\1\1\1\1\1$s \* dup \* val \* d_{192}=val \* x$	\> $\la [x:=d_{192}:i]$	\\
\1\1\1\1$s (0+dup \* val \* d_{192})=val \* x$	\> $\la [x:=d_{192}:i]$	\\
\1\1\1$0+s \* dup \* val \* d_{192}=val \* x$	\> $\la [x:=d_{192}:i]$	\\
\1\1$0+val (d_{192}:i)=val \* x$	\> $\la [x:=d_{192}:i]$	\\
\1$val \* nil+val (d_{192}:i)=val \* x$	\> $\la [x:=d_{192}:i]$	\\
\1$val (c_{189}:o)+val \* nil=val \* x$	\\
\1\1$val (c_{189}:o)+0=val \* x$	\\
\1\1\1$dup \* val \* c_{189}+0=val \* x$	\\
\1\1\1\1$dup \* val \* c_{189}=val \* x$	\\
\1\1\1\1$dup \* val \* c_{189}=val \* x$	\> $\la [x:=c_{189}:o]$	\\
\1\1\1$dup \* val \* c_{189}+0=val \* x$	\> $\la [x:=c_{189}:o]$	\\
\1\1$val (c_{189}:o)+0=val \* x$	\> $\la [x:=c_{189}:o]$	\\
\1$val (c_{189}:o)+val \* nil=val \* x$	\> $\la [x:=c_{189}:o]$	\\
\1$val (c_{189}:o)+val (d_{191}:o)=val \* x$	\\
\1\1$dup \* val \* c_{189}+val (d_{191}:o)=val \* x$	\\
\1\1\1$dup \* val \* c_{189}+dup \* val \* d_{191}=val \* x$	\\
\1\1\1\1$dup (val \* c_{189}+val \* d_{191})=val \* x$	\\
\1\1\1\1\1$dup \* val \* f_{188}(c_{189},d_{191})=val \* x$	\\
\1\1\1\1\1$dup \* val \* f_{188}(c_{189},d_{191})=val \* x$	\> $\la [x:=f_{188}(c_{189},d_{191}):o]$	\\
\1\1\1\1$dup (val \* c_{189}+val \* d_{191})=val \* x$	\> $\la [x:=f_{188}(c_{189},d_{191}):o]$	\\
\1\1\1$dup \* val \* c_{189}+dup \* val \* d_{191}=val \* x$	\> $\la [x:=f_{188}(c_{189},d_{191}):o]$	\\
\1\1$dup \* val \* c_{189}+val (d_{191}:o)=val \* x$	\> $\la [x:=f_{188}(c_{189},d_{191}):o]$	\\
\1$val (c_{189}:o)+val (d_{191}:o)=val \* x$	\> $\la [x:=f_{188}(c_{189},d_{191}):o]$	\\
\1$val (c_{189}:o)+val (d_{192}:i)=val \* x$	\\
\1\1$dup \* val \* c_{189}+val (d_{192}:i)=val \* x$	\\
\1\1\1$dup \* val \* c_{189}+s \* dup \* val \* d_{192}=val \* x$	\\
\1\1\1\1$s (dup \* val \* c_{189}+dup \* val \* d_{192})=val \* x$	\\
\1\1\1\1\1$s \* dup (val \* c_{189}+val \* d_{192})=val \* x$	\\
\1\1\1\1\1\1$s \* dup \* val \* f_{188}(c_{189},d_{192})=val \* x$	\\
\1\1\1\1\1\1$s \* dup \* val \* f_{188}(c_{189},d_{192})=val \* x$	\> $\la [x:=f_{188}(c_{189},d_{192}):i]$	\\
\1\1\1\1\1$s \* dup (val \* c_{189}+val \* d_{192})=val \* x$	\> $\la [x:=f_{188}(c_{189},d_{192}):i]$	\\
\1\1\1\1$s (dup \* val \* c_{189}+dup \* val \* d_{192})=val \* x$	\> $\la [x:=f_{188}(c_{189},d_{192}):i]$	\\
\1\1\1$dup \* val \* c_{189}+s \* dup \* val \* d_{192}=val \* x$	\> $\la [x:=f_{188}(c_{189},d_{192}):i]$	\\
\1\1$dup \* val \* c_{189}+val (d_{192}:i)=val \* x$	\> $\la [x:=f_{188}(c_{189},d_{192}):i]$	\\
\1$val (c_{189}:o)+val (d_{192}:i)=val \* x$	\> $\la [x:=f_{188}(c_{189},d_{192}):i]$	\\
\1$val (c_{190}:i)+val \* nil=val \* x$	\\
\1\1$val (c_{190}:i)+0=val \* x$	\\
\1\1\1$s \* dup \* val \* c_{190}+0=val \* x$	\\
\1\1\1\1$s \* dup \* val \* c_{190}=val \* x$	\\
\1\1\1\1$s \* dup \* val \* c_{190}=val \* x$	\> $\la [x:=c_{190}:i]$	\\
\1\1\1$s \* dup \* val \* c_{190}+0=val \* x$	\> $\la [x:=c_{190}:i]$	\\
\1\1$val (c_{190}:i)+0=val \* x$	\> $\la [x:=c_{190}:i]$	\\
\1$val (c_{190}:i)+val \* nil=val \* x$	\> $\la [x:=c_{190}:i]$	\\
\1$val (c_{190}:i)+val (d_{191}:o)=val \* x$	\\
\1\1$val (c_{190}:i)+dup \* val \* d_{191}=val \* x$	\\
\1\1\1$s \* dup \* val \* c_{190}+dup \* val \* d_{191}=val \* x$	\\
\1\1\1\1$s (dup \* val \* c_{190}+dup \* val \* d_{191})=val \* x$	\\
\1\1\1\1\1$s \* dup (val \* c_{190}+val \* d_{191})=val \* x$	\\
\1\1\1\1\1\1$s \* dup \* val \* f_{188}(c_{190},d_{191})=val \* x$	\\
\1\1\1\1\1\1$s \* dup \* val \* f_{188}(c_{190},d_{191})=val \* x$	\> $\la [x:=f_{188}(c_{190},d_{191}):i]$	\\
\1\1\1\1\1$s \* dup (val \* c_{190}+val \* d_{191})=val \* x$	\> $\la [x:=f_{188}(c_{190},d_{191}):i]$	\\
\1\1\1\1$s (dup \* val \* c_{190}+dup \* val \* d_{191})=val \* x$	\> $\la [x:=f_{188}(c_{190},d_{191}):i]$	\\
\1\1\1$s \* dup \* val \* c_{190}+dup \* val \* d_{191}=val \* x$	\> $\la [x:=f_{188}(c_{190},d_{191}):i]$	\\
\1\1$val (c_{190}:i)+dup \* val \* d_{191}=val \* x$	\> $\la [x:=f_{188}(c_{190},d_{191}):i]$	\\
\1$val (c_{190}:i)+val (d_{191}:o)=val \* x$	\> $\la [x:=f_{188}(c_{190},d_{191}):i]$	\\
\1$val (c_{190}:i)+val (d_{192}:i)=val \* x$	\\
\1\1$s \* dup \* val \* c_{190}+val (d_{192}:i)=val \* x$	\\
\1\1\1$s \* dup \* val \* c_{190}+s \* dup \* val \* d_{192}=val \* x$	\\
\1\1\1\1$s (s \* dup \* val \* c_{190}+dup \* val \* d_{192})=val \* x$	\\
\1\1\1\1\1$s \* s (dup \* val \* c_{190}+dup \* val \* d_{192})=val \* x$	\\
\1\1\1\1\1\1$s \* s \* dup (val \* c_{190}+val \* d_{192})=val \* x$	\\
\1\1\1\1\1\1\1$s \* s \* dup \* val \* f_{188}(c_{190},d_{192})=val \* x$	\\
\1\1\1\1\1\1\1\1$dup \* val \* x_{248}=s \* s \* dup \* val \* f_{188}(c_{190},d_{192})$	\\
\1\1\1\1\1\1\1\1\1$val \* f_{24}(f_{188}(c_{190},d_{192}))=val \* x_{248}$	\\
\1\1\1\1\1\1\1\1\1$val \* f_{24}(f_{188}(c_{190},d_{192}))=val \* x_{248}$	\> $\la [x_{248}:=f_{24}(f_{188}(c_{190},d_{192}))]$	\\
\1\1\1\1\1\1\1\1$dup \* val \* x_{248}=s \* s \* dup \* val \* f_{188}(c_{190},d_{192})$	\> $\la [x_{248}:=f_{24}(f_{188}(c_{190},d_{192}))]$	\\
\1\1\1\1\1\1\1$s \* s \* dup \* val \* f_{188}(c_{190},d_{192})=val \* x$	\> $\la [x:=f_{24}(f_{188}(c_{190},d_{192})):o]$	\\
\1\1\1\1\1\1$s \* s \* dup (val \* c_{190}+val \* d_{192})=val \* x$	\> $\la [x:=f_{24}(f_{188}(c_{190},d_{192})):o]$	\\
\1\1\1\1\1$s \* s (dup \* val \* c_{190}+dup \* val \* d_{192})=val \* x$	\> $\la [x:=f_{24}(f_{188}(c_{190},d_{192})):o]$	\\
\1\1\1\1$s (s \* dup \* val \* c_{190}+dup \* val \* d_{192})=val \* x$	\> $\la [x:=f_{24}(f_{188}(c_{190},d_{192})):o]$	\\
\1\1\1$s \* dup \* val \* c_{190}+s \* dup \* val \* d_{192}=val \* x$	\> $\la [x:=f_{24}(f_{188}(c_{190},d_{192})):o]$	\\
\1\1$s \* dup \* val \* c_{190}+val (d_{192}:i)=val \* x$	\> $\la [x:=f_{24}(f_{188}(c_{190},d_{192})):o]$	\\
\1$val (c_{190}:i)+val (d_{192}:i)=val \* x$	\> $\la [x:=f_{24}(f_{188}(c_{190},d_{192})):o]$	\\
$val \* c+val \* d=val \* x$	\> $\la [x:=f_{188}(c,d)]$	\\
\\
$S = [x:=f_{188}(c,d)] $	\\
\\
?- $solve(a+b+b = a+dup(b),S).$	\\
$S = [ \; ] $	\\
\\
?- $solve(a*dup(b) = dup(a*b),S).$	\\
$S = [ \; ] $	\\
\\
?- $solve(dup(val(c))+val(d) = val(f_{188}(c:o,d)),S).$	\\
$S = [ \; ] $	\\
\\
?- $solve(val(c)*val(d) = val(x),S).$	\\
$(val \* c)* (val \* d)=val \* x$	\\
\\
$1$	$[0= (val \* c)* (val \* d)]	\la [x:=nil]$	\\
$2$	$[dup \* val \* x_{412}= (val \* c)* (val \* d)]	\la [x:=x_{412}:o]$	\\
$3$	$[s \* dup \* val \* x_{419}= (val \* c)* (val \* d)]	\la [x:=x_{419}:i]$	\\
$4$	$[0=val \* x,0=val \* d]	\la [u_{420}:=val \* c]$	\\
$5$	$[u_{427}*v_{428}+u_{427}=val \* x,s \* v_{428}=val \* d]	\la [u_{427}:=val \* c]$	\> ind[d].	\\
\\
\1$(val \* c)* (val \* nil)=val \* x$	\\
\1\1$(val \* c)*0=val \* x$	\\
\1\1\1$0=val \* x$	\\
\1\1\1$0=val \* x$	\> $\la [x:=nil]$	\\
\1\1$(val \* c)*0=val \* x$	\> $\la [x:=nil]$	\\
\1$(val \* c)* (val \* nil)=val \* x$	\> $\la [x:=nil]$	\\
\1$(val \* c)* (val (d_{430}:o))=val \* x$	\\
\1\1$(val \* c)* (dup \* val \* d_{430})=val \* x$	\\
\1\1\1$dup (val \* c)* (val \* d_{430})=val \* x$	\\
\1\1\1\1$dup \* val \* f_{429}(c,d_{430})=val \* x$	\\
\1\1\1\1$dup \* val \* f_{429}(c,d_{430})=val \* x$	\> $\la [x:=f_{429}(c,d_{430}):o]$	\\
\1\1\1$dup (val \* c)* (val \* d_{430})=val \* x$	\> $\la [x:=f_{429}(c,d_{430}):o]$	\\
\1\1$(val \* c)* (dup \* val \* d_{430})=val \* x$	\> $\la [x:=f_{429}(c,d_{430}):o]$	\\
\1$(val \* c)* (val (d_{430}:o))=val \* x$	\> $\la [x:=f_{429}(c,d_{430}):o]$	\\
\1$(val \* c)* (val (d_{431}:i))=val \* x$	\\
\1\1$(val \* c)* (s \* dup \* val \* d_{431})=val \* x$	\\
\1\1\1$(val \* c)* (dup \* val \* d_{431})+val \* c=val \* x$	\\
\1\1\1\1$dup (val \* c)* (val \* d_{431})+val \* c=val \* x$	\\
\1\1\1\1\1$dup \* val \* f_{429}(c,d_{431})+val \* c=val \* x$	\\
\1\1\1\1\1\1$val \* f_{188}(f_{429}(c,d_{431}):o,c)=val \* x$	\\
\1\1\1\1\1\1$val \* f_{188}(f_{429}(c,d_{431}):o,c)=val \* x$	\> $\la [x:=f_{188}(f_{429}(c,d_{431}):o,c)]$	\\
\1\1\1\1\1$dup \* val \* f_{429}(c,d_{431})+val \* c=val \* x$	\> $\la [x:=f_{188}(f_{429}(c,d_{431}):o,c)]$	\\
\1\1\1\1$dup (val \* c)* (val \* d_{431})+val \* c=val \* x$	\> $\la [x:=f_{188}(f_{429}(c,d_{431}):o,c)]$	\\
\1\1\1$(val \* c)* (dup \* val \* d_{431})+val \* c=val \* x$	\> $\la [x:=f_{188}(f_{429}(c,d_{431}):o,c)]$	\\
\1\1$(val \* c)* (s \* dup \* val \* d_{431})=val \* x$	\> $\la [x:=f_{188}(f_{429}(c,d_{431}):o,c)]$	\\
\1$(val \* c)* (val (d_{431}:i))=val \* x$	\> $\la [x:=f_{188}(f_{429}(c,d_{431}):o,c)]$	\\
$(val \* c)* (val \* d)=val \* x$	\> $\la [x:=f_{429}(c,d)]$	\\
\\
$S = [x:=f_{429}(c,d)] $	\\
\\
?- $eval\_term(f_{429}(nil:i:o:i,nil:i:i:o),T).$	\\
% $eval(f_{429}(nil:i:o:i,nil:i:i:o)) b$	\\
% \1$eval(f_{429}(nil:i:o:i,nil:i:i))$	\\
% \1\1$eval(f_{188}(f_{429}(nil:i:o:i,nil:i):o,nil:i:o:i))$	\\
% \1\1\1$eval(f_{429}(nil:i:o:i,nil:i))$	\\
% \1\1\1\1$eval(f_{188}(f_{429}(nil:i:o:i,nil):o,nil:i:o:i))$	\\
% \1\1\1\1\1$eval(f_{429}(nil:i:o:i,nil))$	\\
% \1\1\1\1\1$eval(f_{429}(nil:i:o:i,nil))=nil$	\\
% \1\1\1\1\1$eval(f_{188}(nil,nil:i:o))$	\\
% \1\1\1\1\1$eval(f_{188}(nil,nil:i:o))=nil:i:o$	\\
% \1\1\1\1$eval(f_{188}(f_{429}(nil:i:o:i,nil):o,nil:i:o:i))=nil:i:o:i$	\\
% \1\1\1$eval(f_{429}(nil:i:o:i,nil:i))=nil:i:o:i$	\\
% \1\1\1$eval(f_{188}(nil:i:o:i,nil:i:o))$	\\
% \1\1\1\1$eval(f_{188}(nil:i:o,nil:i))$	\\
% \1\1\1\1\1$eval(f_{188}(nil:i,nil))$	\\
% \1\1\1\1\1$eval(f_{188}(nil:i,nil))=nil:i$	\\
% \1\1\1\1$eval(f_{188}(nil:i:o,nil:i))=nil:i:i$	\\
% \1\1\1$eval(f_{188}(nil:i:o:i,nil:i:o))=nil:i:i:i$	\\
% \1\1$eval(f_{188}(f_{429}(nil:i:o:i,nil:i):o,nil:i:o:i))=nil:i:i:i:i$	\\
% \1$eval(f_{429}(nil:i:o:i,nil:i:i))=nil:i:i:i:i$	\\
% $eval(f_{429}(nil:i:o:i,nil:i:i:o))=nil:i:i:i:i:o$	\\
% \\
$T = nil:i:i:i:i:o $	\\
\\
?- $listing(\sortdef), \; listing({\bf law}), \; listing({\bf def}).$\\
\\
$nat \sortdef 0 \mid s \* nat.$	\\
$bin \sortdef nil \mid bin:o \mid bin:i.$	\\
$sort_{1} \sortdef 0 \mid s \* s \* sort_{1}.$	\\
\\
law $s \* val \* v_{43}=val \* f_{24}(v_{43}).$	\\
law $0+v_{50}=v_{50}.$	\\
law $s \* v_{87}+v_{88}=s (v_{87}+v_{88}).$	\\
law $dup \* v_{117}+dup \* v_{118}=dup (v_{117}+v_{118}).$	\\
law $val \* v_{254}+val \* v_{255}=val \* f_{188}(v_{254},v_{255}).$	\\
law $v_{340}+v_{341}+v_{341}=v_{340}+dup \* v_{341}.$	\\
law $v_{356}* (dup \* v_{357})=dup \* v_{356}*v_{357}.$	\\
law $dup \* val \* v_{397}+val \* v_{398}=val \* f_{188}(v_{397}:o,v_{398}).$	\\
law $(val \* v_{446})* (val \* v_{447})=val \* f_{429}(v_{446},v_{447}).$	\\
\\
def $val \* nil=0.$	\\
def $val (x:o)=dup \* val \* x.$	\\
def $val (x:i)=s \* dup \* val \* x.$	\\
def $dup \* 0=0.$	\\
def $dup \* s \* u=s \* s \* dup \* u.$	\\
def $u+0=u.$	\\
def $u+s \* v=s (u+v).$	\\
def $u*0=0.$	\\
def $u* (s \* v)=u*v+u.$	\\
def $f_{24}(nil)=nil:i.$	\\
def $f_{24}(v_{35}:o)=v_{35}:i.$	\\
def $f_{24}(v_{42}:i)=f_{24}(v_{42}):o.$	\\
def $f_{188}(nil,nil)=nil.$	\\
def $f_{188}(nil,v_{201}:o)=v_{201}:o.$	\\
def $f_{188}(nil,v_{207}:i)=v_{207}:i.$	\\
def $f_{188}(v_{215}:o,nil)=v_{215}:o.$	\\
def $f_{188}(v_{224}:o,v_{225}:o)=f_{188}(v_{224},v_{225}):o.$	\\
def $f_{188}(v_{231}:o,v_{232}:i)=f_{188}(v_{231},v_{232}):i.$	\\
def $f_{188}(v_{238}:i,nil)=v_{238}:i.$	\\
def $f_{188}(v_{244}:i,v_{245}:o)=f_{188}(v_{244},v_{245}):i.$	\\
def $f_{188}(v_{252}:i,v_{253}:i)=f_{24}(f_{188}(v_{252},v_{253})):o.$	\\
def $f_{429}(v_{433},nil)=nil.$	\\
def $f_{429}(v_{442},v_{443}:o)=f_{429}(v_{442},v_{443}):o.$	\\
def $f_{429}(v_{444},v_{445}:i)=f_{188}(f_{429}(v_{444},v_{445}):o,v_{444}).$	\\

\\
\end{tabbing}

}

\newpage

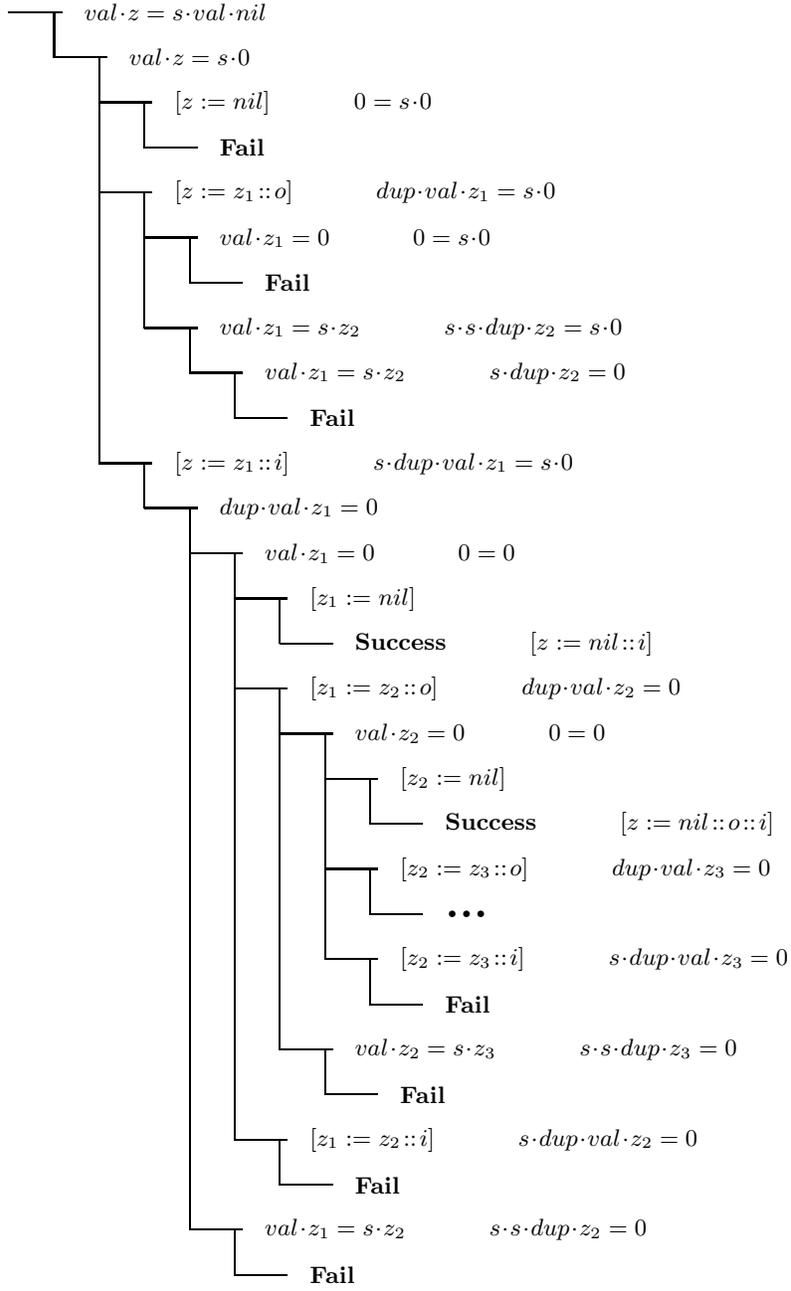
\begin{figure}
\begin{center}
\begin{picture}(10.3,17.3)
% {Fall 1 ohne Sorten
% !2}textree hg 0.6 fs 0.6 lt 0.3 ll 0.1 | tc yo 7.3

% \tree{$val\*z = s\*val\*nil$}
%   {\tree{$val\*z = s\*0$}
%     {\tree{$[z:=nil]$ \tab $0 = s\*0$}
%	{\tree{{\bf Fail}}
%	}
%     }
%     {\tree{$[z:=z_1\.{::}o]$ \tab $dup\*val\*z_1 = s\*0$}
%	{\tree{$val\*z_1=0$ \tab $0 = s\*0$}
%	  {\tree{{\bf Fail}}
%	  }
%	}
%	{\tree{$val\*z_1=s\*z_2$ \tab $s\*s\*dup\*z_2 = s\*0$}
%	  {\tree{$val\*z_1=s\*z_2$ \tab $s\*dup\*z_2 = 0$}
%	    {\tree{{\bf Fail}}
%	    }
%	  }
%	}
%     }
%     {\tree{$[z:=z_1\.{::}i]$ \tab $s\*dup\*val\*z_1 = s\*0$}
%	{\tree{$dup\*val\*z_1 = 0$}
%	  {\tree{$val\*z_1 = 0$ \tab $0=0$}
%	    {\tree{$[z_1 := nil]$}
%	      {\tree{{\bf Success} \tab $[z := nil\.{::}i]$}
%	      }
%	    }
%	    {\tree{$[z_1:=z_2\.{::}o]$ \tab $dup\*val\*z_2 = 0$}
%	      {\tree{$val\*z_2 = 0$ \tab $0 = 0$}
%		{\tree{$[z_2 := nil]$}
%		  {\tree{{\bf Success} \tab $[z := nil\.{::}o\.{::}i]$}
%		  }
%		}
%		{\tree{$[z_2:=z_3\.{::}o]$ \tab $dup\*val\*z_3 = 0$}
%		  {\tree{{\LARGE\bf  ... }}
%		  }
%		}
%		{\tree{$[z_2:=z_3\.{::}i]$ \tab $s\*dup\*val\*z_3 = 0$}
%		  {\tree{{\bf Fail}}
%		  }
%		}
%	      }
%	      {\tree{$val\*z_2=s\*z_3$ \tab $s\*s\*dup\*z_3 = 0$}
%		{\tree{{\bf Fail}}
%		}
%	      }
%	    }
%	    {\tree{$[z_1:=z_2\.{::}i]$ \tab $s\*dup\*val\*z_2 = 0$}
%	      {\tree{{\bf Fail}}
%	      }
%	    }
%	  }
%	  {\tree{$val\*z_1=s\*z_2$ \tab $s\*s\*dup\*z_2 = 0$}
%	    {\tree{{\bf Fail}}
%	    }
%	  }
%	}
%     }
%   }

\ % text after tree ignored
\put(0.000,17.300){\line(1,0){0.700}}
\put(1.000,17.300){\makebox(0.000,0.000)[l]{$val\*z = s\*val\*nil$}}
\put(0.600,16.700){\line(1,0){0.700}}
\put(1.600,16.700){\makebox(0.000,0.000)[l]{$val\*z = s\*0$}}
\put(1.200,16.100){\line(1,0){0.700}}
\put(2.200,16.100){\makebox(0.000,0.000)[l]{$[z:=nil]$ \tab $0 = s\*0$}}
\put(1.800,15.500){\line(1,0){0.700}}
\put(2.800,15.500){\makebox(0.000,0.000)[l]{{\bf Fail}}}
\put(1.800,16.100){\line(0,-1){0.600}}
\put(1.200,14.900){\line(1,0){0.700}}
\put(2.200,14.900){\makebox(0.000,0.000)[l]{$[z:=z_1\.{::}o]$ \tab $dup\*val\*z_1 = s\*0$}}
\put(1.800,14.300){\line(1,0){0.700}}
\put(2.800,14.300){\makebox(0.000,0.000)[l]{$val\*z_1=0$ \tab $0 = s\*0$}}
\put(2.400,13.700){\line(1,0){0.700}}
\put(3.400,13.700){\makebox(0.000,0.000)[l]{{\bf Fail}}}
\put(2.400,14.300){\line(0,-1){0.600}}
\put(1.800,13.100){\line(1,0){0.700}}
\put(2.800,13.100){\makebox(0.000,0.000)[l]{$val\*z_1=s\*z_2$ \tab $s\*s\*dup\*z_2 = s\*0$}}
\put(2.400,12.500){\line(1,0){0.700}}
\put(3.400,12.500){\makebox(0.000,0.000)[l]{$val\*z_1=s\*z_2$ \tab $s\*dup\*z_2
= 0$}}
\put(3.000,11.900){\line(1,0){0.700}}
\put(4.000,11.900){\makebox(0.000,0.000)[l]{{\bf Fail}}}
\put(3.000,12.500){\line(0,-1){0.600}}
\put(2.400,13.100){\line(0,-1){0.600}}
\put(1.800,14.900){\line(0,-1){1.800}}
\put(1.200,11.300){\line(1,0){0.700}}
\put(2.200,11.300){\makebox(0.000,0.000)[l]{$[z:=z_1\.{::}i]$ \tab $s\*dup\*val\*z_1 = s\*0$}}
\put(1.800,10.700){\line(1,0){0.700}}
\put(2.800,10.700){\makebox(0.000,0.000)[l]{$dup\*val\*z_1 = 0$}}
\put(2.400,10.100){\line(1,0){0.700}}
\put(3.400,10.100){\makebox(0.000,0.000)[l]{$val\*z_1 = 0$ \tab $0=0$}}
\put(3.000,9.500){\line(1,0){0.700}}
\put(4.000,9.500){\makebox(0.000,0.000)[l]{$[z_1 := nil]$}}
\put(3.600,8.900){\line(1,0){0.700}}
\put(4.600,8.900){\makebox(0.000,0.000)[l]{{\bf Success} \tab $[z := nil\.{::}i]$}}
\put(3.600,9.500){\line(0,-1){0.600}}
\put(3.000,8.300){\line(1,0){0.700}}
\put(4.000,8.300){\makebox(0.000,0.000)[l]{$[z_1:=z_2\.{::}o]$ \tab $dup\*val\*z_2 = 0$}}
\put(3.600,7.700){\line(1,0){0.700}}
\put(4.600,7.700){\makebox(0.000,0.000)[l]{$val\*z_2 = 0$ \tab $0 = 0$}}
\put(4.200,7.100){\line(1,0){0.700}}
\put(5.200,7.100){\makebox(0.000,0.000)[l]{$[z_2 := nil]$}}
\put(4.800,6.500){\line(1,0){0.700}}
\put(5.800,6.500){\makebox(0.000,0.000)[l]{{\bf Success} \tab $[z := nil\.{::}o\.{::}i]$}}
\put(4.800,7.100){\line(0,-1){0.600}}
\put(4.200,5.900){\line(1,0){0.700}}
\put(5.200,5.900){\makebox(0.000,0.000)[l]{$[z_2:=z_3\.{::}o]$ \tab $dup\*val\*z_3 = 0$}}
\put(4.800,5.300){\line(1,0){0.700}}
\put(5.800,5.300){\makebox(0.000,0.000)[l]{{\LARGE\bf  ... }}}
\put(4.800,5.900){\line(0,-1){0.600}}
\put(4.200,4.700){\line(1,0){0.700}}
\put(5.200,4.700){\makebox(0.000,0.000)[l]{$[z_2:=z_3\.{::}i]$ \tab $s\*dup\*val\*z_3 = 0$}}
\put(4.800,4.100){\line(1,0){0.700}}
\put(5.800,4.100){\makebox(0.000,0.000)[l]{{\bf Fail}}}
\put(4.800,4.700){\line(0,-1){0.600}}
\put(4.200,7.700){\line(0,-1){3.000}}
\put(3.600,3.500){\line(1,0){0.700}}
\put(4.600,3.500){\makebox(0.000,0.000)[l]{$val\*z_2=s\*z_3$ \tab $s\*s\*dup\*z_3 = 0$}}
\put(4.200,2.900){\line(1,0){0.700}}
\put(5.200,2.900){\makebox(0.000,0.000)[l]{{\bf Fail}}}
\put(4.200,3.500){\line(0,-1){0.600}}
\put(3.600,8.300){\line(0,-1){4.800}}
\put(3.000,2.300){\line(1,0){0.700}}
\put(4.000,2.300){\makebox(0.000,0.000)[l]{$[z_1:=z_2\.{::}i]$ \tab $s\*dup\*val\*z_2 = 0$}}
\put(3.600,1.700){\line(1,0){0.700}}
\put(4.600,1.700){\makebox(0.000,0.000)[l]{{\bf Fail}}}
\put(3.600,2.300){\line(0,-1){0.600}}
\put(3.000,10.100){\line(0,-1){7.800}}
\put(2.400,1.100){\line(1,0){0.700}}
\put(3.400,1.100){\makebox(0.000,0.000)[l]{$val\*z_1=s\*z_2$ \tab $s\*s\*dup\*z_2 = 0$}}
\put(3.000,0.500){\line(1,0){0.700}}
\put(4.000,0.500){\makebox(0.000,0.000)[l]{{\bf Fail}}}
\put(3.000,1.100){\line(0,-1){0.600}}
\put(2.400,10.700){\line(0,-1){9.600}}
\put(1.800,11.300){\line(0,-1){0.600}}
\put(1.200,16.700){\line(0,-1){5.400}}
\put(0.600,17.300){\line(0,-1){0.600}}
\end{picture}
\end{center}

\caption{Brute-force search space for equation
	$val \* z = s \* val \* nil$}
\label{Brute Force Search Space for Equation val z = s val nil F}

\end{figure}

\begin{figure}
\begin{center}
\begin{picture}(12.8,10.1)
% {Fall 2 ohne Sorten
% !2}textree hg 0.6 fs 0.6 lt 0.3 ll 0.1 | tc yo 0.1

% \tree{$val\*z = s\*val(c_1\.{::}o)$}
%   {\tree{$val\*z = s\*dup\*val\*c_1$}
%     {\tree{$[z:=nil]$ \tab $0 = s\*dup\*val\*c_1$}
%	{\tree{{\bf Fail}}
%	}
%     }
%     {\tree{$[z:=z_1\.{::}o]$ \tab $dup\*val\*z_1 = s\*dup\*val\*c_1$}
%	{\tree{$val\*z_1=0$ \tab $0 = s\*dup\*val\*c_1$}
%	  {\tree{{\bf Fail}}
%	  }
%	}
%	{\tree{$val\*z_1=s\*z_2$\tab $s\*s\*dup\*z_2=s\*dup\*val\*c_1$}
%	  {\tree{$val\*z_1=s\*z_2$\tab $s\*dup\*z_2 = dup\*val\*c_1$}
%	    {\tree{$val\*z_1=s\*z_2$ \tab $val\*c_1=0$
%						  \tab $s\*dup\*z_2=0$}
%	      {\tree{{\bf Fail}}
%	      }
%	    }
%	    {\tree{$val\*z_1=s\*z_2$ \tab $val\*c_1=s\*z_3$
%				     \tab $s\*dup\*z_2=s\*s\*dup\*z_3$}
%	      {\tree{$val\*z_1=s\*z_2$ \tab $val\*c_1=s\*z_3$
%					 \tab $dup\*z_2 = s\*dup\*z_3$}
%		{\tree{{\LARGE\bf  ... }}
%		}
%	      }
%	    }
%	  }
%	}
%     }
%     {\tree{$[z:=z_1\.{::}i]$ \tab $s\*dup\*val\*z_1=s\*dup\*val\*c_1$}
%	{\tree{$[z_1 := c_1]$}
%	  {\tree{{\bf Success} \tab $[z := c_1\.{::}i]$}
%	  }
%	}
%     }
%   }

\ % text after tree ignored
\put(0.000,10.100){\line(1,0){0.700}}
\put(1.000,10.100){\makebox(0.000,0.000)[l]{$val\*z = s\*val(c_1\.{::}o)$}}
\put(0.600,9.500){\line(1,0){0.700}}
\put(1.600,9.500){\makebox(0.000,0.000)[l]{$val\*z = s\*dup\*val\*c_1$}}
\put(1.200,8.900){\line(1,0){0.700}}

\put(2.200,8.900){\makebox(0.000,0.000)[l]{$[z:=nil]$ \tab $0 = s\*dup\*val\*c_1$}}
\put(1.800,8.300){\line(1,0){0.700}}
\put(2.800,8.300){\makebox(0.000,0.000)[l]{{\bf Fail}}}
\put(1.800,8.900){\line(0,-1){0.600}}
\put(1.200,7.700){\line(1,0){0.700}}
\put(2.200,7.700){\makebox(0.000,0.000)[l]{$[z:=z_1\.{::}o]$ \tab $dup\*val\*z_1 = s\*dup\*val\*c_1$}}
\put(1.800,7.100){\line(1,0){0.700}}
\put(2.800,7.100){\makebox(0.000,0.000)[l]{$val\*z_1=0$ \tab $0 = s\*dup\*val\*c_1$}}
\put(2.400,6.500){\line(1,0){0.700}}
\put(3.400,6.500){\makebox(0.000,0.000)[l]{{\bf Fail}}}
\put(2.400,7.100){\line(0,-1){0.600}}
\put(1.800,5.900){\line(1,0){0.700}}
\put(2.800,5.900){\makebox(0.000,0.000)[l]{$val\*z_1=s\*z_2$\tab $s\*s\*dup\*z_2=s\*dup\*val\*c_1$}}
\put(2.400,5.300){\line(1,0){0.700}}
\put(3.400,5.300){\makebox(0.000,0.000)[l]{$val\*z_1=s\*z_2$\tab $s\*dup\*z_2 =
dup\*val\*c_1$}}
\put(3.000,4.700){\line(1,0){0.700}}
\put(4.000,4.700){\makebox(0.000,0.000)[l]{$val\*z_1=s\*z_2$ \tab $val\*c_1=0$
						 \tab $s\*dup\*z_2=0$}}
\put(3.600,4.100){\line(1,0){0.700}}
\put(4.600,4.100){\makebox(0.000,0.000)[l]{{\bf Fail}}}
\put(3.600,4.700){\line(0,-1){0.600}}
\put(3.000,3.500){\line(1,0){0.700}}
\put(4.000,3.500){\makebox(0.000,0.000)[l]{$val\*z_1=s\*z_2$ \tab $val\*c_1=s\*z_3$
				    \tab $s\*dup\*z_2=s\*s\*dup\*z_3$}}
\put(3.600,2.900){\line(1,0){0.700}}
\put(4.600,2.900){\makebox(0.000,0.000)[l]{$val\*z_1=s\*z_2$ \tab $val\*c_1=s\*z_3$
					\tab $dup\*z_2 = s\*dup\*z_3$}}
\put(4.200,2.300){\line(1,0){0.700}}
\put(5.200,2.300){\makebox(0.000,0.000)[l]{{\LARGE\bf ... }}}
\put(4.200,2.900){\line(0,-1){0.600}}
\put(3.600,3.500){\line(0,-1){0.600}}
\put(3.000,5.300){\line(0,-1){1.800}}
\put(2.400,5.900){\line(0,-1){0.600}}
\put(1.800,7.700){\line(0,-1){1.800}}
\put(1.200,1.700){\line(1,0){0.700}}
\put(2.200,1.700){\makebox(0.000,0.000)[l]{$[z:=z_1\.{::}i]$ \tab $s\*dup\*val\*z_1=s\*dup\*val\*c_1$}}
\put(1.800,1.100){\line(1,0){0.700}}
\put(2.800,1.100){\makebox(0.000,0.000)[l]{$[z_1 := c_1]$}}
\put(2.400,0.500){\line(1,0){0.700}}
\put(3.400,0.500){\makebox(0.000,0.000)[l]{{\bf Success} \tab $[z := c_1\.{::}i]$}}
\put(2.400,1.100){\line(0,-1){0.600}}
\put(2.400,1.100){\line(0,-1){0.600}}
\put(1.800,1.700){\line(0,-1){0.600}}
\put(1.200,9.500){\line(0,-1){7.800}}
\put(0.600,10.100){\line(0,-1){0.600}}

\end{picture}
\end{center}

\caption{Brute-force search space for equation
	$val \* z = s \* val(c_1::o)$}
\label{Brute Force Search Space for Equation val z = s val c1::o F}

\end{figure}

\begin{figure}
\begin{center}
\begin{picture}(14.9,13.1)
% {Fall 3 ohne Sorten
% !2}textree hg 0.6 fs 0.6 lt 0.3 ll 0.1 | tc yo 3.1

% \tree{$val\*z = s\*val(c_2\.{::}i)$}
%   {\tree{$val\*z = s\*s\*dup\*val\*c_2$}
%     {\tree{$[z:=nil]$ \tab $0 = s\*s\*dup\*val\*c_2$}
%	{\tree{{\bf Fail}}
%	}
%     }
%     {\tree{$[z:=z_1\.{::}o]$ \tab $dup\*val\*z_1=s\*s\*dup\*val\*c_2$}
%	{\tree{$val\*z_1=0$ \tab $0 = s\*s\*dup\*val\*c_2$}
%	  {\tree{{\bf Fail}}
%	  }
%	}
%	{\tree{$val\*z_1=s\*z_2$ \tab
%				  $s\*s\*dup\*z_2=s\*s\*dup\*val\*c_2$}
%	  {\tree{$val\*z_1 = s\*z_2$ \tab $[z_2:=val\*c_2]$}
%	    {\tree{$val\*z_1 = s\*val\*c_2$}
%	      {\tree{{\bf Induction} \tab $[z := incr(c_2)::o]$}
%	      }
%	    }
%	  }
%	}
%     }
%     {\tree{$[z:=z_1\.{::}i]$\tab
%				$s\*dup\*val\*z_1=s\*s\*dup\*val\*c_2$}
%	{\tree{$dup\*val\*z_1=s\*dup\*val\*c_2$}
%	  {\tree{$val\*z_1=0$ \tab $0 = s\*dup\*val\*c_2$}
%	    {\tree{{\bf Fail}}
%	    }
%	  }
%	  {\tree{$val\*z_1=s\*z_2$ \tab
%				     $s\*s\*dup\*z_2=s\*dup\*val\*c_2$}
%	    {\tree{$val\*z_1=s\*z_2$ \tab $s\*dup\*z_2=dup\*val\*c_2$}
%	      {\tree{$val\*z_1=s\*z_2$ \tab $val\*c_2=0$ \tab
%						       $s\*dup\*z_2=0$}
%		{\tree{{\bf Fail}}
%		}
%	      }
%	      {\tree{$val\*z_1=s\*z_2$ \tab
%				$val\*c_2=s\*z_3$ \tab
%				$s\*dup\*z_2 = s\*s\*dup\*z_3$}
%		{\tree{$val\*z_1=s\*z_2$ \tab
%				$val\*c_2=s\*z_3$ \tab
%				$dup\*z_2 = s\*dup\*z_3$}
%		  {\tree{{\LARGE\bf  ... }}
%		  }
%		}
%	      }
%	    }
%	  }
%	}
%     }
%   }

\ % text after tree ignored
\put(0.000,13.100){\line(1,0){0.700}}
\put(1.000,13.100){\makebox(0.000,0.000)[l]{$val\*z = s\*val(c_2\.{::}i)$}}
\put(0.600,12.500){\line(1,0){0.700}}
\put(1.600,12.500){\makebox(0.000,0.000)[l]{$val\*z = s\*s\*dup\*val\*c_2$}}
\put(1.200,11.900){\line(1,0){0.700}}
\put(2.200,11.900){\makebox(0.000,0.000)[l]{$[z:=nil]$ \tab $0 = s\*s\*dup\*val\*c_2$}}
\put(1.800,11.300){\line(1,0){0.700}}
\put(2.800,11.300){\makebox(0.000,0.000)[l]{{\bf Fail}}}
\put(1.800,11.900){\line(0,-1){0.600}}
\put(1.200,10.700){\line(1,0){0.700}}
\put(2.200,10.700){\makebox(0.000,0.000)[l]{$[z:=z_1\.{::}o]$ \tab $dup\*val\*z_1=s\*s\*dup\*val\*c_2$}}
\put(1.800,10.100){\line(1,0){0.700}}
\put(2.800,10.100){\makebox(0.000,0.000)[l]{$val\*z_1=0$ \tab $0 = s\*s\*dup\*val\*c_2$}}
\put(2.400,9.500){\line(1,0){0.700}}
\put(3.400,9.500){\makebox(0.000,0.000)[l]{{\bf Fail}}}
\put(2.400,10.100){\line(0,-1){0.600}}
\put(1.800,8.900){\line(1,0){0.700}}
\put(2.800,8.900){\makebox(0.000,0.000)[l]{$val\*z_1=s\*z_2$ \tab
				  $s\*s\*dup\*z_2=s\*s\*dup\*val\*c_2$}}
\put(2.400,8.300){\line(1,0){0.700}}
\put(3.400,8.300){\makebox(0.000,0.000)[l]{$val\*z_1 = s\*z_2$ \tab $[z_2:=val\*c_2]$}}
\put(3.000,7.700){\line(1,0){0.700}}
\put(4.000,7.700){\makebox(0.000,0.000)[l]{$val\*z_1 = s\*val\*c_2$}}
\put(3.600,7.100){\line(1,0){0.700}}
\put(4.600,7.100){\makebox(0.000,0.000)[l]{{\bf Induction} \tab $[z := incr(c_2)::o]$}}
\put(3.600,7.700){\line(0,-1){0.600}}
\put(3.000,8.300){\line(0,-1){0.600}}
\put(2.400,8.900){\line(0,-1){0.600}}
\put(1.800,10.700){\line(0,-1){1.800}}
\put(1.200,6.500){\line(1,0){0.700}}
\put(2.200,6.500){\makebox(0.000,0.000)[l]{$[z:=z_1\.{::}i]$\tab
				$s\*dup\*val\*z_1=s\*s\*dup\*val\*c_2$}}
\put(1.800,5.900){\line(1,0){0.700}}
\put(2.800,5.900){\makebox(0.000,0.000)[l]{$dup\*val\*z_1=s\*dup\*val\*c_2$}}
\put(2.400,5.300){\line(1,0){0.700}}
\put(3.400,5.300){\makebox(0.000,0.000)[l]{$val\*z_1=0$ \tab $0 = s\*dup\*val\*c_2$}}
\put(3.000,4.700){\line(1,0){0.700}}
\put(4.000,4.700){\makebox(0.000,0.000)[l]{{\bf Fail}}}
\put(3.000,5.300){\line(0,-1){0.600}}
\put(2.400,4.100){\line(1,0){0.700}}
\put(3.400,4.100){\makebox(0.000,0.000)[l]{$val\*z_1=s\*z_2$ \tab
				     $s\*s\*dup\*z_2=s\*dup\*val\*c_2$}}
\put(3.000,3.500){\line(1,0){0.700}}
\put(4.000,3.500){\makebox(0.000,0.000)[l]{$val\*z_1=s\*z_2$ \tab $s\*dup\*z_2=dup\*val\*c_2$}}
\put(3.600,2.900){\line(1,0){0.700}}
\put(4.600,2.900){\makebox(0.000,0.000)[l]{$val\*z_1=s\*z_2$ \tab $val\*c_2=0$ \tab
						       $s\*dup\*z_2=0$}}
\put(4.200,2.300){\line(1,0){0.700}}
\put(5.200,2.300){\makebox(0.000,0.000)[l]{{\bf Fail}}}
\put(4.200,2.900){\line(0,-1){0.600}}
\put(3.600,1.700){\line(1,0){0.700}}
\put(4.600,1.700){\makebox(0.000,0.000)[l]{$val\*z_1=s\*z_2$ \tab
				$val\*c_2=s\*z_3$ \tab
				$s\*dup\*z_2 = s\*s\*dup\*z_3$}}
\put(4.200,1.100){\line(1,0){0.700}}
\put(5.200,1.100){\makebox(0.000,0.000)[l]{$val\*z_1=s\*z_2$ \tab
				$val\*c_2=s\*z_3$ \tab
				$dup\*z_2 = s\*dup\*z_3$}}
\put(4.800,0.500){\line(1,0){0.700}}
\put(5.800,0.500){\makebox(0.000,0.000)[l]{{\LARGE\bf  ... }}}
\put(4.800,1.100){\line(0,-1){0.600}}
\put(4.200,1.700){\line(0,-1){0.600}}
\put(3.600,3.500){\line(0,-1){1.800}}
\put(3.000,4.100){\line(0,-1){0.600}}
\put(2.400,5.900){\line(0,-1){1.800}}
\put(1.800,6.500){\line(0,-1){0.600}}
\put(1.200,12.500){\line(0,-1){6.000}}
\put(0.600,13.100){\line(0,-1){0.600}}

\end{picture}
\end{center}

\caption{Brute-force search space for equation
	$val \* z = s \* val(c_2::i)$}
\label{Brute Force Search Space for Equation val z = s val c2::i F}

\end{figure}

\begin{figure}
\begin{center}
\begin{picture}(10.9,7.7)
% {Fall 1 mit Sorten
% !2}textree hg 0.6 fs 0.6 lt 0.3 ll 0.1 | tc yo -2.3

% \tree{$val\*z = s\*val\*nil$}
%   {\tree{$val\*z = s\*0$}
%     {\tree{$[z:=z_1\.{::}i]$ \tab $s\*dup\*val\*z_1 = s\*0$}
%	{\tree{$dup\*val\*z_1 = 0$}
%	  {\tree{$val\*z_1 = 0$}
%	    {\tree{$[z_1 := nil]$}
%	      {\tree{{\bf Success} \tab $[z := nil\.{::}i]$}
%	      }
%	    }
%	    {\tree{$[z_1:=z_2\.{::}o]$ \tab $dup\*val\*z_2 = 0$}
%	      {\tree{$val\*z_2 = 0$}
%		{\tree{$[z_2 := nil]$}
%		  {\tree{{\bf Success} \tab $[z := nil\.{::}o\.{::}i]$}
%		  }
%		}
%		{\tree{$[z_2:=z_3\.{::}o]$ \tab $dup\*val\*z_3 = 0$}
%		  {\tree{{\LARGE\bf \ldots }}
%		  }
%		}
%	      }
%	    }
%	  }
%	}
%     }
%   }

\ % text after tree ignored
\put(0.000,7.700){\line(1,0){0.700}}
\put(1.000,7.700){\makebox(0.000,0.000)[l]{$val\*z = s\*val\*nil$}}
\put(0.600,7.100){\line(1,0){0.700}}
\put(1.600,7.100){\makebox(0.000,0.000)[l]{$val\*z = s\*0$}}
\put(1.200,6.500){\line(1,0){0.700}}
\put(2.200,6.500){\makebox(0.000,0.000)[l]{$[z:=z_1\.{::}i]$ \tab $s\*dup\*val\*z_1 = s\*0$}}
\put(1.800,5.900){\line(1,0){0.700}}
\put(2.800,5.900){\makebox(0.000,0.000)[l]{$dup\*val\*z_1 = 0$}}
\put(2.400,5.300){\line(1,0){0.700}}
\put(3.400,5.300){\makebox(0.000,0.000)[l]{$val\*z_1 = 0$}}
\put(3.000,4.700){\line(1,0){0.700}}
\put(4.000,4.700){\makebox(0.000,0.000)[l]{$[z_1 := nil]$}}
\put(3.600,4.100){\line(1,0){0.700}}
\put(4.600,4.100){\makebox(0.000,0.000)[l]{{\bf Success} \tab $[z := nil\.{::}i]$}}
\put(3.600,4.700){\line(0,-1){0.600}}
\put(3.000,3.500){\line(1,0){0.700}}
\put(4.000,3.500){\makebox(0.000,0.000)[l]{$[z_1:=z_2\.{::}o]$ \tab $dup\*val\*z_2 = 0$}}
\put(3.600,2.900){\line(1,0){0.700}}
\put(4.600,2.900){\makebox(0.000,0.000)[l]{$val\*z_2 = 0$}}
\put(4.200,2.300){\line(1,0){0.700}}
\put(5.200,2.300){\makebox(0.000,0.000)[l]{$[z_2 := nil]$}}
\put(4.800,1.700){\line(1,0){0.700}}
\put(5.800,1.700){\makebox(0.000,0.000)[l]{{\bf Success} \tab $[z := nil\.{::}o\.{::}i]$}}
\put(4.800,2.300){\line(0,-1){0.600}}
\put(4.200,1.100){\line(1,0){0.700}}
\put(5.200,1.100){\makebox(0.000,0.000)[l]{$[z_2:=z_3\.{::}o]$ \tab $dup\*val\*z_3 = 0$}}
\put(4.800,0.500){\line(1,0){0.700}}
\put(5.800,0.500){\makebox(0.000,0.000)[l]{{\LARGE\bf \ldots }}}
\put(4.800,1.100){\line(0,-1){0.600}}
\put(4.200,2.900){\line(0,-1){1.800}}
\put(3.600,3.500){\line(0,-1){0.600}}
\put(3.000,5.300){\line(0,-1){1.800}}
\put(2.400,5.900){\line(0,-1){0.600}}
\put(1.800,6.500){\line(0,-1){0.600}}
\put(1.200,7.100){\line(0,-1){0.600}}
\put(0.600,7.700){\line(0,-1){0.600}}
\end{picture}
\vspace{0.5cm}

\begin{picture}(9.9,2.9)
% {Fall 2 mit Sorten
% !2}textree hg 0.6 fs 0.6 lt 0.3 ll 0.1 | tc yo -7.1

% \tree{$val\*z = s\*val(c_1\.{::}o)$}
%   {\tree{$val\*z = s\*dup\*val\*c_1$}
%     {\tree{$[z:=z_1\.{::}i]$ \tab $s\*dup\*val\*z_1=s\*dup\*val\*c_1$}
%	{\tree{$[z_1 := c_1]$}
%	  {\tree{{\bf Success} \tab $[z := c_1\.{::}i]$}
%	  }
%	}
%     }
%   }

\ % text after tree ignored
\put(0.000,2.900){\line(1,0){0.700}}
\put(1.000,2.900){\makebox(0.000,0.000)[l]{$val\*z = s\*val(c_1\.{::}o)$}}
\put(0.600,2.300){\line(1,0){0.700}}
\put(1.600,2.300){\makebox(0.000,0.000)[l]{$val\*z = s\*dup\*val\*c_1$}}
\put(1.200,1.700){\line(1,0){0.700}}
\put(2.200,1.700){\makebox(0.000,0.000)[l]{$[z:=z_1\.{::}i]$ \tab $s\*dup\*val\*z_1=s\*dup\*val\*c_1$}}
\put(1.800,1.100){\line(1,0){0.700}}
\put(2.800,1.100){\makebox(0.000,0.000)[l]{$[z_1 := c_1]$}}
\put(2.400,0.500){\line(1,0){0.700}}
\put(3.400,0.500){\makebox(0.000,0.000)[l]{{\bf Success} \tab $[z := c_1\.{::}i]$}}
\put(2.400,1.100){\line(0,-1){0.600}}
\put(1.800,1.700){\line(0,-1){0.600}}
\put(1.200,2.300){\line(0,-1){0.600}}
\put(0.600,2.900){\line(0,-1){0.600}}

\end{picture}
\vspace{0.5cm}

\begin{picture}(10.9,4.1)
% {Fall 3 mit Sorten
% !2}textree hg 0.6 fs 0.6 lt 0.3 ll 0.1 | tc yo -5.9

% \tree{$val\*z = s\*val(c_2\.{::}i)$}
%   {\tree{$val\*z = s\*s\*dup\*val\*c_2$}
%     {\tree{$[z:=z_1\.{::}o]$ \tab $dup\*val\*z_1=s\*s\*dup\*val\*c_2$}
%	{\tree{$val\*z_1=s\*z_2$ \tab $s\*s\*dup\*z_2=s\*s\*dup\*val\*c_2$}
%	  {\tree{$val\*z_1 = s\*z_2$ \tab $[z_2:=val\*c_2]$}
%	    {\tree{$val\*z_1 = s\*val\*c_2$}
%	      {\tree{{\bf Induction} \tab $[z := incr(c_2)::o]$}
%	      }
%	    }
%	  }
%	}
%     }
%   }

\ % text after tree ignored
\put(0.000,4.100){\line(1,0){0.700}}
\put(1.000,4.100){\makebox(0.000,0.000)[l]{$val\*z = s\*val(c_2\.{::}i)$}}
\put(0.600,3.500){\line(1,0){0.700}}
\put(1.600,3.500){\makebox(0.000,0.000)[l]{$val\*z = s\*s\*dup\*val\*c_2$}}
\put(1.200,2.900){\line(1,0){0.700}}
\put(2.200,2.900){\makebox(0.000,0.000)[l]{$[z:=z_1\.{::}o]$ \tab $dup\*val\*z_1=s\*s\*dup\*val\*c_2$}}
\put(1.800,2.300){\line(1,0){0.700}}
\put(2.800,2.300){\makebox(0.000,0.000)[l]{$val\*z_1=s\*z_2$ \tab $s\*s\*dup\*z_2=s\*s\*dup\*val\*c_2$}}
\put(2.400,1.700){\line(1,0){0.700}}
\put(3.400,1.700){\makebox(0.000,0.000)[l]{$val\*z_1 = s\*z_2$ \tab $[z_2:=val\*c_2]$}}
\put(3.000,1.100){\line(1,0){0.700}}
\put(4.000,1.100){\makebox(0.000,0.000)[l]{$val\*z_1 = s\*val\*c_2$}}
\put(3.600,0.500){\line(1,0){0.700}}
\put(4.600,0.500){\makebox(0.000,0.000)[l]{{\bf Induction} \tab $[z := incr(c_2)::o]$}}
\put(3.600,1.100){\line(0,-1){0.600}}
\put(3.000,1.700){\line(0,-1){0.600}}
\put(2.400,2.300){\line(0,-1){0.600}}
\put(1.800,2.900){\line(0,-1){0.600}}
\put(1.200,3.500){\line(0,-1){0.600}}
\put(0.600,4.100){\line(0,-1){0.600}}
\end{picture}
\end{center}

\caption{Search space of $incr$ synthesis using sorts}
\label{Search space of $incr$ synthesis using sorts}

\end{figure}
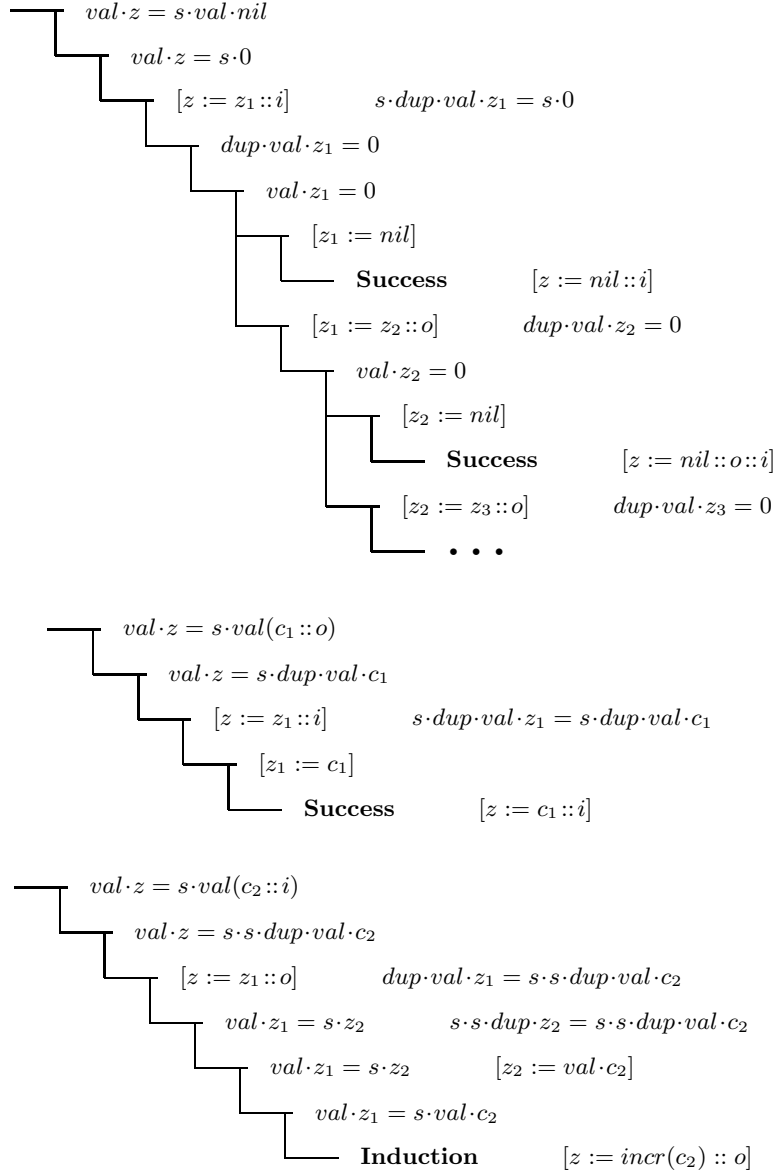

\clearpage

\renewcommand{\eqra}[1]{#1}
\renewcommand{\eqda}[1]{#1}

\section{Case Study ``Comb Vector Construction''}
\label{Case Study ``Comb Vector Construction''}

In this section, we demonstrate the use of the sort discipline by
applying it in a paper
case study from the area of compiler construction.
The parser generating system PGS is a tool for generating a syntax
analyzer for a programming language or, in general, any structured
input \cit{Klein 1989}{Klein.Martin.1989}.
The user of PGS has to specify the language to be analyzed by a grammar.
The main applications of PGS are
in the area of compiler construction,
e.g.\ parsing, syntax analysis or syntax-directed translation.

PGS uses a comb vector technique to compress the two-dimensional
array representation of parse tables. A parse table can be
merged into one array, called $cont$, where the beginning of
each original row is indicated by an entry in an additional
array called $base$. In order to be able to distinguish between
error and non-error entries, an array called $row$ is introduced
in parallel to $cont$, containing the row number from which the
associated entry in the $cont$ array originated. 

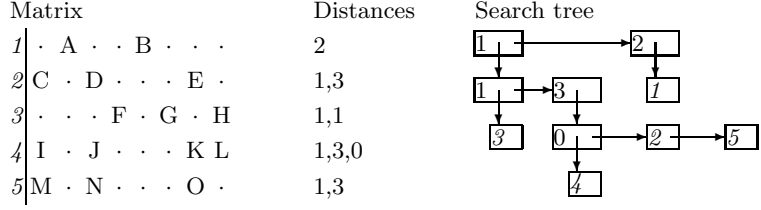
\begin{figure}
\begin{center}
\begin{tabular}{l|ccccccccp{0.9cm}lp{0.6cm}l}
\multicolumn{9}{l}{Matrix} & $\;$ & Distances &$\;$& Search tree \\
\em 1 & $\cdot$ & A	  & $\cdot$ & $\cdot$ & B	& $\cdot$ &
	$\cdot$ & $\cdot$ & $\;$    & 2
&& \begin{picture}(3.8,0.3)
\put(0.000,-0.025){\makebox(0,0)[bl]{\framebox(0.625,0.3125)[l]{1}}}
\put(2.083,-0.025){\makebox(0,0)[bl]{\framebox(0.625,0.3125)[l]{2}}}
\put(0.000,-0.650){\makebox(0,0)[bl]{\framebox(0.625,0.3125)[l]{1}}}
\put(1.042,-0.650){\makebox(0,0)[bl]{\framebox(0.625,0.3125)[l]{3}}}
\put(2.292,-0.650){\makebox(0,0)[bl]{\framebox(0.416,0.3125)[l]{\em 1}}}
\put(1.042,-1.275){\makebox(0,0)[bl]{\framebox(0.625,0.3125)[l]{0}}}
\put(0.208,-1.275){\makebox(0,0)[bl]{\framebox(0.416,0.3125)[l]{\em 3}}}
\put(2.292,-1.275){\makebox(0,0)[bl]{\framebox(0.416,0.3125)[l]{\em 2}}}
\put(3.333,-1.275){\makebox(0,0)[bl]{\framebox(0.416,0.3125)[l]{\em 5}}}
\put(1.250,-1.900){\makebox(0,0)[bl]{\framebox(0.416,0.3125)[l]{\em 4}}}
\put(0.312,0.132){\vector(0,-1){0.469}}
\put(0.312,-0.493){\vector(0,-1){0.469}}
\put(1.354,-0.493){\vector(0,-1){0.469}}
\put(1.354,-1.118){\vector(0,-1){0.469}}
\put(2.396,0.132){\vector(0,-1){0.469}}
\put(0.521,0.132){\vector(1,0){1.562}}
\put(0.521,-0.493){\vector(1,0){0.521}}
\put(1.562,-1.118){\vector(1,0){0.729}}
\put(2.604,-1.118){\vector(1,0){0.729}}
\end{picture}	\\
\em 2 & C	& $\cdot$ & D	    & $\cdot$ & $\cdot$ & $\cdot$ &
	E	& $\cdot$ & $\;$    & 1,3    \\
\em 3 & $\cdot$ & $\cdot$ & $\cdot$ & F	      & $\cdot$ & G	  &
	$\cdot$ & H	  & $\;$    &1,1     \\
\em 4 & I	& $\cdot$ & J	    & $\cdot$ & $\cdot$ & $\cdot$ &
	K	& L	  & $\;$    & 1,3,0  \\
\em 5 & M	& $\cdot$ & N	    & $\cdot$ & $\cdot$ & $\cdot$ &
	O	& $\cdot$ & $\;$    & 1,3    \\
\end{tabular}

\caption{Search-tree construction for comb vectors}
\label{Search-tree construction for comb vectors}
\end{center}

\end{figure}

Given a two-dimensional array, one way of constructing a
comb vector is to enter each row into a search tree,
lexicographically sorted by the list of its distances. The tree
is a son-brother tree, a vertical link pointing to the first son
of a node, a horizontal link to the next brother. There are two
kinds of nodes, depending on whether a vertical link is necessary
or not. A node which has a vertical link corresponds to a
distance; brother nodes of this kind are in ascending
order with respect to it. 
A node without a
vertical link corresponds to a row number (shown in italics in
Fig.~\ref{Search-tree construction for comb vectors}).

The tree is then traversed in post order, and the corresponding
rows are entered into the comb vector. Figure \ref{Search-tree
construction for comb vectors} shows an example two-dimensional
array together with the constructed search tree. For example, the
path $down$, $right$, $down$, $down$ corresponds to the distance list
$1,3,0$ of row {\em 4}. Figure \ref{Comb Vector and Access
Function} shows the constructed comb vector and its access
function. 

Our aim is to define the data structure of a search tree and to
construct an algorithm for inserting a list of distances into a
search tree. For the sake of simplicity, we do not distinguish
between distances and row numbers, representing both by natural
numbers. 
\vspace{0.3cm}

\noindent
\vbox{
Assume the constructors
	\display{
	\begin{tabular}[t]{@{}lll@{}}
	for search trees:
	& $nil_t$ & empty search tree,	\\
	& $node1(\cdot,\cdot,\cdot)$
		& node with vertical and horizontal link, \\
	& $node2(\cdot,\cdot)$ & node with horizontal link only, \\
	for distance lists:
	& $nil_l$ & empty list,	\\
	& $(\cdot) \m (\cdot)$ & list ``cons'',	\\
	for sets of distance lists:
	& $mt$ & empty set, and	\\
	& $add(\cdot,\cdot)$ & add an element	\\
	\end{tabular}
	}
}

\noindent
The search tree in Fig.\ \ref{Search-tree construction for comb
vectors} is represented by the term
\display{
\renewcommand{\arraystretch}{0.8}
$\begin{array}{@{}l@{}l@{}l@{}l@{}l@{}}
node1(1, & node1(1, & node2(3, & nil_t),			\\
	 &	    & node1(3, & node1(0, & node2(4,nil_t),	\\
	 &	    &	       &	  & node2(2,node2(5,nil_t))), \\
	 &	    &	       & nil_t)),			\\
	 & node1(2, & node2(1, & nil_t),			\\
	 &	    & nil_t)) \; ;				\\
\end{array}$
}

its set of distance lists can be represented by
\\
$add(1 \m 1 \m 3 \m nil_l,
add(1 \m 3 \m 0 \m 4 \m nil_l,
add(1 \m 3 \m 2 \m nil_l,
add(1 \m 3 \m 5 \m nil_l,
add(2 \m 1 \m nil_l,
mt)))))$.

\begin{figure}
\begin{center}
\begin{tabular}[t]{@{}l|*{18}{r}}
&  1 &	2 &  3 &  4 &  5 &  6 &	 7 &  8 &  9 & 10
& 11 & 12 & 13 & 14 & 15 & 16 & 17 & \ldots	\\
\hline
row	&  3 &	4 &  3 &  4 &  3 &  . &	 . &  4 &  4 &	2
&  5 &	2 &  5 &  . &  . &  2 &	 5	\\
cont	& F & I & G & J & H &  . &	 . & K & L & C
& M & D & N &  . &  . & E & O	\\
\end{tabular}
\\[0.5cm]

\begin{tabular}[t]{@{}l|*{5}{r|}@{}}
       &   1 &	2 &  3 &  4 &	5	\\
\hline
base & $+16$ & $+9$ & $-3$ & $+1$ & $+10$	\\
\end{tabular}
\\[0.5cm]

\begin{tabular}[t]{@{}ll@{$\;$}l@{}}
$get(i,j) =$ & \bf if & $row[base[i]+j] == i$ \\
	& \bf then & $cont[base[i]+j]$	\\
	& \bf else & $0$	\\
	& \bf fi	\\
\end{tabular}

\caption{Comb vector and access function}
\label{Comb Vector and Access Function}
\end{center}

\end{figure}

\noindent
To form a valid search tree, a term has to
satisfy the following conditions:
\bi
\item a vertical link may not be $nil_t$ (\eqRa{38},\eqRa{39}),
\item the horizontal link of a $node2$ never points to a $node1$
	(\eqRa{39}),
\item each horizontal chain of $node1$s is in ascending
	order (first line of \eqRa{38}).
\ei

\noindent
This leads to the sort definitions:
\display{
\begin{tabular}{@{}l@{$\;$}r@{$\;$}lr@{}}
$Tree$	& $\sortdef$ & $nil_t \mid Tree1 \mid Tree2$ & (\eqda{37}) \\
$Tree1$ & $\sortdef$ & $node1(n\.:Nat,t_1\.:Tree1 \.\mid Tree2,
	t_2\.:Tree1) \lhd n<t_2$ & (\eqda{38})	\\
	& $\mid$ & $node1(n\.:Nat,t_1\.:Tree1 \.\mid Tree2,
		t_2\.:Tree2 \.\mid nil_t)$	\\
$Tree2$ & $\sortdef$ & $node2(Nat,Tree2 \.\mid nil_t)$ & (\eqda{39}) \\
$List$	& $\sortdef$ & $nil_l \mid Nat \m List$ & (\eqda{40}) \\
$Set$	& $\sortdef$ & $mt \mid add(List,Set)$ & (\eqda{41}) \\
\end{tabular}
}

The definition of $Tree1$ makes use of a constraint predicate, cf.\ the
remarks at the end of Sect.~\ref{Regular Sorts}.
$Tree1$ and $Tree2$ denote the sort of all search trees starting
with a $node1$ and a $node2$, respectively.
The constraint predicate
is defined by the axiom $n_1 < node1(n_2,t_3,t_4) \lra n_1 < n_2$.

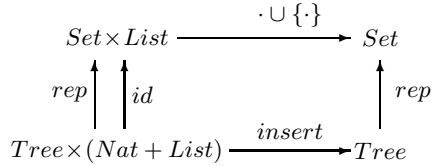
\begin{figure}
\begin{center}
\begin{picture}(4.8,2.2)
	%\put(0,0){\makebox(0,0){+}}
\put(1.000,0.200){\makebox(0,0){$Tree \.\times (Nat+List)$}}
\put(1.000,1.700){\makebox(0,0){$Set \.\times List$}}
\put(0.700,0.500){\vector(0,1){0.900}}
\put(1.100,0.500){\vector(0,1){0.900}}
\put(4.500,0.200){\makebox(0,0){$Tree$}}
\put(4.500,1.700){\makebox(0,0){$Set$}}
\put(4.500,0.500){\vector(0,1){0.900}}
\put(2.500,0.200){\vector(1,0){1.600}}
\put(3.300,0.350){\makebox(0,0)[b]{$insert$}}
\put(0.600,0.950){\makebox(0,0)[r]{$rep$}}
\put(1.200,0.950){\makebox(0,0)[l]{$id$}}
\put(4.700,0.950){\makebox(0,0)[l]{$rep$}}
\put(1.800,1.700){\vector(1,0){2.300}}
\put(3.300,1.850){\makebox(0,0)[b]{$\cdot \cup \{\cdot\}$}}
\end{picture}

\caption{Specification of the insert algorithm}
\label{Specification of the Insert Algorithm}
\end{center}
\end{figure}

\begin{figure}
\begin{center}
\begin{tabular}{@{}r@{$\;$}l@{\tab}rl@{}}
\mca{4}{$rep:Tree \ra Set$ 
	gives the set of distance lists represented by a (sub)tree:} \\
$rep(nil_t)$ & $= mt$ & (\eqda{42})	\\
$rep(node1(n,t_1,t_2))$
	& $= \union{n \ptm rep \* t_1}{rep \* t_2}$ & (\eqda{43}) \\
$rep(node2(n,t))$ & $= add(n \m nil_l,rep \* t)$ & (\eqda{44})	\\
[0.2cm]
\mca{4}{$\ptm: Nat \times Set \ra Set$
	pointwise prefixes a set of distance lists by a new distance:}\\
$n \ptm mt$ & $= mt$ & (\eqda{45})	\\
$n \ptm add(l,s)$ & $= add(n \m l,n \ptm s)$ & (\eqda{46}) \\
[0.2cm]
\mca{4}{$\unioN: Set \times Set \ra Set$
	is the ordinary set union:} \\
$\union{mt}{s}$ & $= s$ & (\eqda{47})	\\
$\union{add(l,s_1)}{s_2}$ & $= add(l,\union{s_1}{s_2})$ & (\eqda{48}) \\
[0.2cm]
\mca{4}{We have the following equations between constructors:} \\
$add(l_1,add(l_2,s))$ & $=add(l_2,add(l_1,s))$ & (\eqda{49})	\\
$add(l,add(l,s))$ & $=add(l,s)$ & (\eqda{50}) \\
[0.2cm]
\mca{4}{Finally, we need the following derived lemma:}	\\
$\union{s_1}{s_2}$ & $= \union{s_2}{s_1}$ & (\eqda{51})	\\
\end{tabular}

\caption{Auxiliary function definitions for search-tree specification}
\label{Auxiliary Function Definitions for Search Tree Specification}
\end{center}

\end{figure}

Using the terminology introduced in Sect.~\ref{Application in Formal
Program Development}, we have
$as_1 = Set \times (Nat+List)$,
$as_2 = Set$,
$ao(s,l) = add(l,s)$,
$cs_1 = Tree \times List$,
$cs_2 = Tree$,
$co(t,l) = insert(t,l)$ is to be synthesized,
$r_1(t,l) = \tpl{rep(t),l}$, and
$r_2(t) = rep(t)$.

The specification uses several auxiliary functions defined in
Fig.\ \ref{Auxiliary Function Definitions for Search Tree
Specification}.
Expressed in informal terms, it says:
``Given a tree $t$ and a non-empty distance list $l$, find a tree $T$
that 
contains the same distance lists as $t$ and additional $l$'';
and in formal terms:
$\forall t \in Tree^M,l \in (Nat \m List)^M \; \exists T \in Tree^M
\;\;\;
rep \* T = add(l,rep \* t)$.
The $insert$ function will be synthesized as Skolem function for $T$.

Using Alg.~\eqr{73},
we obtain the following range sorts of $rep$, cf.\ Fig.~\ref{Range sort
computation for $rep$}:
	\display{
	\begin{tabular}[t]{@{}l@{$\;$}lr@{}}
	$rg([x \.{:=} nil_t],rep \* x)$ & $= mt$	\\
	$rg([x \.{:=} Tree1],rep \* x)$ & $= Sort_{\eqRa{52}}$	\\
	$rg([x \.{:=} Tree2],rep \* x)$ & $= Sort_{\eqRa{53}}$	\\
	\end{tabular}
	}

Since the data type $Set$ is built up from unfree constructors (cf.\
Eqns.\ (\eqra{49}) and (\eqra{50})), we have to somehow compute the
normal form sorts, cf.\ the remarks at the end of Sect.~\ref{Equational
Theories}.
Setting $Sort_{\eqRa{54}} = nf_c(Sort_{\eqRa{52}})$ and
$Sort_{\eqRa{55}} = nf_c(Sort_{\eqRa{53}})$, we may get:
	\display{
	\begin{tabular}[t]{@{}l@{$\;$}lr@{}}
	$Sort_{\eqda{52}}$ & $\sortdef add(Nat \m Nat \m List,Set)$ \\
	$Sort_{\eqda{53}}$ & $\sortdef add(Nat \m nil_l,mt)
		\;\;\mid\;\;
		add(Nat \m nil_l,Sort_{\eqra{53}})$	\\
	$Sort_{\eqda{54}}$ & $\sortdef add(Nat \m Nat \m List,Set)
		\mid add(List,Sort_{\eqra{54}})$	\\
	$Sort_{\eqda{55}}$ & $\sortdef Sort_{\eqra{53}}$ \\
	\end{tabular}
	}

Intuitively, a term of sort
$Sort_{\eqra{54}}$ denotes a set of distance sequences of
which at least one has a length $\geq 2$, while a term of sort
$Sort_{\eqra{55}}$
denotes a set of distance sequences of length 1.
$Sort_{\eqra{52}}$, $Sort_{\eqra{53}}$, 
and $mt$ are pairwise disjoint,
as are $Sort_{\eqra{54}}$, $Sort_{\eqra{55}}$, and $mt$.
These signatures are too complex
for there to be much likelihood of their being
declared by a user
who does not know the proof in advance.
The estimation of range sorts, especially of $rep$,
with such precision
that inputs starting with different constructors result in disjoint
output sorts is the main
contribution of the sort discipline to search-space reduction in this
example.

When verifying by hand, without use of the sort discipline,
some intuition {\em is}
needed to find out which values $rep(node1(n,t_1,t_2))$
can have:

\begin{quote}
{\small
First, we always have 
$rep(node2(n,t)) = add(n \m nil_l,rep(t)) \neq mt$.

Then, $rep(node1(n,t_1,t_2)) = \union{n \ptm rep(t_1)}{ rep(t_2)}$,
where $t_2$ may be $nil_t$ and thus $rep(t_2) = mt$, but $t_1$ has
again the form $node1(n',t'_1,t'_2)$
	\footnote{Or $node2(n',t')$, see above.}
and thus (by I.H.) $rep(t_1) \neq mt$, hence also $n \ptm
rep(t_1) \neq mt$. Thus, we always have $rep(node1(n,t_1,t_2)) \neq
mt$.

Finally, $rep(t_1)$ contains at least one distance sequence of length
$\geq 1$
(for $t_1 = node2(n',t')$ trivial, for $t_1 = node1(n',t'_1,t'_2)$ by
I.H.); that is why $n \ptm rep(t_1) \subset rep(node1(n,t_1,t_2))$
has to contain at least one distance sequence of length $\geq 2$.
}
\end{quote}

The ``intuition'' in this argumentation consists in recognizing two
induction hypotheses and verifying them as valid.
The main difficulty here consists in recognizing suitable
hypotheses; checking of their validity could probably be
carried out by an arbitrary induction prover.
It is precisely this task of recognition that is performed
by the sort discipline.
The two implicitly made inductions in the intuitive argumentation
correspond to applications of the global transformation rules
from Lemmas~\eqr{68} and~\eqr{69},
cf.\ Fig.\ \ref{Range sort computation for $rep$}.

\begin{figure}
\begin{center}

\begin{tabular}[t]{@{}r@{$\;$}lr@{}}
& $(Tree1\.:x) \; (rep \* x\.:z)$ \\
$=$ & $(Nat\.:n_1) \; (Tree1\.:t_1) \; (Tree\.:t_2) \;
	(\union{n_1 \ptm rep \* t_1}{rep \* t_2}\.:z)$	\\
$\mid$ & $(Nat\.:n_1) \; (Tree2\.:t_1) \; (Tree\.:t_2) \;
	(\union{n_1 \ptm rep \* t_1}{rep \* t_2}\.:z)$	\\
$=$ & \ldots	\\
$=$ & $(Nat\.:n_1) \; (Tree1\.:t_1) \; (Tree\.:t_2) \;
	(rep \* t_1\.:mt) \; (\union{mt}{rep \* t_2}\.:z)$	\\
$\mid$ & \mca{2}{$(Nat\.:n_1) \; (Tree1\.:t_1) \; (Tree\.:t_2) \;
	(rep \* t_1\.:add(l_1,s_1)) %\;
	(add(n_1 \m l_1,\union{n_1 \ptm s_1}{rep \* t_2})\.:z)$} \\
$\mid$ & $(Nat\.:n_1) \; (Tree2\.:t_1) \; (Tree\.:t_2) \;
	(add(Nat \m Nat \m nil_l,Set)\.:z)$	\\
$=$ & \mca{2}{$(Nat\.:n_1) \; (Tree1\.:t_1) \; (Tree\.:t_2) \;
	(rep \* t_1\.:add(l_1,s_1)) %\;
	(add(n_1 \m l_1,\union{n_1 \ptm s_1}{rep \* t_2})\.:z)$} \\
$\mid$ & $(Nat\.:n_1) \; (Tree2\.:t_1) \; (Tree\.:t_2) \;
	(add(Nat \m Nat \m nil_l,Set)\.:z)$
	& $(*)$	\\
$\stackrel{(\subset)}{=}$
	& $(Nat\.:n_1) \; (Tree1\.:t_1) \; (Tree\.:t_2) \;
	(rep \* t_1\.:add(l_1,s_1)) \;
	(add(n_1 \m l_1,max_{\unioN})\.:z)$ & $(**)$ \\
$\mid$ & $(Nat\.:n_1) \; (Tree2\.:t_1) \; (Tree\.:t_2) \;
	(add(Nat \m Nat \m nil_l,Set)\.:z)$	\\
$=$ & \mca{2}{$(Nat\.:n_1) \; (Tree1\.:t_1) \; (Tree\.:t_2) \;
	(rep \* t_1\.:add(l_1,s_1)) \;
	(add(n_1 \m l_1,max_{\unioN})\.:z) \; 
	(l_1\.:Sort_{\eqRa{56}})$} \\
$\mid$ & $(Nat\.:n_1) \; (Tree2\.:t_1) \; (Tree\.:t_2) \;
	(add(Nat \m Nat \m nil_l,Set)\.:z)$
	& $(***)$	\\
$\stackrel{(\subset)}{=}$
	& $(Nat\.:n_1) \; (Tree1\.:t_1) \; (Tree\.:t_2) \;
	(add(n_1 \m l_1,max_{\unioN})\.:z) \;
	(l_1\.:Sort_{\eqRa{56}})$	\\
$\mid$ & $(Nat\.:n_1) \; (Tree2\.:t_1) \; (Tree\.:t_2) \;
	(add(Nat \m Nat \m nil_l,Set)\.:z)$	\\
\end{tabular}
\vspace{0.2cm}

\begin{tabular}{@{}rp{10.7cm}@{}}
$(*)$: &
Constructor term deletion (cf.\ Lemma~\eqr{68}):
the first alternative can be removed since
$rep \* x$ does not produce $mt$ outside of it, but it requires,
in turn,
the production of $mt$ by $rep \* t_1$.
Note that the constructors $mt$ and $add(\cdot,\cdot)$ are regarded
as free such that $add(x,y) \neq mt$ always holds.
It has been shown that it is sufficient to consider
Eqns.\ (\eqra{49}) and (\eqra{50}) only outside the $rg$
computation. 
\\
$(**)$: &
Estimation by trivial upper bound, cf.\ Alg.~\eqr{70}, and $(**)$ in
Fig.\ \ref{Range sort computation for $val$}; $max_{\unioN} = Set$.
\\
$(***)$: &
Constructor argument estimation (cf.\ Lemma~\eqr{69}):
if $rep \* x$ yields $add(l_1,s_1)$, $l_1$ has the form $Nat \m Nat
\m nil_l$ from the second alternative or $add(n'_1 \m l'_1,\ldots)$
from the first one, where the same holds, in turn, for $l'_1$.
\\ &
Hence, $l_1$ belongs to $Sort_{\eqRa{56}}$,
where
$Sort_{\eqda{56}} \sortdef
Nat \m Nat \m nil_l \mid Nat \m Sort_{\eqra{56}}$. \\
\end{tabular}
\vspace{0.2cm}

\begin{tabular}{@{}l@{}}
The result we get is
	$(Tree1\.:x) \; (rep \* x\.:z) = (Sort_{\eqra{52}}\.:z)$ \\
where
	$Sort_{\eqra{52}} \sortdef add(Nat \m Sort_{\eqra{56}},Set)
	\mid add(Nat \m Nat \m nil_l,Set)$,	\\
i.e.\ $Sort_{\eqra{52}} = add(Nat \m Nat \m List,Set)$.	\\
\end{tabular}

\caption{Range sort computation for $rep$}
\label{Range sort computation for $rep$}
\end{center}

\end{figure}

Figures \ref{Case 1.1} to \ref{Case 3.2.3}
show the synthesis proof.
Variables are denoted by upper-case letters, constants by lower-case
letters.
A number in the right-hand column refers to the equation that has been
used for narrowing (rule (ln) in Thm.~\eqr{75}),
an exponent ``$^-$'' denoting the reversed
equation; ``dec'' and ``I.H.'' mean the
application of the decomposition
rule (rule (d) in Thm.~\eqr{75}),
and the induction hypothesis, respectively.
Narrowing steps that are {\em not} uniquely determined by the sort
discipline are marked with ``$*$''.
They all occur as a series of
backward applications of laws for $\ptm$ or $\unioN$
in order to get the right-hand side close to the syntactic structure
of the left-hand side and then perform a decomposition.
Application of the induction hypothesis is marked with ``$(*)$'' since
it need only be taken into account if the actual equation's sort is too
large to determine a narrowing step uniquely.

In cases that do not use the induction hypothesis, the sort
restrictions enable us to find the solution automatically.
For example, in case 2.1 ($t = node2(n_1,t_2), l = n_4 \m nil_l$, cf.\
Fig.~\ref{Case 2.1}),
owing to sort restrictions, only Eqn.\ (\eqra{44}) can
be used in the narrowing step, since the right-hand side has the sort
$add(Nat \m nil_l,Sort_{\eqra{53}}) \subset Sort_{\eqra{53}}$.
In cases that actually use the induction hypothesis, the sort
restrictions prune the search space to the size of a verification
proof, solving the additional problems of synthesis.
Moreover, the use of the sort discipline allows us to perform the
crucial proper narrowing step as the very first one, providing
syntactic information at the equations' left-hand side,
which can be used
by subsequent steps concerned with E-unification wrt.\ the $Set$
equations.

The breaking-down of case 3.2 ($t = node1(n_1,t_2,t_3)$, $l = n_4 \m n_5
\m l_6$) into three subcases 3.2.1 -- 3.2.3 can be done automatically.
Each solution has to fulfill the additional requirement that
$insert(\ldots) \in Tree^M$, since this is not ensured by the narrowing
process itself.
After having found the solution $T = node1(n_4,insert(nil_t,n_5 \m
l_6),t)$ shown in Fig.\ \ref{Case 3.2.1}, it can be determined that
$T \in Tree^M$ only if $insert(nil_t,n_5 \m l_6) \in Tree1^M \cup
Tree2^M$ and $n_4 < n_1$.
The former condition is delayed until the synthesis of the algorithm is
complete and can then be verified by an easy induction.
The latter condition
is intended to be passed to a prover in which the sort algorithms are
embedded; it must be able to detect that $n_4 < n_1 \not\Lra true$ and
to initiate the search for further solutions of case 3.2.
In this way, the solutions shown in 
Figs.\ \ref{Case 3.2.2} and \ref{Case 3.2.3} are found.
Finally, the prover must be able to detect that all subcases have been
covered, i.e.\ 
$n_4 \.< n_1 \vee n_1 \.< n_4 \vee n_4 \.= n_1 \Lra true$.

The synthesized algorithm is shown in Fig.\
\ref{Synthesized comb vector insertion algorithm}.

\begin{figure}
\begin{center}

\begin{tabular}[t]{@{}l@{$\;$}l@{$\;$}l@{$\;$}l@{}}
$insert(nil_t,n_4\m nil_l)$
	& \mca{2}{$= node2(n_4,nil_t)$} \\
$insert(nil_t,n_4\m n_5\m l_6)$
	& \mca{2}{$= node1(n_4,insert(nil_t,n_5\m l_6),nil_t)$} \\
$insert(node2(n_1,t_2),n_4\m nil_l)$
	& \mca{2}{$= node2(n_4,node2(n_1,t_2))$}	\\
$insert(node2(n_1,t_2),n_4\m n_5\m l_6)$
	& \mca{2}{$= node1(n_4,insert(nil_t,n_5\m l_6),
	node2(n_1,t_2))$} \\
$insert(node1(n_1,t_2,t_3),n_4\m nil_l)$
	& \mca{2}{$= node1(n_1,t_2,insert(t_3,n_4\m nil_l))$} \\
$insert(node1(n_1,t_2,t_3),n_4\m n_5\m l_6)$
	& \mca{2}{$= node1(n_4,insert(nil_t,n_5\m l_6),
	node1(n_1,t_2,t_3))$}	\\
	&& $\laa n_4 < n_1$	\\
$insert(node1(n_1,t_2,t_3),n_4\m n_5\m l_6)$
	& $= node1(n_1,t_2,insert(t_3,n_4\m n_5\m l_6))$
	& $\laa n_1 < n_4$	\\
$insert(node1(n_1,t_2,t_3),n_4\m n_5\m l_6)$
	& $= node1(n_1,insert(t_2,n_5\m l_6),t_3)$
	& $\laa n_4 = n_1$	\\
\end{tabular}

\caption{Synthesized comb vector insertion algorithm}
\label{Synthesized comb vector insertion algorithm}
\end{center}
\end{figure}

\newpage

\begin{figure}
\begin{center}

\begin{tabular}[t]{@{}|r@{$\;$}l|l|@{}l@{\hspace*{0.05cm}}c@{}}
\cline{1-3}
$rep \* T$ & $= add(n_4 \m nil_l,rep \* nil_t)$ &	\\ \cline{1-3}
$add(N_7 \m nil_l,rep \* T_8)$ & $= add(n_4 \m nil_l,rep \* nil_t)
	\;\;\wedge$ & (\eqra{44}) \\
    $T$ & $= node2(N_7,T_8)$ &	\\ \cline{1-3}
$add(N_7 \m nil_l,rep \* T_8)$ & $= add(n_4 \m nil_l,mt)$
	& (\eqra{42}) \\ \cline{1-3}
$N_7$ & $= n_4 \;\;\wedge$ & dec.	\\
    $rep \* T_8$ & $= mt$ &	\\ \cline{1-3}
$T_8$ & $= nil_t$ & (\eqra{42})	\\ \cline{1-3}
\end{tabular}
\vspace{0.2cm}

Answer substitution: \tab
\begin{tabular}[c]{@{}l@{$\;$}l@{}}
& $[N_7 \.{:=} n_4,T_8 \.{:=} nil_t]$	\\
$\circ$ & $[T \.{:=} node2(N_7,T_8)]$	\\
\end{tabular}

\caption{Case 1.1 --- $t = nil_t$, $l = n_4 \m nil_l$}
\label{Case 1.1}

\end{center}

\end{figure}

\begin{figure}
\begin{center}
\begin{tabular}[t]{@{}|r@{$\;$}l|l|@{}l@{\hspace*{0.05cm}}c@{}}
\cline{1-3}
$rep \* T$ & $= add(n_4 \m n_5 \m l_6,rep \* nil_t)$ &	\\ \cline{1-3}
$\union{N_7 \ptm rep \* T_8}{rep \* T_9}$
	& $= add(n_4 \m n_5 \m l_6,rep \* nil_t) \;\;\wedge$
	& (\eqra{43})	\\
    $T$ & $= node1(N_7,T_8,T_9)$ &	\\ \cline{1-3}
$\union{N_7 \ptm rep \* T_8}{rep \* T_9}$
	& $= add(n_4 \m n_5 \m l_6,\union{mt}{rep \* nil_t})$ 
	& (\eqra{47})$^-$ && $*$	\\ \cline{1-3}
$\union{N_7 \ptm rep \* T_8}{rep \* T_9}$
	& $= \union{add(n_4 \m n_5 \m l_6,mt)}{rep \* nil_t}$ 
	& (\eqra{48})$^-$ && $*$	\\ \cline{1-3}
$\union{N_7 \ptm rep \* T_8}{rep \* T_9}$
	& $= \union{add(n_4 \m n_5 \m l_6,n_4 \ptm mt)}{rep \* nil_t}$ 
	& (\eqra{45})$^-$ && $*$	\\ \cline{1-3}
$\union{N_7 \ptm rep \* T_8}{rep \* T_9}$
	& $= \union{n_4 \ptm add(n_5 \m l_6,mt)}{rep \* nil_t}$ 
	& (\eqra{46})$^-$ && $*$	\\ \cline{1-3}
$N_7$ & $= n_4 \;\;\wedge$ & dec. && $*$	\\
    $rep \* T_8$ & $= add(n_5 \m l_6,mt) \;\;\wedge$ &	\\
    $T_9$ & $= nil_t$ & \\ \cline{1-3}
$add(L_{11},rep \* T_{10})$
	& $= add(n_5 \m l_6,mt) \;\;\wedge$ & I.H. && $(*)$	\\
    $T_8$ & $= insert(T_{10},L_{11})$ &	\\ \cline{1-3}
$L_{11}$ & $= n_5 \m l_6 \;\;\wedge$ & dec.	\\
    $rep \* T_{10}$ & $= mt$ &	\\ \cline{1-3}
$T_{10}$ & $= nil_t$ & (\eqra{42})	\\ \cline{1-3}
\end{tabular}
\vspace{0.2cm}

Answer substitution: \tab
\begin{tabular}[c]{@{}l@{$\;$}l@{}}
& $[L_{11} \.{:=} n_5 \m l_6,
	T_{10} \.{:=} nil_t]$	\\
$\circ$ & $[N_7 \.{:=} n_4,
	T_9 \.{:=} nil_t,
	T_8 \.{:=} insert(T_{10},L_{11})]$	\\
$\circ$ & $[T \.{:=} node1(N_7,T_8,T_9)]$	\\
\end{tabular}

\caption{Case 1.2 --- $t = nil_t$, $l = n_4 \m n_5 \m l_6$}
\label{Case 1.2}

\end{center}

\end{figure}

\begin{figure}
\begin{center}

\begin{tabular}[t]{@{}|r@{$\;$}l|l|@{}l@{\hspace*{0.05cm}}c@{}}
\cline{1-3}
$rep \* T$ & $= add(n_4 \m nil_l,rep \* node2(n_1,t_2))$ &\\ \cline{1-3}
$add(N_7 \m nil_l,rep \* T_8)$ 
	& $= add(n_4 \m nil_l,rep \* node2(n_1,t_2)) \;\;\wedge$ 
	& (\eqra{44})	\\
    $T$ & $= node2(N_7,T_8)$ &	\\ \cline{1-3}
$N_7$ & $= n_4 \;\;\wedge$ & dec.	\\
    $T_8$ & $= node2(n_1,t_2)$ &	\\ \cline{1-3}
\end{tabular}
\vspace{0.2cm}

Answer substitution: \tab
\begin{tabular}[c]{@{}l@{$\;$}l@{}}
& $[N_7 \.{:=} n_4,T_8 \.{:=} node2(n_1,t_2)]$	\\
$\circ$ & $[T \.{:=} node2(N_7,T_8)]$	\\
\end{tabular}

\caption{Case 2.1 --- $t = node2(n_1,t_2)$, $l = n_4 \m nil_l$}
\label{Case 2.1}

\end{center}

\end{figure}

\begin{figure}
\begin{center}

\begin{tabular}[t]{@{}|r@{$\;$}l|l|@{}l@{\hspace*{0.05cm}}c@{}}
\cline{1-3}
$rep \* T$ & $= add(n_4 \m n_5 \m l_6,rep \* node2(n_1,t_2))$
    & \\ \cline{1-3}
$\union{N_7 \ptm rep \* T_8}{rep \* T_9}$
    & $= add(n_4 \m n_5 \m l_6,rep \* node2(n_1,t_2))
    \;\;\wedge$	& (\eqra{43})	\\
$T$ & $= node1(N_7,T_8,T_9)$ & \\ \cline{1-3}
$\union{N_7 \ptm rep \* T_8}{rep \* T_9}$
	& $= add(n_4 \m n_5 \m l_6,\union{mt}{rep \* node2(n_1,t_2)})$
	& (\eqra{47})$^-$ && $*$	 \\ \cline{1-3}
$\union{N_7 \ptm rep \* T_8}{rep \* T_9}$
	& $= \union{add(n_4 \m n_5 \m l_6,mt)}{rep \* node2(n_1,t_2)}$
	& (\eqra{48})$^-$ && $*$	 \\ \cline{1-3}
$\union{N_7 \ptm rep \* T_8}{rep \* T_9}$
	& $= \union{add(n_4 \m n_5 \m l_6,n_4 \ptm mt)}{
	rep \* node2(n_1,t_2)}$
	& (\eqra{45})$^-$ && $*$	 \\ \cline{1-3}
$\union{N_7 \ptm rep \* T_8}{rep \* T_9}$
	& $= \union{n_4 \ptm add(n_5 \m l_6,mt)}{
	rep \* node2(n_1,t_2)}$
	& (\eqra{46})$^-$ && $*$	 \\ \cline{1-3}
$N_7$ & $= n_4 \;\;\wedge$ & dec. && $*$	 \\
    $rep \* T_8$ & $= add(n_5 \m l_6,mt) \;\;\wedge$ & 	\\
    $T_9$ & $= node2(n_1,t_2)$ &	\\ \cline{1-3}
$add(L_{11},rep \* T_{10})$ & $= add(n_5 \m l_6,mt) \;\;\wedge$ 
	& I.H. && $(*)$	\\
    $T_8$ & $= insert(T_{10},L_{11})$ &	\\ \cline{1-3}
$L_{11}$ & $= n_5 \m l_6 \;\;\wedge$ & dec.	\\
    $rep \* T_{10}$ & $= mt$  & \\ \cline{1-3}
$T_{10}$ & $= nil_t$  & (\eqra{42})	\\ \cline{1-3}
\end{tabular}
\vspace{0.2cm}

Answer substitution: \tab
\begin{tabular}[c]{@{}l@{$\;$}l@{}}
& $[L_{11} \.{:=} n_5 \m l_6,
	T_{10} \.{:=} nil_t]$	\\
$\circ$ & $[N_7 \.{:=} n_4,
	T_9 \.{:=} node2(n_1,t_2),
	T_8 \.{:=} insert(T_{10},L_{11})]$	\\
$\circ$ & $[T \.{:=} node1(N_7,T_8,T_9)]$	\\
\end{tabular}

\caption{Case 2.2 --- $t = node2(n_1,t_2)$, $l = n_4 \m n_5 \m l_6$}
\label{Case 2.2}

\end{center}

\end{figure}

\begin{figure}
\begin{center}
\begin{tabular}[t]{@{}|r@{$\;$}l|l|@{}l@{\hspace*{0.05cm}}c@{}}
\cline{1-3}
$rep \* T$ & $= add(n_4 \m nil_l,rep \* node1(n_1,t_2,t_3))$ 
	& \\ \cline{1-3}
$\union{N_7 \ptm rep \* T_8}{rep \* T_9}$
	& $= add(n_4 \m nil_l,rep \* node1(n_1,t_2,t_3)) \;\;\wedge$
	& (\eqra{43})	\\
$T$ & $= node1(N_7,T_8,T_9)$ & \\ \cline{1-3}
$\union{N_7 \ptm rep \* T_8}{rep \* T_9}$
	& $= add(n_4 \m nil_l,\union{n_1 \ptm rep \* t_2}{rep \* t_3})$
	& (\eqra{43})	\\ \cline{1-3}
$\union{N_7 \ptm rep \* T_8}{rep \* T_9}$
	& $= add(n_4 \m nil_l,\union{rep \* t_3}{n_1 \ptm rep \* t_2})$
	& (\eqra{51}) && $*$	\\ \cline{1-3}
$\union{N_7 \ptm rep \* T_8}{rep \* T_9}$
	& $= \union{add(n_4 \m nil_l,rep \* t_3)}{n_1 \ptm rep \* t_2}$
	& (\eqra{48})$^-$ && $*$	\\ \cline{1-3}
$\union{N_7 \ptm rep \* T_8}{rep \* T_9}$
	& $= \union{n_1 \ptm rep \* t_2}{add(n_4 \m nil_l,rep \* t_3)}$
	& (\eqra{51}) && $*$	\\ \cline{1-3}
$N_7$ & $= n_1 \;\;\wedge$ & dec. && $*$	\\
    $T_8$ & $= t_2 \;\;\wedge$ &	\\
    $rep \* T_9$ & $= add(n_4 \m nil_l,rep \* t_3)$ 
	&	\\ \cline{1-3}
$add(L_{11},rep \* T_{10})$ & $= add(n_4 \m nil_l,rep \* t_3)
	\;\;\wedge$ & I.H. && $(*)$	\\
    $T_9$ & $= insert(T_{10},L_{11})$ &	\\ \cline{1-3}
$L_{11}$ & $= n_4 \m nil_l \;\;\wedge$ & dec.	\\
    $T_{10}$ & $= t_3$ & \\ \cline{1-3}
\end{tabular}
\vspace{0.2cm}

Answer substitution: \tab
\begin{tabular}[c]{@{}l@{$\;$}l@{}}
& $[L_{11} \.{:=} n_4 \m nil_l,
	T_{10} \.{:=} t_3]$	\\
$\circ$ & $[N_7 \.{:=} n_1,
	T_8 \.{:=} t_2,
	T_9 \.{:=} insert(T_{10},L_{11})]$	\\
$\circ$ & $[T \.{:=} node1(N_7,T_8,T_9)]$	\\
\end{tabular}

\caption{Case 3.1 --- $t = node1(n_1,t_2,t_3)$, $l = n_4 \m nil_l$}
\label{Case 3.1}

\end{center}

\end{figure}

\begin{figure}
\begin{center}
\begin{tabular}[t]
{@{}|r@{$\;$}l@{\hspace*{0.050cm}}|l|@{}l@{\hspace*{0.05cm}}c@{}}
\cline{1-3}
$rep \* T$ & $= add(n_4 \m n_5 \m l_6,rep \* node1(n_1,t_2,t_3))$
	&	\\ \cline{1-3}
$\union{N_7 \ptm rep \* T_8}{rep \* T_9}$
	& $= add(n_4 \m n_5 \m l_6,rep \* node1(n_1,t_2,t_3))
	\;\;\wedge$ & (\eqra{43})	\\
    $T$ & $= node1(N_7,T_8,T_9)$ &	\\ \cline{1-3}
$\union{N_7 \ptm rep \* T_8}{rep \* T_9}$
	& $= add(n_4 \m n_5 \m l_6,\union{mt}{
	rep \* node1(n_1,t_2,t_3)})$
	& (\eqra{47})$^-$ && $*$	\\ \cline{1-3}
$\union{N_7 \ptm rep \* T_8}{rep \* T_9}$
	& $= \union{add(n_4 \m n_5 \m l_6,mt)}{
	rep \* node1(n_1,t_2,t_3)}$
	& (\eqra{48})$^-$ && $*$	\\ \cline{1-3}
$\union{N_7 \ptm rep \* T_8}{rep \* T_9}$
	& $= \union{add(n_4 \.\m n_5 \.\m l_6,n_4 \ptm mt)}{
	rep \* node1(n_1,t_2,t_3)}$
	& (\eqra{45})$^-$ && $*$	\\ \cline{1-3}
$\union{N_7 \ptm rep \* T_8}{rep \* T_9}$
	& $= \union{n_4 \ptm add(n_5 \m l_6,mt)}{
	rep \* node1(n_1,t_2,t_3)}$
	& (\eqra{46})$^-$ && $*$	\\ \cline{1-3}
$N_7$ & $= n_4 \;\;\wedge$ & dec. && $*$	\\
    $rep \* T_8$ & $= add(n_5 \m l_6,mt) \;\;\wedge$ &	\\
    $T_9$ & $= node1(n_1,t_2,t_3)$ &	\\ \cline{1-3}
$add(L_{11},rep \* T_{10})$
	& $= add(n_5 \m l_6,mt) \;\;\wedge$ & I.H. && $(*)$	\\
    $T_8$ & $= insert(T_{10},L_{11})$ &	\\ \cline{1-3}
$L_{11}$ & $= n_5 \m l_6 \;\;\wedge$ & dec.	\\
    $rep \* T_{10}$ & $= mt$ &	\\ \cline{1-3}
$T_{10}$ & $= nil_t$ & (\eqra{42})	\\ \cline{1-3}
\end{tabular}
\vspace{0.2cm}

Answer substitution: \tab
\begin{tabular}[c]{@{}l@{$\;$}lr@{}}
& $[L_{11} \.{:=} n_5 \m l_6,
	T_{10} \.{:=} nil_t]$	\\
$\circ$ & $[N_7 \.{:=} n_4,
	T_9 \.{:=} node1(n_1,t_2,t_3),
	T_8 \.{:=} insert(T_{10},L_{11})]$	\\
$\circ$ & $[T \.{:=} node1(N_7,T_8,T_9)]$	\\
\end{tabular}

\caption{Case 3.2.1 --- $t = node1(n_1,t_2,t_3)$, 
	$l = n_4 \m n_5 \m l_6$, $n_4 < n_1$}
\label{Case 3.2.1}

\end{center}

\end{figure}

\begin{figure}
\begin{center}
\begin{tabular}[t]{@{}|r@{$\;$}l|l|@{}l@{\hspace*{0.05cm}}c@{}}
\cline{1-3}
$rep \* T$ & $= add(n_4 \m n_5 \m l_6,rep \* node1(n_1,t_2,t_3))$
	&	\\ \cline{1-3}
$\union{N_7 \ptm rep \* T_8}{rep \* T_9}$
	& $= add(n_4 \m n_5 \m l_6,rep \* node1(n_1,t_2,t_3))
	\;\;\wedge$ & (\eqra{43})	\\
    $T$ & $= node1(N_7,T_8,T_9)$ &	\\ \cline{1-3}
$\union{N_7 \ptm rep \* T_8}{rep \* T_9}$
	& $= add(n_4 \m n_5 \m l_6,\union{n_1 \ptm rep \* t_2}{
	rep \* t_3})$ & (\eqra{43})	\\ \cline{1-3}
$\union{N_7 \ptm rep \* T_8}{rep \* T_9}$
	& $= add(n_4 \m n_5 \m l_6,\union{rep \* t_3}{
	n_1 \ptm rep \* t_2})$ & (\eqra{51}) && $*$ \\ \cline{1-3}
$\union{N_7 \ptm rep \* T_8}{rep \* T_9}$
	& $= \union{add(n_4 \m n_5 \m l_6,rep \* t_3)}{
	n_1 \ptm rep \* t_2}$ & (\eqra{48})$^-$ && $*$	\\ \cline{1-3}
$\union{N_7 \ptm rep \* T_8}{rep \* T_9}$
	& $= \union{n_1 \ptm rep \* t_2}{add(n_4 \m n_5 \m l_6,
	rep \* t_3)}$ & (\eqra{51}) && $*$	\\ \cline{1-3}
$N_7$ & $= n_1 \;\;\wedge$ & dec. && $*$	\\
    $T_8$ & $= t_2 \;\;\wedge$ &	\\
    $rep \* T_9$ & $= add(n_4 \m n_5 \m l_6, rep \* t_3)$ 
	& \\ \cline{1-3}
$add(L_{11},rep \* T_{10})$ & $= add(n_4 \m n_5 \m l_6, rep \* t_3)
	\;\;\wedge$ & I.H. && $(*)$	\\
    $T_9$ & $= insert(T_{10},L_{11})$ &	\\ \cline{1-3}
$L_{11}$ & $= n_4 \m n_5 \m l_6 \;\;\wedge$ & dec.	\\
    $T_{10}$ & $= t_3$ &	\\ \cline{1-3}
\end{tabular}
\vspace{0.2cm}

Answer substitution: \tab
\begin{tabular}[c]{@{}l@{$\;$}lr@{}}
& $[L_{11} \.{:=} n_r \m n_5 \m l_6,
	T_{10} \.{:=} t_3]$	\\
$\circ$ & $[N_7 \.{:=} n_1,
	T_8 \.{:=} t_2,
	T_9 \.{:=} insert(T_{10},L_{11})]$	\\
$\circ$ & $[T \.{:=} node1(N_7,T_8,T_9)]$	\\
\end{tabular}

\caption{Case 3.2.2 --- $t = node1(n_1,t_2,t_3)$, 
	$l = n_4 \m n_5 \m l_6$, $n_1 < n_4$}
\label{Case 3.2.2}

\end{center}

\end{figure}

\begin{figure}
\begin{center}
\begin{tabular}[t]{@{}|r@{$\;$}l|l|@{}l@{\hspace*{0.05cm}}c@{}}
\cline{1-3}
$rep \* T$ & $= add(n_1 \m n_5 \m l_6,rep \* node1(n_1,t_2,t_3))$
	&	\\ \cline{1-3}
$\union{N_7 \ptm rep \* T_8}{rep \* T_9}$
	& $= add(n_1 \m n_5 \m l_6,rep \* node1(n_1,t_2,t_3))
	\;\;\wedge$ & (\eqra{43})	\\
    $T$ & $= node1(N_7,T_8,T_9)$ &	\\ \cline{1-3}
$\union{N_7 \ptm rep \* T_8}{rep \* T_9}$
	& $= add(n_1 \m n_5 \m l_6,\union{n_1 \ptm rep \* t_2}{
	rep \* t_3})$ & (\eqra{43})	\\ \cline{1-3}
$\union{N_7 \ptm rep \* T_8}{rep \* T_9}$
	& $= \union{add(n_1 \m n_5 \m l_6,n_1 \ptm rep \* t_2)}{
	rep \* t_3}$ & (\eqra{48})$^-$ && $*$	\\ \cline{1-3}
$\union{N_7 \ptm rep \* T_8}{rep \* T_9}$
	& $= \union{n_1 \ptm add(n_5 \m l_6,rep \* t_2)}{rep \* t_3}$
	& (\eqra{46})$^-$ && $*$	\\ \cline{1-3}
$N_7$ & $= n_1 \;\;\wedge$ & dec. && $*$	\\
    $rep \* T_8$ & $= add(n_5 \m l_6,rep \* t_2) \;\;\wedge$ & \\
    $T_9$ & $= t_3$ &	\\ \cline{1-3}
$add(L_{11},rep \* T_{10})$ 
	& $= add(n_5 \m l_6,rep \* t_2) \;\;\wedge$ & I.H. && $(*)$ \\
    $T_8$ & $= insert(T_{10},L_{11})$ &	\\ \cline{1-3}
$L_{11}$ & $= n_5 \m l_6 \;\;\wedge$ & dec.	\\
    $T_{10}$ & $= t_2$ &	\\ \cline{1-3}
\end{tabular}
\vspace{0.2cm}

Answer substitution: \tab
\begin{tabular}[c]{@{}l@{$\;$}lr@{}}
& $[L_{11} \.{:=} n_5 \m l_6,
	T_{10} \.{:=} t_2]$	\\
$\circ$ & $[N_7 \.{:=} n_1,
	T_9 \.{:=} t_3,
	T_8 \.{:=} insert(T_{10},L_{11})]$	\\
$\circ$ & $[T \.{:=} node1(N_7,T_8,T_9)]$	\\
\end{tabular}

\caption{Case 3.2.3 --- $t = node1(n_1,t_2,t_3)$, 
	$l = n_4 \m n_5 \m l_6$, $n_4 = n_1$}
\label{Case 3.2.3}

\end{center}

\end{figure}

\newpage

\twocolumn

\section{Index}

\begin{tabbing}
MMMMMMMMMMMMMMMMMM\=MM\=\kill
{\bf NOTION} \> {\bf Nr.} \> {\bf Page} \\
\\
% {
% !}sort -bdf
% :.,$s/&/\\>/g 
% :.,$s/{eqd/{/
% :.,$s/\$\$/$/
% }}

\EQi{$\ablcdi$} \> \eqr{77} \> \pageref{77} \\
\EQi{$\ablci$} \> \eqr{77} \> \pageref{77} \\
\EQi{$\Ablds$} \> \eqr{61} \> \pageref{61} \\
\EQi{$\ablds$} \> \eqr{61} \> \pageref{61} \\
\EQi{$\vs \sigma v_1 \abldi \vs \sigma v_2$} \> \eqr{61} \> \pageref{61} \\
\EQi{$\vs {\mu_i} f(u_{i1},\ldots,u_{in})$} \> \eqr{60} \> \pageref{60} \\

\EQi{$\psubset$} \> \eqr{1} \> \pageref{1} \\
\EQi{$\subset$} \> \eqr{1} \> \pageref{1} \\
\EQi{$A \times B$} \> \eqr{2} \> \pageref{2} \\

\EQi{$\sortdef$} \> \eqr{4} \> \pageref{4} \\
\EQi{$\stackrel{\mbox{\Large\bf ..}}{<}$} \> \eqr{4} \> \pageref{4} \\
\EQi{$\lg$} \> \eqr{4} \> \pageref{4} \\
\EQi{$\bigmid_{i=1}^n S_i$} \> \eqr{2} \> \pageref{2} \\
\EQi{$\mid$} \> \eqr{4} \> \pageref{4} \\

\EQi{$\bot$} \> \eqr{36} \> \pageref{36} \\
\EQi{$\top$} \> \eqr{36} \> \pageref{36} \\
\EQi{$\top_V$} \> \eqr{36} \> \pageref{36} \\

\EQi{$\func{V}{\cal CR}$} \> \eqr{13} \> \pageref{13} \\
\EQi{$\pfunc{V}{\cal CR}$} \> \eqr{17} \> \pageref{17} \\

\EQi{$\varepsilon$} \> \eqr{13} \> \pageref{13} \\
\EQi{$\beta$} \> \eqr{3} \> \pageref{3} \\
\EQi{$\beta \rs V$} \> \eqr{3} \> \pageref{3} \\
\EQi{$\beta v$} \> \eqr{3} \> \pageref{3} \\
\EQi{$\beta_1 \pscomp \beta_2$} \> \eqr{3} \> \pageref{3} \\
\EQi{$\gamma$} \> \eqr{3} \> \pageref{3} \\
\EQi{$\mu$} \> \eqr{15} \> \pageref{15} \\
\EQi{$\mu'$} \> \eqr{15} \> \pageref{15} \\
\EQi{$\sigma$} \> \eqr{15} \> \pageref{15} \\
\EQi{$\sigma'$} \> \eqr{15} \> \pageref{15} \\
\EQi{$\sigma \by \beta$} \> \eqr{26} \> \pageref{26} \\
\EQi{$\sigma' \by \beta$} \> \eqr{26} \> \pageref{26} \\
\EQi{$\sigma \circ \beta$} \> \eqr{24} \> \pageref{24} \\
\EQi{$\sigma' \circ \beta$} \> \eqr{24} \> \pageref{24} \\
\EQi{$\sigma \pecomp \tau$} \> \eqr{22} \> \pageref{22} \\
\EQi{$\sigma' \pecomp \tau'$} \> \eqr{22} \> \pageref{22} \\
\EQi{$\sigma \rs V$} \> \eqr{21} \> \pageref{21} \\
\EQi{$\sigma' \rs V$} \> \eqr{21} \> \pageref{21} \\
\EQi{$\sigma u$} \> \eqr{18} \> \pageref{18} \\
\EQi{$\sigma' u$} \> \eqr{18} \> \pageref{18} \\
\EQi{$\tau$} \> \eqr{15} \> \pageref{15} \\
\EQi{$\tau'$} \> \eqr{15} \> \pageref{15} \\
\EQi{$\vs \sigma v$} \> \eqr{56} \> \pageref{56} \\
\EQi{$::$} \> \eqr{4} \> \pageref{4} \\
\EQi{$0_x s_y$} \> \eqr{14} \> \pageref{14} \\

\EQi{$abstract(S,x)$} \> \eqr{42} \> \pageref{42} \\
\EQi{admissible t-substitutions} \> \eqr{15} \> \pageref{15} \\
\EQi{alternatives} \> \eqr{64} \> \pageref{64} \\
\EQi{annotated term} \> \eqr{56} \> \pageref{56} \\
\EQi{application} \> \eqr{13} \> \pageref{13} \\
\EQi{application} \> \eqr{18} \> \pageref{18} \\
\EQi{$apply(\sigma,x)$} \> \eqr{37} \> \pageref{37} \\
\EQi{approximation rule} \> \eqr{64} \> \pageref{64} \\
\EQi{$ar(g)$} \> \eqr{1} \> \pageref{1} \\
\EQi{arity} \> \eqr{1} \> \pageref{1} \\
\EQi{$ar(\vec{cr})$} \> \eqr{13} \> \pageref{13} \\
\EQi{$Bin$} \> \eqr{4} \> \pageref{4} \\
\EQi{$Bin$} \> \eqr{78} \> \pageref{78} \\
\EQi{$compose(\sigma,\tau)$} \> \eqr{41} \> \pageref{41} \\
\EQi{computation path} \> \eqr{64} \> \pageref{64} \\
\EQi{computation tree} \> \eqr{64} \> \pageref{64} \\
\EQi{constraint} \> \eqr{12} \> \pageref{12} \\
\EQi{constructor symbols} \> \eqr{1} \> \pageref{1} \\
\EQi{constructor terms} \> \eqr{1} \> \pageref{1} \\
\EQi{constructor-matching rules} \> \eqr{9} \> \pageref{9} \\
\EQi{constructors for t-substitutions} \> \eqr{13} \> \pageref{13} \\
\EQi{$cr$} \> \eqr{1} \> \pageref{1} \\
\EQi{$\vec{cr}$} \> \eqr{13} \> \pageref{13} \\
\EQi{$\vec{cr} \pecomp \vec{cr}'$} \> \eqr{13} \> \pageref{13} \\
\EQi{$\vec{cr} \rs {V'}$} \> \eqr{13} \> \pageref{13} \\
\EQi{$\vec{cr}_x$} \> \eqr{13} \> \pageref{13} \\
\EQi{${\cal CR}$} \> \eqr{1} \> \pageref{1} \\
\EQi{(d)} \> \eqr{75} \> \pageref{75} \\
\EQi{defining equations} \> \eqr{59} \> \pageref{59} \\
\EQi{defining equations} \> \eqr{60} \> \pageref{60} \\
\EQi{dependent} \> \eqr{38} \> \pageref{38} \\
\EQi{depth} \> \eqr{3} \> \pageref{3} \\
\EQi{$diff(S_1,S_2 \mid \ldots \mid S_m)$} \> \eqr{11} \> \pageref{11} \\
\EQi{distributivity rules} \> \eqr{9} \> \pageref{9} \\
\EQi{$div(\sigma,\beta)$} \> \eqr{55} \> \pageref{55} \\
\EQi{$dom(\beta)$} \> \eqr{3} \> \pageref{3} \\
\EQi{$dom(f)$} \> \eqr{60} \> \pageref{60} \\
\EQi{$dom(f,I)$} \> \eqr{60} \> \pageref{60} \\
\EQi{$dom(\sigma')$} \> \eqr{17} \> \pageref{17} \\
\EQi{$dom(\sigma)$} \> \eqr{36} \> \pageref{36} \\
\EQi{$dup(\sigma,\beta)$} \> \eqr{44} \> \pageref{44} \\
\EQi{$dup(x)$} \> \eqr{78} \> \pageref{78} \\
\EQi{elementwise extension} \> \eqr{2} \> \pageref{2} \\
\EQi{equations between constructors} \> \eqr{77} \> \pageref{77} \\
\EQi{$Even$} \> \eqr{78} \> \pageref{78} \\
\EQi{extended sort} \> \eqr{57} \> \pageref{57} \\
\EQi{$f$} \> \eqr{1} \> \pageref{1} \\
\EQi{${\cal F}$} \> \eqr{1} \> \pageref{1} \\
\EQi{$f \* x$} \> \eqr{2} \> \pageref{2} \\
\EQi{$f[A']$} \> \eqr{2} \> \pageref{2} \\
\EQi{factorization} \> \eqr{26} \> \pageref{26} \\
\EQi{$fact(\sigma,\beta)$} \> \eqr{47} \> \pageref{47} \\
\EQi{$fact(\sigma,\beta)$} \> \eqr{49} \> \pageref{49} \\
\EQi{$g$} \> \eqr{1} \> \pageref{1} \\
\EQi{ground constructor terms} \> \eqr{1} \> \pageref{1} \\
\EQi{homogeneous} \> \eqr{48} \> \pageref{48} \\
\EQi{$i$} \> \eqr{4} \> \pageref{4} \\
\EQi{independent} \> \eqr{38} \> \pageref{38} \\
\EQi{Induction Principle} \> \eqr{6} \> \pageref{6} \\
\EQi{$inf(S_1,S_2)$} \> \eqr{10} \> \pageref{10} \\
\EQi{inhabitance} \> \eqr{9} \> \pageref{9} \\
\EQi{$inh(S,Occ)$} \> \eqr{12} \> \pageref{12} \\
\EQi{intersection} \> \eqr{9} \> \pageref{9} \\
\EQi{junk terms} \> \eqr{61} \> \pageref{61} \\
\EQi{lazy narrowing} \> \eqr{75} \> \pageref{75} \\
\EQi{$Lex_{x<y}$} \> \eqr{42} \> \pageref{42} \\
\EQi{lifting} \> \eqr{23} \> \pageref{23} \\
\EQi{linear} \> \eqr{3} \> \pageref{3} \\
\EQi{linear} \> \eqr{3} \> \pageref{3} \\
\EQi{(ln)} \> \eqr{75} \> \pageref{75} \\
\EQi{local transformation rules} \> \eqr{64} \> \pageref{64} \\
\EQi{loop-checking rules} \> \eqr{9} \> \pageref{9} \\
\EQi{$M$} \> \eqr{4} \> \pageref{4} \\
\EQi{$max_f$} \> \eqr{70} \> \pageref{70} \\
\EQi{$max'_f$} \> \eqr{71} \> \pageref{71} \\
\EQi{$mgu(v_1,v_2)$} \> \eqr{3} \> \pageref{3} \\
\EQi{most general unifier} \> \eqr{3} \> \pageref{3} \\
\EQi{$Mtch_{x,y}$} \> \eqr{42} \> \pageref{42} \\
\EQi{$Nat$} \> \eqr{78} \> \pageref{78} \\
\EQi{$Nat_x$} \> \eqr{36} \> \pageref{36} \\
\EQi{$Nat_{x,y}$} \> \eqr{36} \> \pageref{36} \\
\EQi{$Nat_{x\.<y}$} \> \eqr{36} \> \pageref{36} \\
\EQi{$Nat_{x\.=y}$} \> \eqr{36} \> \pageref{36} \\
\EQi{$Nat_y$} \> \eqr{36} \> \pageref{36} \\
\EQi{$nf[A]$} \> \eqr{61} \> \pageref{61} \\
\EQi{$nf_c$} \> \eqr{78} \> \pageref{78} \\
\EQi{$nf(v)$} \> \eqr{61} \> \pageref{61} \\
\EQi{non-constructor functions} \> \eqr{1} \> \pageref{1} \\
\EQi{$o$} \> \eqr{4} \> \pageref{4} \\
\EQi{ordinary substitution} \> \eqr{3} \> \pageref{3} \\
\EQi{parallel composition} \> \eqr{13} \> \pageref{13} \\
\EQi{parallel composition} \> \eqr{22} \> \pageref{22} \\
\EQi{parallel composition} \> \eqr{3} \> \pageref{3} \\
\EQi{partial mappings} \> \eqr{17} \> \pageref{17} \\
\EQi{$Pref_{x,y,z}$} \> \eqr{42} \> \pageref{42} \\
\EQi{pseudolinear} \> \eqr{3} \> \pageref{3} \\
\EQi{pseudolinear} \> \eqr{3} \> \pageref{3} \\
\EQi{$ran(\beta)$} \> \eqr{3} \> \pageref{3} \\
\EQi{Rank of a T-Substitution} \> \eqr{64} \> \pageref{64} \\
\EQi{$rank(\sigma',(w_1:u_1))$} \> \eqr{65} \> \pageref{65} \\
\EQi{$rank(\sigma',(w_1:u_1) \; \ldots \; (w_n:u_n))$} \> \eqr{65} \> \pageref{65} \\
\EQi{regular} \> \eqr{6} \> \pageref{6} \\
\EQi{relative complement} \> \eqr{9} \> \pageref{9} \\
\EQi{renaming substitution} \> \eqr{3} \> \pageref{3} \\
\EQi{restriction} \> \eqr{13} \> \pageref{13} \\
\EQi{restriction} \> \eqr{21} \> \pageref{21} \\
\EQi{restriction} \> \eqr{3} \> \pageref{3} \\
\EQi{$restrict(\sigma,V)$} \> \eqr{40} \> \pageref{40} \\
\EQi{rewrite relation} \> \eqr{61} \> \pageref{61} \\
\EQi{$rg$} \> \eqr{59} \> \pageref{59} \\
\EQi{$rg_c$} \> \eqr{78} \> \pageref{78} \\
\EQi{$rg(\sigma,v)$} \> \eqr{73} \> \pageref{73} \\
\EQi{$S$} \> \eqr{1} \> \pageref{1} \\
\EQi{${\cal S}$} \> \eqr{1} \> \pageref{1} \\
\EQi{semantics} \> \eqr{4} \> \pageref{4} \\
\EQi{semi-independent} \> \eqr{38} \> \pageref{38} \\
\EQi{semilinear} \> \eqr{3} \> \pageref{3} \\
\EQi{$(\sigma)$} \> \eqr{62} \> \pageref{62} \\
\EQi{$single(S)$} \> \eqr{12} \> \pageref{12} \\
\EQi{$S^M$} \> \eqr{4} \> \pageref{4} \\
\EQi{$snoc$} \> \eqr{4} \> \pageref{4} \\
\EQi{solution of an equation} \> \eqr{74} \> \pageref{74} \\
\EQi{sort definitions} \> \eqr{4} \> \pageref{4} \\
\EQi{sort equivalence} \> \eqr{9} \> \pageref{9} \\
\EQi{sort expressions} \> \eqr{4} \> \pageref{4} \\
\EQi{sort names} \> \eqr{1} \> \pageref{1} \\
\EQi{sort system} \> \eqr{4} \> \pageref{4} \\
\EQi{Sorted Narrowing} \> \eqr{74} \> \pageref{74} \\
\EQi{Sorted Rewriting} \> \eqr{60} \> \pageref{60} \\
\EQi{$SortName$} \> \eqr{4} \> \pageref{4} \\
\EQi{subsort} \> \eqr{9} \> \pageref{9} \\
\EQi{substitution} \> \eqr{3} \> \pageref{3} \\
\EQi{$Sum_{x,y,z}$} \> \eqr{42} \> \pageref{42} \\
\EQi{$S^X$} \> \eqr{4} \> \pageref{4} \\
\EQi{${\cal T}_{\cal CR}$} \> \eqr{1} \> \pageref{1} \\
\EQi{${\cal T}_{{\cal CR},{\cal F}}/{\Ablds}$} \> \eqr{61} \> \pageref{61} \\
\EQi{${\cal T}_{{\cal CR},{\cal F},{\cal V}}$} \> \eqr{1} \> \pageref{1} \\
\EQi{${\cal T}_{{\cal CR},{\cal F},{\cal V},{\cal S}}$} \> \eqr{1} \> \pageref{1} \\
\EQi{${\cal T}_{{\cal CR},{\cal V}}$} \> \eqr{1} \> \pageref{1} \\
\EQi{${\cal T}_{\func{V}{\cal CR}}$} \> \eqr{15} \> \pageref{15} \\
\EQi{${\cal T}_X$} \> \eqr{1} \> \pageref{1} \\
\EQi{${\cal T}_{X,Y}$} \> \eqr{1} \> \pageref{1} \\
\EQi{$\TT_{\func{V}{\cal CR}}$} \> \eqr{15} \> \pageref{15} \\
\EQi{$\TT_{\pfunc{V}{\cal CR}}$} \> \eqr{17} \> \pageref{17} \\
\EQi{terms} \> \eqr{1} \> \pageref{1} \\
\EQi{t-sets} \> \eqr{15} \> \pageref{15} \\
\EQi{t-substitutions} \> \eqr{15} \> \pageref{15} \\
\EQi{tuple} \> \eqr{2} \> \pageref{2} \\
\EQi{$u$} \> \eqr{1} \> \pageref{1} \\
\EQi{$use(S)$} \> \eqr{6} \> \pageref{6} \\
\EQi{$v$} \> \eqr{1} \> \pageref{1} \\
\EQi{${\cal V}$} \> \eqr{1} \> \pageref{1} \\
\EQi{$v_1 \psubterm v_2$} \> \eqr{3} \> \pageref{3} \\
\EQi{$v_1 \subterm v_2$} \> \eqr{3} \> \pageref{3} \\
\EQi{$\tpl{v_1,\ldots,v_n}$} \> \eqr{2} \> \pageref{2} \\
\EQi{$\tpl{v_i \mid p(v_i), \; i=1,\ldots,n}$} \> \eqr{2} \> \pageref{2} \\
\EQi{$val$} \> \eqr{78} \> \pageref{78} \\
\EQi{variables} \> \eqr{1} \> \pageref{1} \\
\EQi{$vars(v_1,\ldots,v_n)$} \> \eqr{3} \> \pageref{3} \\
\EQi{$w$} \> \eqr{1} \> \pageref{1} \\
\EQi{$((w_1:u_1) \; (w_2:u_2))^M$} \> \eqr{62} \> \pageref{62} \\
\EQi{well-defined} \> \eqr{61} \> \pageref{61} \\
\EQi{$w^M$} \> \eqr{62} \> \pageref{62} \\
\EQi{$(w:u)^M$} \> \eqr{62} \> \pageref{62} \\
\EQi{$x$} \> \eqr{1} \> \pageref{1} \\
\EQi{$[x \.{:=} S]$} \> \eqr{23} \> \pageref{23} \\
\EQi{$[x \.{:=} u]$} \> \eqr{23} \> \pageref{23} \\
\EQi{$[x_1 \.{:=} v_1,\ldots,x_n \.{:=} v_n]$} \> \eqr{3} \> \pageref{3} \\
\EQi{$x\ty{S}$} \> \eqr{2} \> \pageref{2} \\
\EQi{$y$} \> \eqr{1} \> \pageref{1} \\
\EQi{$z$} \> \eqr{1} \> \pageref{1} \\
\EQi{$Zip_{x,y,z}$} \> \eqr{42} \> \pageref{42} \\

\end{tabbing}

%\onecolumn
%
%\tableofcontents
%
%
%\newpage
%
%
%\listoffigures

\end{document}